\documentclass[prb,preprint,aps,showpacs,superscriptaddress,amsmath,amssymb]{revtex4-1}

\usepackage{amsmath}
\usepackage{amssymb}
\usepackage{amsthm}
\usepackage{amsfonts}
\usepackage{psfrag}
\usepackage{bm}
\usepackage{easybmat}
\usepackage[pdftex]{graphicx}
\usepackage{hyperref}
\hypersetup{colorlinks=true, linkcolor=blue, citecolor=blue, filecolor=blue, urlcolor=blue, pdftitle=, pdfauthor=, pdfsubject=, pdfkeywords=}

\def\ket#1{\left|#1 \right\rangle}
\def\bra#1{\left\langle #1 \right|}
\def\braket#1#2{\left\langle #1 | #2 \right\rangle}
\def\matrix22#1#2#3#4{\left(\begin{array}{cc}#1&#2\\#3&#4\end{array}\right)}

\begin{document}

\title{Entanglement Spectroscopy and its Application to the Quantum Hall Effects}
\author{N. Regnault}
\affiliation{Department of Physics, Princeton University, Princeton, NJ 08544
Laboratoire Pierre Aigrain, Ecole Normale Sup\'erieure-PSL Research University, CNRS, Universit\'e Pierre et Marie Curie-Sorbonne Universit\'es, Universit\'e Paris Diderot-Sorbonne Paris Cité\'e, 24 rue Lhomond, 75231 Paris Cedex 05, France}

\begin{abstract}
The entanglement spectroscopy, initially introduced by Li and Haldane in the context of the fractional quantum Hall effects, has stimulated an extensive range of studies. The entanglement spectrum is the spectrum of the reduced density matrix, when we partition the system into two. For many quantum systems, it unveils a unique feature: Computed from the bulk ground state wave function, the entanglement spectrum give access to the physics of edge excitations. Using this property, the entanglement spectroscopy has proved to be a highly valuable tool to diagnose topological ordering. 

These lectures intend to provide an overview of the entanglement spectroscopy, mainly in the context of the fractional quantum Hall effect. We introduce the basic concepts through the case of the quantum spin chains. We discuss the connection with the entanglement entropy and the matrix product state representation. We show how the entanglement spectrum can be computed for non-interacting topological phases and how it reveals the edge excitation from the ground state. We then present an extensive review of the entanglement spectra applied to the fractional quantum Hall phases, showing how much information is encoded within the ground state and how different partitions probe different type of excitations. Finally, we discuss the application of this tool to study the fractional Chern insulators.

These lectures were in parts held at:  
\begin{itemize}
\item
Les Houches Summer School ``Topological Aspects of Condensed Matter Physics", 
4--29~August 2014,
\'{E}cole de Physique des Houches, Les Houches, France
\item
SFT 2015 - ``Lectures on Statistical Field Theories'', 
2--13 February 2015,
The Galileo Galilei Institute for Theoretical Physics, Florence, Italy
\end{itemize}
\end{abstract}

\maketitle

\tableofcontents

\section{Introduction}

In the past decade it has become clear that Landau's theory of phase transitions which involves the appearance of a broken-symmetry order parameter does not apply to a series of phases of matter with so-called topological order. Topological phases exhibit the surprising property that their quantum ground state might be degenerate and that no local measurement can distinguish these degenerate states. This feature is the key of topological quantum computing: Using this robustness to solve the problem of local decoherence (for example due to disorder) by construction instead of quantum error correction (hardware vs software approach). This inherent robustness is also the source of a major issue: The absence of a local order parameter makes the identification of a topological order a difficult task.

Most of the intrinsic topologically ordered phases are strongly correlated systems. Thus numerical simulations have been an important tool to understand the emergence of these phases from microscopic models. The absence of local order parameter coupled to the few finite sizes that can be reached through simulations, restrict our ability to characterize these systems. From this perspective, we would like to achieve the two following goals:

\begin{itemize}
\item Convince ourself that the phase is indeed emerging. The more signatures we have, the better we might claim that we have strong evidence of this emergence.
\item Minimize our effort, meaning only compute the ground state of your system. This is actually more a limitation of the algorithm (or technique) or a limitation of  the computational power than a consequence of being lazy. If we work with quantum many body systems, the Hilbert space dimension grows exponentially with the system size (think about a spin $1/2$ system). This will be our bottleneck sooner or later (and most probably sooner).
\end{itemize}

So if we only have access to the ground state, we can wonder how much information about the system can be extracted. If we are lucky, we might have a model ground state we can compare with. The simplest way to compare the ground state from your simulation and this model ground state would be to compute an overlap (meaning a scalar product). Unfortunately this might be tricky to do (for example, the two states might be expressed in two different basis). Moreover, we end up with a number between 0 and 1, scaling with the system size. Thus we have to decide what is a good overlap.

Other routes would be to use a global order parameter, looking for the ground state degeneracy (playing with the genus of the surface you are working on, which may affect the ground state degeneracy for a topological phase), play with the boundary conditions (twist the boundary, flux insertion,...)

A promising tool to extract topological information from the ground state wave function is the entanglement entropy\cite{Levin-PhysRevLett.96.110405,Kitaev-PhysRevLett.96.110404,calabrese-04jsmp06002}.  The key idea is to break the system ground state into pieces and look at the entanglement between these pieces. In the simplest case, we consider the bipartite entanglement between two parts $A$ and $B$. As we will see, this technique will reveal lots of information about the phase itself such as its excitations. In many examples, this provide an in-depth view about the information encoded within the wave function of a topological phase. As a corollary, it means that we can store  a wave function in a more efficient way when we perform numerical simulations, just by keeping the relevant information.

In the simplest case, we consider the bipartite entanglement between two parts $A$ and $B$ of the system in its ground state $\ket{\Psi}$. This partition is characterized by the reduced density matrix $\rho_A = {\rm Tr}_B \ket{\Psi}\bra{\Psi}$ of subsystem $A$, obtained by tracing out all the $B$ degrees of freedom. Among the various entropies that have been considered as an entanglement measurement, the entanglement entropy is the most popular one (see Ref.~\cite{Amico-RevModPhys.80.517} for an extensive review). It is defined as the Von Neumann entropy associated with $\rho_A$ i.e. ${\mathcal S}_A= -{\rm Tr}_A \left[ \rho_A \ln \rho_A \right]$. For a system in $d$ dimensions with a finite correlation length $l$, the entanglement entropy satisfies the area law\cite{Srednicki-PhysRevLett.71.666}
\begin{eqnarray}
{\mathcal S}_A&\simeq&\alpha {\mathcal L}^{d-1}\label{eq:AreaLaw}
\end{eqnarray}
${\mathcal L} \gg l$ is the typical length that defines the size of the region $A$ and $\alpha$ is a non-universal constant. The area law indicates that the dominant part of the entanglement entropy is controlled by the area (${\mathcal L}^{d-1}$) that separates the two domains. Physically, it means that the entanglement between $A$ and $B$ is located at the interface of the two regions.

For two dimensional topological phases, Refs.~\cite{Levin-PhysRevLett.96.110405} and \cite{Kitaev-PhysRevLett.96.110404} showed that the first correction to the area law is a topological term: ${\mathcal S}_A \sim \alpha {\mathcal L} - \gamma$. The sub-leading term $\gamma$ is called the topological entanglement entropy: It is a constant for a given topologically ordered phase, $\gamma = \ln {\mathcal D}$. Here ${\mathcal D}$ is the total quantum dimension characterizing the topological field theory describing the phase and thus the nature of the system excitations. The topological entanglement entropy appears as a way to characterize the topological order of a phase. However, its practical calculation depends on scaling arguments, which might be hard to obtain to sufficient accuracy from numerical calculations\cite{Haque-PhysRevLett.98.060401,lauchli-10njp075004}. Moreover, it does not uniquely determine the topological order in the state.

While the topological entanglement entropy compresses the information contained in the reduced density matrix into a single number, the concept of entanglement spectroscopy have been shown to be a powerful tool to probe the topological order. Indeed, the entanglement spectrum (ES) aims to have a deeper look at $\rho_A$ by analyzing its full spectrum. The ES have been initially introduced by Li and Haldane\cite{li-08prl010504} in the context of the Fractional Quantum Hall Effect (FQHE), stimulating an extensive range of studies.\cite{Regnault-PhysRevLett.103.016801,Zozulya-PhysRevB.79.045409,lauchli-PhysRevLett.104.156404,Thomale-PhysRevLett.104.180502,Thomale-PhysRevB.84.045127,Hermanns-PhysRevB.84.121309,Chandran-PhysRevB.84.205136,Ardonne-PhysRevB.84.205134,Sterdyniak-NewJourPhys-13-10-105001,Liu-PhysRevB.85.045119,Sterdyniak-PhysRevB.85.125308,Dubail:2012p2980,Rodriguez-PhysRevLett.108.256806,Schliemann-PhysRevB.83.115322,Alba-PhysRevLett.108.227201} They have also been studied and applied to a large scale of topological and non-topological phases including spin systems,\cite{Calabrese-PhysRevA.78.032329,Pollmann-1367-2630-12-2-025006,Pollmann-PhysRevB.81.064439,Thomale-PhysRevLett.105.116805,Lauchli-PhysRevB.85.054403,Yao-PhysRevLett.105.080501,Cirac-PhysRevB.83.245134,Peschel-0295-5075-96-5-50006,Huang-PhysRevB.84.125110,Lou-PhysRevB.84.245128,Schliemann-jstatmech-2012-11-11021} as well as topological insulators,\cite{Fidkowski-PhysRevLett.104.130502,Prodan-PhysRevLett.105.115501,Turner-PhysRevB.82.241102} Bose-Hubbard models\cite{Liu-PhysRevA.83.013620} or complex paired superfluids.\cite{Dubail-PhysRevLett.107.157001}. Moreover, the partition of the system has to be thought in a broad sense: It can be done in the real space, in the momentum or Fourier space, or in the particle space. For many model states such as the Laughlin wave function\cite{Laughlin:1983p301} or the AKLT spin chain\cite{Affleck-PhysRevLett.59.799,Affleck-CommMathPhys.115.3}, the counting (number of non-zero eigenvalues) is exponentially lower than expected. This counting is related to the nature of the system excitations. The salient feature is that this information about the excitations is obtained only from the ground state. The ES is a way to extract this information and each type of cut reveals different aspects of these excitations.

These lecture notes will try to give an overview of the entanglement spectroscopy, devoting a large part to its application to the FQHE and similar phases. The extensive studies of the ES for these phases and the detailed understanding of ES that have been gained through these studies, motivate this bias. These notes is organized as follow. Sec.~\ref{sec:EntanglementSpectrum} provides an introduction to the notations and concepts of the entanglement spectra. We exemplify these notions on simple spin systems and relate ES to the matrix product state (MPS) representation. In Sec.~\ref{sec:EdgeandES}, we show how the chiral edge mode can be observed from the ES of a non-interacting system, illustrating this feature on the integer quantum Hall effect (IQHE) and the Chern insulator (the simplest example of a topological insulator). In Sec.~\ref{sec:FQHEandES}, we present an extensive overview of the ES for the fractional quantum Hall effect. We show the different bipartite partitions that were considered for these systems and the kind of information that were revealed performing the ES. Finally in Sec.~\ref{sec:FCI}, we discuss how the ES was used as tool to probe the phases that emerge in fractional Chern insulators (FCI).

\section{Entanglement spectrum and entanglement entropy}\label{sec:EntanglementSpectrum}

As a first step, we discuss the concept of entanglement spectroscopy in some simple cases. We also briefly cover the definition and the relevant properties of the entanglement entropy. We introduce the Li-Haldane conjecture in the case of the AKLT spin chain. We discuss the important situation where the number of reduced density matrix non-zero eigenvalues is massively reduced. In particular, we show the relation between the latter property and the matrix product state representation.

\subsection{Definitions}\label{subsubsection:definitions}

Let consider a generic $n$-body quantum state $\ket{\Psi}$ that can be decomposed on the orthonormal basis $\{\ket{\lambda}\}$. We now assume that this basis can be written as the tensor product of two orthonormal basis $\{\ket{\mu_A}\}$ and $\{\ket{\mu_B}\}$ i.e. $\{\ket{\lambda}=\ket{\mu_A}\otimes \ket{\mu_B}\}$, providing a natural bipartition of the system into $A$ and $B$. The decomposition of the state $\ket{\Psi}$ reads
\begin{eqnarray}
\ket{\Psi}&=&\sum_{\mu_A, \mu_B} c_{\mu_A, \mu_B}\ket{\mu_A}\otimes \ket{\mu_B}\label{eq:GenericPsiDecomposition}
\end{eqnarray}
The entanglement matrix $M$ is defined such that its matrix elements are given by $M_{\mu_A, \mu_B}=c_{\mu_A, \mu_B}$. The size of $M$ is given by the dimension of the subspaces $A$ and $B$ that we denote respectively ${\rm dim}_A$ and ${\rm dim}_B$. Note that we do not assume that ${\rm dim}_A={\rm dim}_B$, and thus $M$ is generically a rectangular matrix. One can perform a singular value decomposition (SVD) of $M$. The SVD allows to write a rectangular matrix 
\begin{eqnarray}
M&=&U D V^{\dagger} \label{eq:DefSVD}
\end{eqnarray}
where $U$ is a ${\rm dim}_A \times {\rm min}\left({\rm dim}_A,{\rm dim}_B\right)$ matrix which satisfies $U^{\dagger} U=1$ (i.e. has orthonormalized columns), $V$ is a ${\rm dim}_B \times {\rm min}\left({\rm dim}_A,{\rm dim}_B\right)$ matrix which satisfies $V V^{\dagger}=1$ (i.e. has orthonormalized rows). $D$ is a diagonal square of dimension ${\rm min}\left({\rm dim}_A,{\rm dim}_B\right)$ where all entries are non-negative and can be expressed as $\{e^{-\xi_i/2}\}$ .

Using the SVD, one can derive the Schmidt decomposition of $\ket{\Psi}$
\begin{eqnarray}
\ket{\Psi}&=&\sum_{i=1}^{{\rm min}\left({\rm dim}_A,{\rm dim}_B\right)} e^{-\xi_i/2} \ket{A:i}\otimes \ket{B:i} \label{eq:SchmidtDecomposition}
\end{eqnarray}
where 
\begin{eqnarray}
&&\ket{A:i}=\sum_{\mu_A} U^{\dagger}_{i,\mu_A}\ket{\mu_A}\label{eq:DefinitionAi}\\
{\rm  and}&\;\; & \ket{B:i}=\sum_{\mu_B} V^{\dagger}_{i,\mu_B}\ket{\mu_B}\label{eq:DefinitionBi}
\end{eqnarray}
To be a Schmidt decomposition, the states $\ket{A:i}$ and $\ket{B:i}$ have to obey $\braket{A:i}{A:j}=\braket{B:i}{B:j}=\delta_{i,j}$. This property is trivially verified using the identities on $U$ and $V$.
The Schmidt decomposition provides a nice and numerically efficient way to compute the spectrum of the reduced density matrix. Consider the density matrix of the pure state $\rho=\ket{\Psi}\bra{\Psi}$, we compute the reduced density matrix of $A$ by tracing out the degree of freedom related to $B$, i.e. $\rho_A={\rm Tr}_B \rho$. Using Eq.~\ref{eq:SchmidtDecomposition}, we deduce that 
\begin{eqnarray}
\rho_A&=&\sum_{i} e^{-\xi_i} \ket{A:i}\bra{A:i}\label{eq:ReducedDensityMatrixA}
\end{eqnarray}
Thus the spectrum of $\rho_A$ can be obtained from the coefficient of the Schmidt decomposition or the SVD of the entanglement matrix and is given by the set $\{e^{-\xi_i}\}$. From a numerical perspective, getting the spectrum of $\rho_A$ is more accurate using the SVD of $M$ than a brute force calculation of $\rho_A$ in the $\{\ket{\mu_A}\}$ basis followed by its diagonalization. In a similar way, we can obtain the reduced density matrix of $B$
\begin{eqnarray}
\rho_B={\rm Tr}_A \rho = \sum_i e^{-\xi_i} \ket{B:i}\bra{B:i}\label{eq:ReducedDensityMatrixB}
\end{eqnarray}
Note that $\rho_A$ and $\rho_B$ have the same spectrum. While these two square matrices might have different dimensions (respectively ${\rm dim}_A$ and ${\rm dim}_B$), they both have the same number of non-zero eigenvalues. This number has to be lower than or equal to ${\rm min}\left({\rm dim}_A,{\rm dim}_B\right)$. Thus studying the properties of $\rho_A$ for various partitions (i.e. choices of $A$ and $B$) can be restricted to the cases where ${\rm dim}_A \leq {\rm dim}_B$.

With these tools and properties, we can now define the entanglement spectrum. The latter corresponds to the set $\{\xi_i\}$, the logarithm of the reduced density matrix eigenvalues. The key idea of the original article of Li and Haldane\cite{li-08prl010504} was not only to look at this whole spectrum, but at a specific subset of these values (or a block of $\rho_A$) with well defined quantum numbers. Assume an operator ${\mathcal O}$ that can be decomposed as ${\mathcal O}_A+{\mathcal O}_B$ where ${\mathcal O}_A$ (resp. ${\mathcal O}_B$) only acts on the $A$ (resp. $B$) subspace. One can think about ${\mathcal O}$ as the projection of the spin operator or the momentum. If $[{\mathcal O}, \rho]=0$, we also have $0=\text{Tr}_B [{\mathcal O}_A, \rho] + \text{Tr}_B [{\mathcal O}_B, \rho] =[{\mathcal O}_A, \text{Tr}_B \rho]= [{\mathcal O}_A, \rho_A] $ as the trace over the $B$ degrees of freedom of a commutator operator in the $B$ part vanishes. If $\ket{\Psi}$ is an eigenstate of ${\mathcal O}$, then the latter commutes with $\rho$. We can simultaneously diagonalize $\rho_A$ and ${\mathcal O}_A$ ,and label the $\{\xi_i\}$ according to the quantum number of ${\mathcal O}_A$.

\subsection{A simple example: Two spin-$\frac{1}{2}$}\label{subsubsection:spin12}

\begin{figure}
\centering
\includegraphics[width=0.2\linewidth]{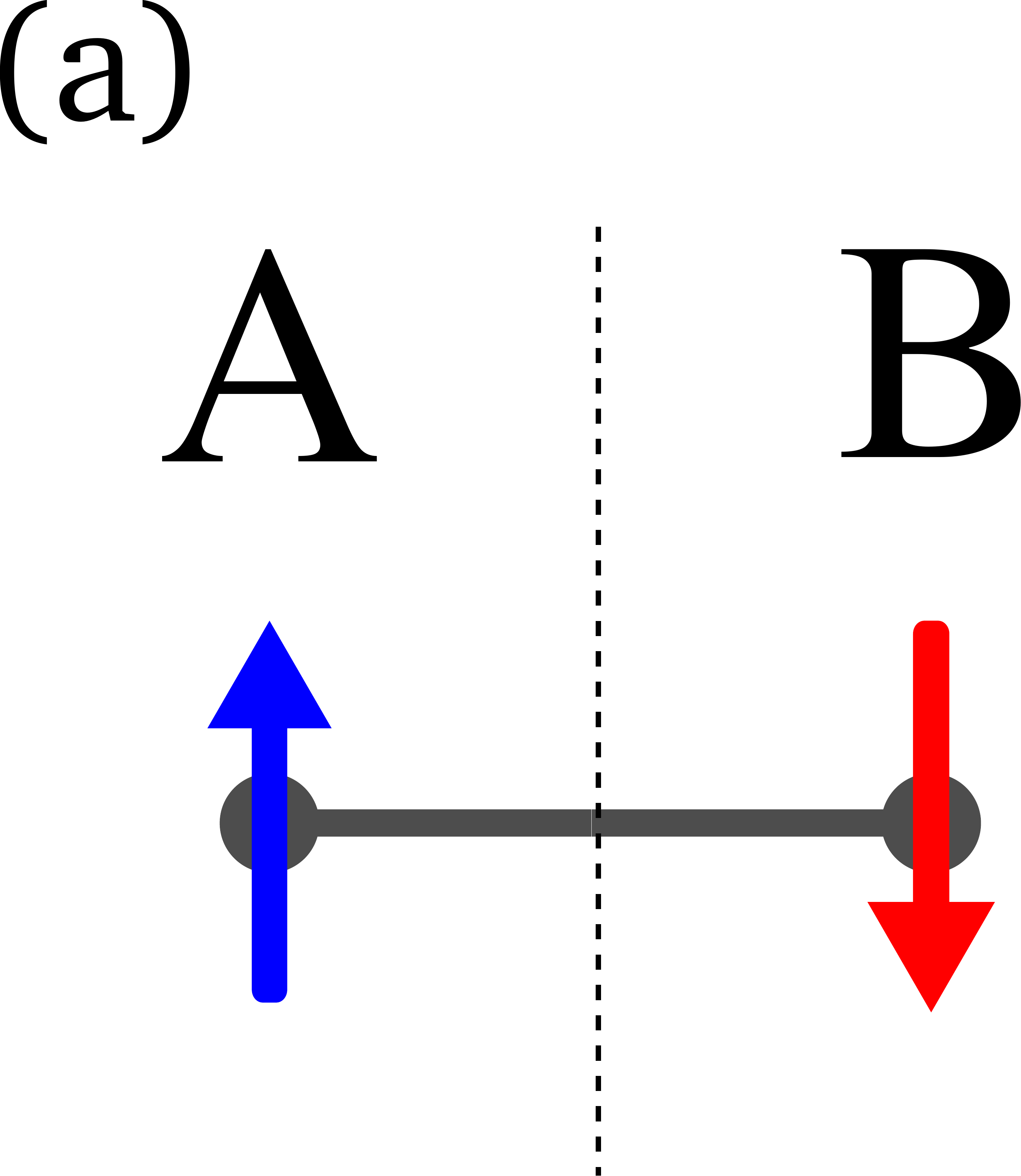}
\includegraphics[width=0.25\linewidth]{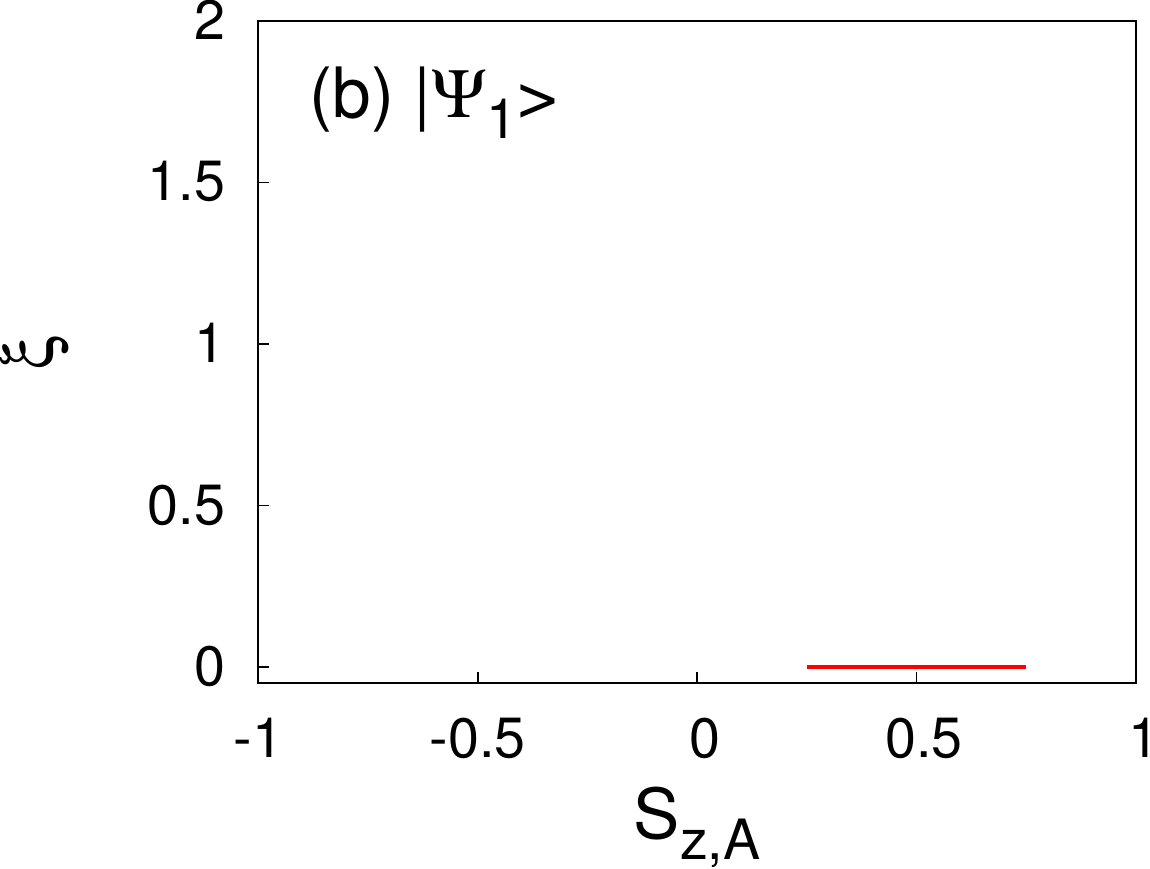}
\includegraphics[width=0.25\linewidth]{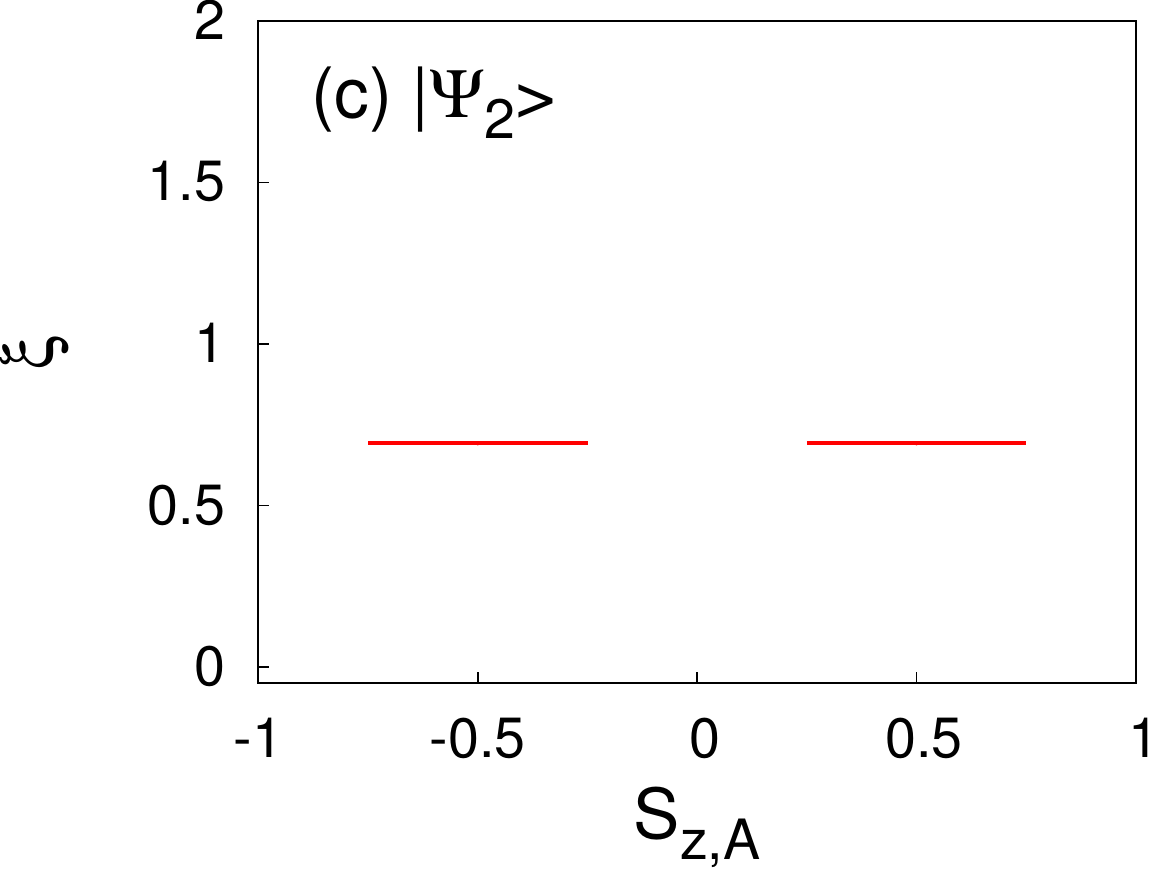}
\includegraphics[width=0.25\linewidth]{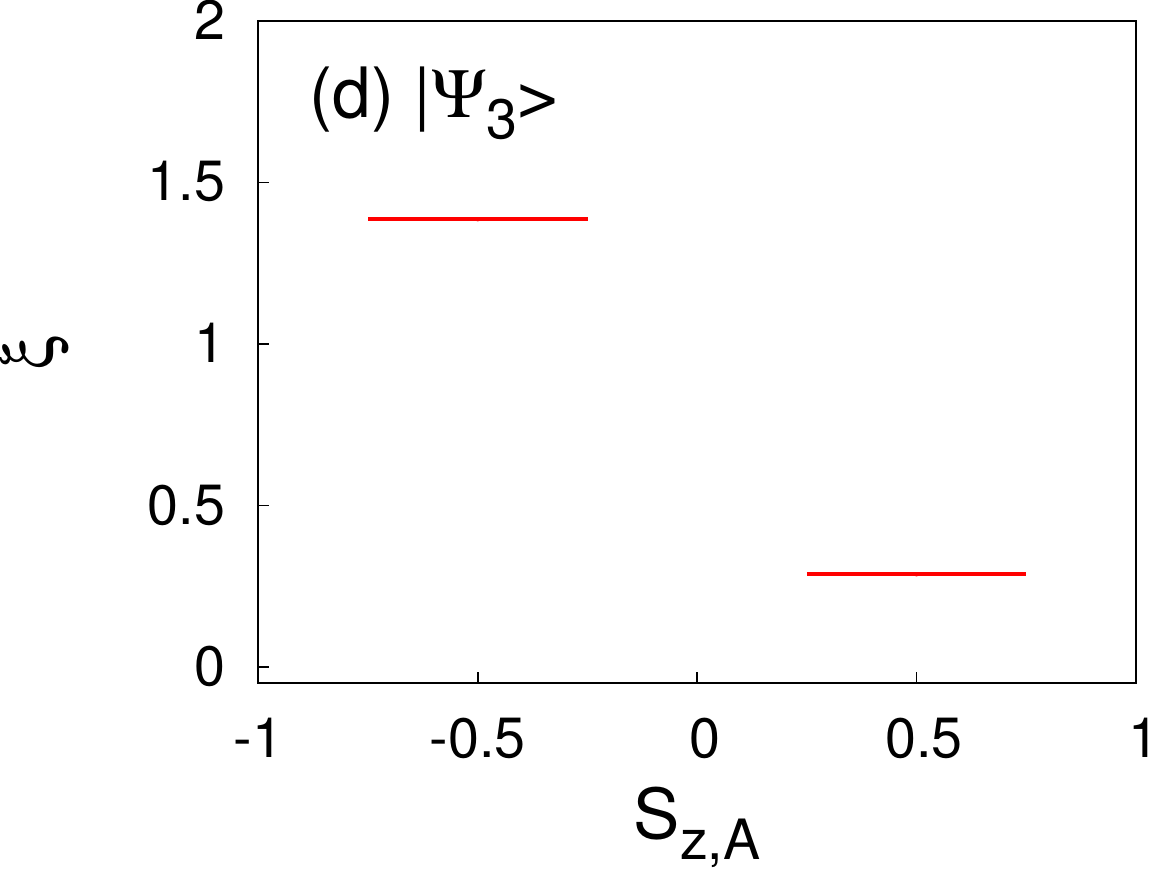}
\caption{{\it From left to right:} (a) schematic picture of the two spin-$\frac{1}{2}$ system. (b) Entanglement spectrum for the state $\ket{\Psi_1}=\ket{\uparrow\uparrow}$. (c) Entanglement spectrum for the state $\ket{\Psi_2}=\frac{1}{\sqrt{2}}\left(\ket{\uparrow\downarrow} - \ket{\downarrow\uparrow}\right)$. (d) Entanglement spectrum for the state $\ket{\Psi_3}=\frac{1}{2}\ket{\uparrow\downarrow} + \frac{\sqrt{3}}{2}\ket{\downarrow\uparrow}$.}\label{fig:spin12}
\end{figure}

To exemplify the previous notations and concepts, we consider a system of two spin-$\frac{1}{2}$ as depicted in Fig.~\ref{fig:spin12}a. Any state $\ket{\Psi}$ can be decomposed onto the four basis states:
\begin{eqnarray}
\ket{\Psi}&=&c_{\uparrow \uparrow}\ket{\uparrow \uparrow}+c_{\uparrow \downarrow}\ket{\uparrow \downarrow}
+c_{\downarrow \uparrow}\ket{\downarrow \uparrow}+c_{\downarrow \downarrow}\ket{\downarrow \downarrow}\label{eq:PsiSpin12}
\end{eqnarray}

A natural way to cut this system into two parts consists of the $A$ (resp. $B$) part being the left (resp. right) spin. The entanglement matrix is given by
\begin{eqnarray}
M&=& \begin{BMAT}{cc}{cc}
\begin{BMAT}(b){cc}{c}
 \ket{B:\uparrow}&\ket{B:\downarrow}
\end{BMAT}
&
\\
\left(\begin{BMAT}(b){cc}{cc}
c_{\uparrow \uparrow}&c_{\uparrow \downarrow}\\c_{\downarrow \uparrow} & c_{\downarrow \downarrow}\\
\end{BMAT}\right)
&
\begin{BMAT}(b){c}{cc}
\ket{A:\uparrow} \\  \ket{A:\downarrow}\\
\end{BMAT}
 \end{BMAT}
\label{eq:PsiSpin12EntMatrix}
\end{eqnarray}
where we have explicitly written which states were associated with each row and column of $M$. We consider three examples: A product state $\ket{\Psi_1}=\ket{\uparrow\uparrow}$, a maximally entangled state $\ket{\Psi_2}=\frac{1}{\sqrt{2}}\left(\ket{\uparrow\downarrow} - \ket{\downarrow\uparrow}\right)$ and a generic entangled state $\ket{\Psi_3}=\frac{1}{2}\ket{\uparrow\downarrow} + \frac{\sqrt{3}}{2}\ket{\downarrow\uparrow}$. The entanglement matrices for these three states are 

\begin{eqnarray}
M_1=\matrix22{1}{0}{0}{0},&M_2=\matrix22{0}{\frac{1}{\sqrt{2}}}{-\frac{1}{\sqrt{2}}}{0},&M_3=\matrix22{0}{\frac{1}{2}}{\frac{\sqrt{3}}{2}}{0}\label{eq:EntMatrixExamples}
\end{eqnarray}

Performing the SVD on the first state $\ket{\Psi_1}$ is trivial: Being a product state, it is already written as a Schmidt decomposition. For $\ket{\Psi_2}$, we can do the SVD 

\begin{eqnarray}
M_2&=&\matrix22{1}{0}{0}{1} \matrix22{\frac{1}{\sqrt{2}}}{0}{0}{\frac{1}{\sqrt{2}}} \matrix22{0}{-1}{1}{0} \label{eq:SVDPsi2}
\end{eqnarray}
such that the Schmidt decomposition is 
\begin{eqnarray}
\ket{\Psi_2}&=&\frac{1}{\sqrt{2}}\left(+\ket{\uparrow}\right)\otimes\left(+\ket{\downarrow}\right)\label{eq:SchmidtDecompositionPsi2}\\
&&+\frac{1}{\sqrt{2}}\left(+\ket{\downarrow}\right)\otimes\left(-\ket{\uparrow}\right)\nonumber
\end{eqnarray}
A similar calculation can be performed for $\ket{\Psi_3}$.

The projection of the total spin along the $z$ axis $S_z$ is the sum of individual components $S_{z,A}$ and $S_{z,B}$. Thus, when performing the cut into the two parts $A$ and $B$, $S_{z,A}$ is a good quantum number that can be used to label the eigenvalues of the entanglement spectrum according to the discussion in Sec.~\ref{subsubsection:definitions}. The entanglement spectra for the three states $\ket{\Psi_1}$, $\ket{\Psi_2}$ and $\ket{\Psi_3}$ are shown in Figs.~\ref{fig:spin12}b-d. For the product state $\ket{\Psi_1}$, there is a single level appearing since the reduced density matrix has a single non-zero eigenvalue. For the two other examples, there are two levels each with a given $S_{z,A}$ value. The calculation of the entanglement entropy, which is a measure of the entanglement, directly tells that $\ket{\Psi_1}$ is a product state. We can derive the same conclusion from the number of levels in the entanglement spectrum. While this example is rather a trivial result obtained from the entanglement spectrum, it stresses one of strong points of this technique. Some properties of the states can be deduced just by counting the non-zero eigenvalues of reduced density matrix.

\subsection{Entanglement entropy}\label{subsubsection:ententropy}

They are several ways to quantify the entanglement between two parts of a system and there is an extensive literature on this topic (see Ref.~\cite{Amico-RevModPhys.80.517} for an extensive review). The goal of these lectures is not to give a detailed introduction to entanglement entropies. So we will restrict to a few useful examples in the context of topological phases. Perhaps the most common measure of entanglement is the Von Neumann entanglement entropy

\begin{eqnarray}
{\mathcal S}_A&=& -{\rm Tr}_A \left[ \rho_A \ln \rho_A \right]\label{eq:VonNeumannEntropy}
\end{eqnarray}

From a practical point of view, the calculation of the Von Neumann entanglement entropy can be easily obtained once the Schmidt decomposition or the spectrum of the reduced density matrix has been obtained.

\begin{eqnarray}
{\mathcal S}_A&=-\sum_i \lambda_i \ln \lambda_i &=\sum_i \xi_i e^{-\xi_i}\label{eq:VonNeumannEntropyFromXi}
\end{eqnarray}

Similarly, we can define the entanglement entropy for the $B$ part of the system $S_B=-{\rm Tr}_B \left[ \rho_B \ln \rho_B \right]$. Using Eq.~\ref{eq:ReducedDensityMatrixB}, we immediately see that $S_A=S_B$. If $A$ and $B$ are not entangled (i.e. $\ket{\Psi}=\ket{\Psi_A} \otimes \ket{\Psi_B}$), we get $S_A=0$. For the full system $A+B$, the entanglement entropy is also zero. As a consequence, we have in general that $S_A+S_B \ne S_{A+B}$ (the entanglement entropy is actually strongly subadditive)

We will now turn to the entanglement entropy of some specific systems. In many situations, it is useful to look at the case of a random state. Especially for people interested in numerical simulations, it is always a good idea to compare with what a random output would give. For example, consider the calculation of the overlap (the simplest way to compare two wave functions). Let's take two random states $\ket{\Psi_1}$ and $\ket{\Psi_2}$ defined in a Hilbert space of dimension ${\mathcal D}$. Then the average overlap $|\braket{\Psi_1}{\Psi_2}|^2\simeq \frac{1}{\mathcal D}$. This result gives a simple bound for what is a bad overlap in finite systems (note that one should not cheat and define ${\mathcal D}$ as the dimension of the Hilbert space with all the symmetries the system has).

For the entanglement entropy, we remind the notations ${\rm dim}_A$ for the dimension of the Hilbert associated to the $A$
part and ${\rm dim}_B$ for the dimension of the Hilbert space of the $B$ part. In the limit
${\rm dim}_B \ge {\rm dim}_A \gg 1$, it was shown\cite{Page-PhysRevLett.71.1291} that 
\begin{eqnarray}
S_A\simeq \ln \left({\rm dim}_A\right) - \frac{{\rm dim}_A}{2 \; {\rm dim}_B}\label{eq:EntEntropyRandom}
\end{eqnarray}
In particular when ${\rm dim}_B \gg {\rm dim}_A \gg 1$, we obtain that $S_A \simeq \ln \left({\rm dim}_A\right)$.

To get a more physical picture of this formula, we can consider that the system
is made of spin-$\frac{1}{2}$, $V_A$ spin-$\frac{1}{2}$ for $A$ and $V_B$ spin-$\frac{1}{2}$ for $B$. The Hilbert space dimensions are ${\rm dim}_A = 2^{V_A}$ and ${\rm dim}_B = 2^{V_B}$, leading to $S_A \simeq V_A \ln 2$. Thus for a random state, the entanglement entropy is proportional to the volume of the subsystem $A$, meaning the entanglement entropy obeys a volume law.

\begin{figure}
\centering
\includegraphics[width=0.95\linewidth]{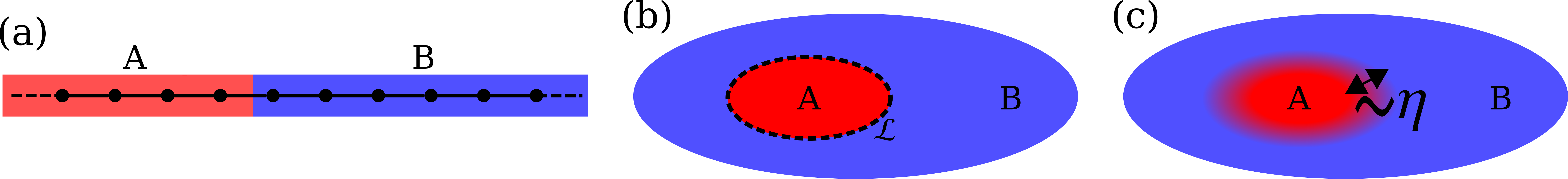}
\caption{A schematic description of the bipartite geometrical partition for a one dimensional system (a) and for a two dimensional system (b). The rightmost panel (c) illustrates the small region around the boundary between $A$ and $B$ (with a thickness of the order of the correlation length $\eta$) that is relevant in the entanglement entropy when considering a gapped phase.}\label{fig:1d2dcut}
\end{figure}

We can now move to the case of gapped phases. We note $\eta$ the correlation length. We consider a geometrical bipartition of the system into $A$ and $B$ as depicted in Fig.~\ref{fig:1d2dcut}. For one dimensional gapped systems,  if the size of $A$ in large enough compared to $\eta$, the entanglement entropy does not depend on the length $V_A$, i. e. $S_A$ is constant. This statement can be proved and an upper bound on the constant can be found\cite{Hastings-1742-5468-2007-08-P08024}.

For higher dimensional systems, it is conjectured that the entanglement entropy (see e.g. Ref.~\cite{Eisert-RevModPhys.82.277} for an extensive discussion) satisfies

\begin{eqnarray}
{\mathcal S}_A&\simeq&\alpha {\mathcal L}\label{eq:AreaLaw2}
\end{eqnarray}

${\mathcal L}\gg \eta$ denotes the area of the surface that separates $A$ from $B$ and $\alpha$ is a constant. Thus the entanglement entropy for a gapped system satisfies an area law (as opposed to the volume law of the random state). In two dimension, ${\mathcal L}$ is just the perimeter of the boundary between $A$ and $B$ (see Fig.~\ref{fig:1d2dcut}b). Here we should make two remarks. First, one dimensional gapped systems also obey the area law (just set ${\mathcal L}$ to $1$, the boundary being just a point). Second, this is a major difference with a random state where one gets a volume law for the entanglement entropy. Intuitively, if one has a finite correlation length, we expect that only the region around the boundary between $A$ and $B$, whose thickness is of the order of a few $\eta$'s (as shown in Fig.~\ref{fig:1d2dcut}c) should matter in the entanglement between $A$ and $B$.

For two dimensional topological phases, we can go beyond the area law contribution. Refs.~\cite{Levin-PhysRevLett.96.110405} and \cite{Kitaev-PhysRevLett.96.110404} showed that the first correction to this area law is a constant term $\gamma$ 

\begin{eqnarray}
{\mathcal S}_A \sim \alpha {\mathcal L} - \gamma\label{eq:AreaLawTopological2dphase}
\end{eqnarray}

While $\alpha$ is non-universal, this is not the case the sub-leading term $\gamma$. This latest is called the topological entanglement entropy: It is a constant for a given topologically ordered phase

\begin{eqnarray}
\gamma &=& \ln \left(\frac{\mathcal D}{d_a}\right)\label{eq:DefTopologicalEntEntropy}
\end{eqnarray}

For a given type of excitations $a$, the quantum dimension $d_a$ defines how the Hilbert space dimension exponentially increases with the number of such excitations. Each type of excitations corresponds to a topological sector. Abelian excitations have a quantum dimension equal to $1$ while non-abelian ones have $d_a > 1$. The total quantum dimension is given by ${\mathcal D}=\sqrt{\sum_a d_a^2}$. These quantum dimensions characterize the topological field theory describing the phase and thus the nature of the system excitations. Note that in Eq.~\ref{eq:DefTopologicalEntEntropy}, the $a$ of the $d_a$ term corresponds to the topological sector of the wave function $\ket{\Psi}$ whose entanglement entropy is computed.

The topological entanglement entropy appears as a way to characterize the topological order of a phase. However, its practical calculation depends on scaling arguments, which might be hard to obtain to sufficient accuracy from numerical calculations\cite{Haque-PhysRevLett.98.060401,lauchli-10njp075004}. Moreover, it does not uniquely determine the topological order in the state. For that reason, it is interesting to look at the full spectrum of the reduced density matrix and not to reduce it to a single number.

\subsection{The AKLT spin chain}\label{subsubsection:spinAKLT}

We now move to a typical example of strongly correlated $n$-body quantum systems: The quantum spin chains. One of the simplest example of a strongly correlated gapped system is the antiferromagnetic spin-$1$ chain. It is also one of the simple example of a symmetry protected topological phase (here protected by spin rotation or time reversal symmetries, see e.g. Ref.~\cite{Senthil-annurev-conmatphys-031214-014740} for a short review) . More precisely, we focus on the Affleck-Kennedy-Lieb-Tasaki (AKLT) model\cite{Affleck-PhysRevLett.59.799,Affleck-CommMathPhys.115.3}. This system is the prototype of a gapped spin-$1$ chain\cite{Haldane-PhysRevLett.50.1153}. The AKLT Hamiltonian of the one dimensional spin-1 chain reads:
\begin{eqnarray}
H_{\rm AKLT}&=&\sum_{j}\vec{S}_j \cdot \vec{S}_{j + 1}\; +\; \frac{1}{3} \sum_{j} \left( \vec{S}_j \cdot \vec{S}_{j + 1}\right)^2\label{eq:AKLTHamiltonian}
\end{eqnarray}

\begin{figure}
\centering
\includegraphics[width=0.7\linewidth]{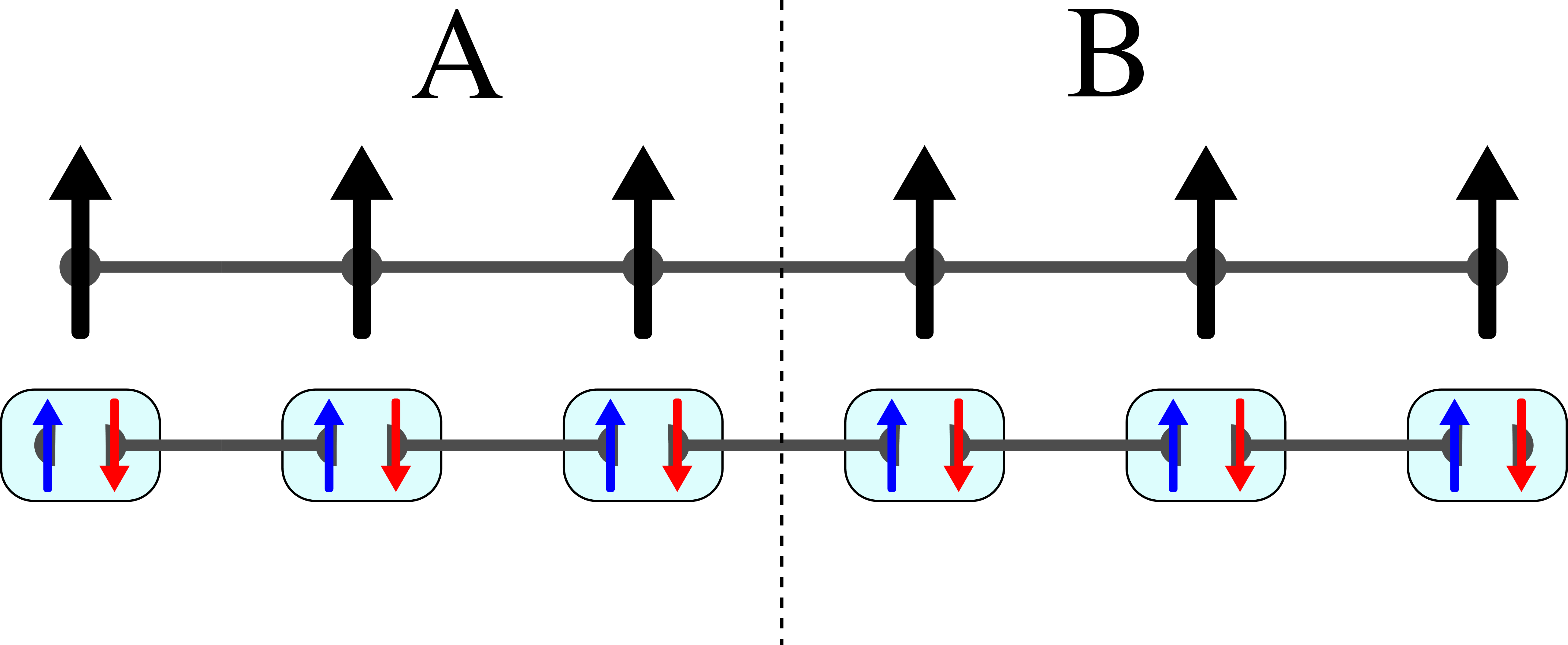}
\caption{A schematic description of the AKLT ground state. The upper chain shows the spin-1 AKLT chain while the lower chain exhibits its valence bond description. Each spin-$1$ is decomposed into two spin-$\frac{1}{2}$, one red and one blue, that are projected on the triplet state (depicted by a box). The AKLT ground state is obtained by projecting one red spin-$\frac{1}{2}$ of one site with one blue spin-$\frac{1}{2}$ of the neighboring site in the singlet state. We observed the two unpaired spin-$\frac{1}{2}$, one at each end of the spin chain. This figure also shows how the system is cut into two parts $A$ and $B$ when performing the entanglement spectrum calculation.}\label{fig:AKLT}
\end{figure}

The ground state of the AKLT Hamiltonian is a valence bond state. It can be understood within a simple picture sketched in Fig.~\ref{fig:AKLT}. Each spin-$1$ can be written as two spin-$\frac{1}{2}$ combined in the triplet state. Between two neighboring sites, two of the four spin-$\frac{1}{2}$ (one per site) are combined in the singlet sector. When an open chain is considered, the two extreme unpaired spin-$\frac{1}{2}$ (see Fig.~\ref{fig:AKLT}) correspond to the edge excitations, leading to a four-fold degenerate ground state (one singlet state and one triplet state).

To compute the entanglement spectrum of the AKLT ground state for an open chain, we first have to decide which of the four degenerate state we would like to analyze. In the sector of total spin $S_z=\pm 1$, there is only one state so the choice is simple, while in the sector $S_z=0$, there are two states. For sake of simplicity we focus on the $S_z=1$ case. To cut the system into two parts, we can follow the same procedure than the one described in the Sec.~\ref{subsubsection:spin12}: The $A$ part will be made of the $l_A$ consecutive leftmost sites and the $B$ part by the remaining rightmost sites (see Fig.~\ref{fig:AKLT}). 

\begin{figure}
\centering
\includegraphics[width=0.9\linewidth]{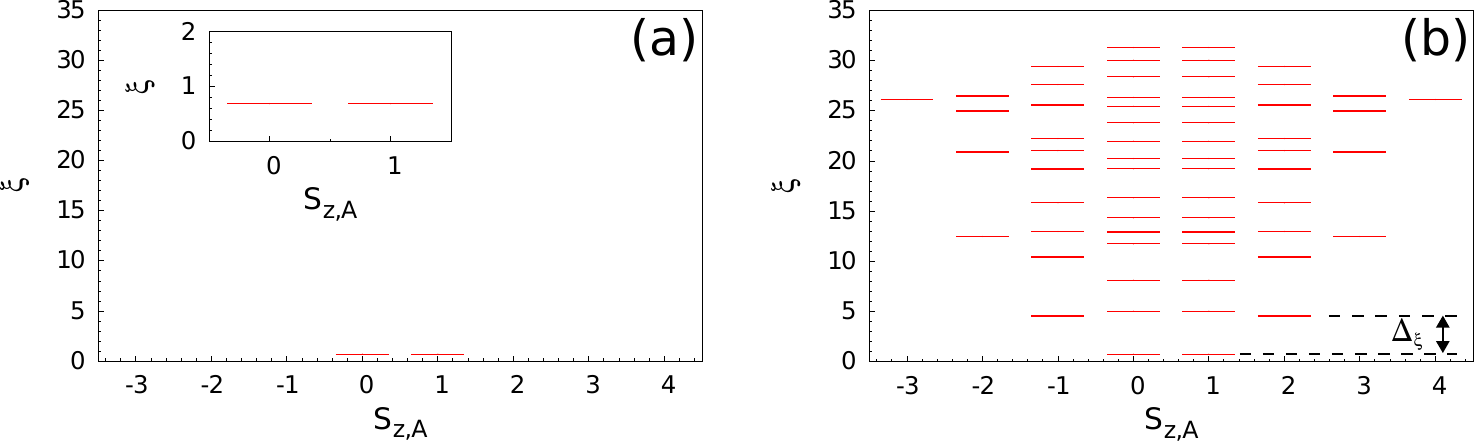}
\caption{{\it Left panel:} The entanglement spectrum for the AKLT ground state with $8$ sites in the $S_z=1$ sector. The system is cut into two equal parts of size $l_A=4$. The entanglement spectrum only contains two levels, i.e. two non zero eigenvalues in the reduced density matrix. This reflects the edge excitation (a spin-$\frac{1}{2}$) of the AKLT ground state. The inset is a zoom on these two levels. {\it Right panel:} The entanglement spectrum for the Heisenberg spin-$1$ chain with $8$ sites in the $S_z=1$ sector. The system is cut in a similar way to the AKLT case. The two lowest entanglement energy states (i.e. the two largest eigenvalue of the reduced density matrix) are similar to those of the AKLT ground state. We show the entanglement gap $\Delta_\xi$ between the similar to the part of entanglement spectrum AKLT ground state and the higher the entanglement energy states.}\label{fig:AKLTES}
\end{figure}

Fig.~\ref{fig:AKLTES}a displays the entanglement spectrum for a AKLT open chain with $8$ sites and $l_A=4$. The entanglement energies $\xi$ are plotted versus $S_{z,A}$, the $z$-projection of the $A$ part total spin. The reduced density matrix has only two non-zero eigenvalues whereas the size of reduced density matrix is $81 \times 81$. This dramatic reduction of the number of non-zero eigenvalues compared to a random state, is a major characteristic that we will observe for many model states. If one thinks about the cut as an artificial edge we have introduced in the system, the physical interpretation becomes obvious: What we observe here is a spin-$\frac{1}{2}$ edge excitation of the AKLT chain. This is the first example where the Li-Haldane conjecture\cite{li-08prl010504} can be observed: For this gapped phase, the entanglement spectrum is directly related to the spectrum of the edge excitation. Note that the true edge excitations of the system do not play any role here since our choice of the AKLT ground state in the $S_z=1$ sector freezes these excitations. 

The AKLT Hamiltonian being the prototype of the gapped quantum spin-$1$ chain, it is interesting to look at the behavior entanglement spectrum away from this specific case. The simplest case one can consider is the Heisenberg spin-$1$ chain where the Hamiltonian is just given by $H=\sum_{j}\vec{S}_j \cdot \vec{S}_{j + 1}$. In Fig.~\ref{fig:AKLTES}b, we consider a similar situation than the one for the AKLT model of Fig.~\ref{fig:AKLTES}a. At the bottom of the entanglement spectrum, we recover two states, with the same quantum numbers than the AKLT case. Contrary to the latter, we also observe some higher entanglement energy levels. The AKLT ground state and the Heisenberg spin-$1$ chain being adiabatically connected, we would like to argue that the low entanglement energy structure in the Heisenberg spin-$1$ entanglement spectrum will characterize the system. We define the entanglement gap $\Delta_\xi$ as the minimum difference of entanglement energy level between the low entanglement energy structure similar to a model state (the AKLT model in this example) and the entanglement energy levels above this structure. The meaning of $\Delta_\xi$ is actually the Li-Haldane conjecture away from model states: If this entanglement gap stays finite in the thermodynamical limit, the edge excitations of the system will be in the same universality class than the model state whose entanglement spectrum reduces to the same low entanglement energy structure. Indeed numerical simulations based on the density matrix renormalization group\cite{capponi} have shown such a property for the Heisenberg model.

To summarize, this example has already been able to show us several features of the entanglement spectrum. For some model states, the number of non-zero eigenvalues might be related to the edge excitation of the system. This number can be exponentially smaller than one can expect from a random state, which is a non-trivial signature. Away from this ideal situation and as long as we stay in the same universality class, we would expect to observe a similar fingerprint than the model state in the low entanglement energy part of the spectrum. This structure should be protected from the higher entanglement energy levels by an entanglement gap.

\subsection{Matrix Product States and entanglement spectrum}\label{sec:MPSES}

The understanding and simulation of quantum many-body states in one space dimension has experienced revolutionary progress with the advent of the density matrix renormalization group~\cite{White-PhysRevLett.69.2863}. In modern language, this method can be viewed as a variational optimization over the set of matrix product states (MPS)~\cite{fannes1992finitely,perez-garcia}. Let consider a quantum state $\ket{\Psi}=\sum_{\{m_i\}} c_{\{m_i\}} \ket{m_1,...,m_{N_{\rm orb}}}$, where the $\{m_i\}=\{m_1,...,m_{N_{\rm orb}}\}$ are a set of physical indices such as a spin up or down, an occupied or empty orbital, ... 

\begin{equation}
\ket{\Psi} = \sum_{\{m_i\}} \; \left(C^{[m_1]} ...  C^{[m_{N_{\rm orb}}]}\right)_{\alpha_L,\alpha_R} \; \ket{m_1,...,m_{N_{\rm orb}}}\label{eq:MPS}
\end{equation}
where the $\{C^{[m]}\}$ is a set of matrices (each orbital might require a different set of matrices) and $\alpha_L$ and $\alpha_R$ are boundary conditions that pick one matrix element of the matrix product (taking the trace being another option). The $C^{[m]}_{\alpha,\beta}$ matrices have two types of indices. $[m]$ is the physical index and $(\alpha,\beta )$ are the bond indices (or auxiliary space indices) where $\alpha,\beta,=1,...,\chi$. $\chi$ is called the bond dimension. Such a rewriting of a state decomposition is always possible. When the bond dimension $\chi$ of the matrix $C^{[m]}$ is much smaller than the size of the $n$-body Hilbert space, this formulation provides a more economical representation of the state. The crucial question is how small can $\chi$ be for Eq.~\ref{eq:MPS} to still be an exact statement. Generic 1-D gapped systems can be approximated by finite $\chi$~\cite{verstraete}. Critical systems however require an MPS with an infinite bond dimension \cite{Cirac-PhysRevB.81.104431,Nielsen-JPhysStat-P11014}.

The AKLT ground state that we have discussed in the previous section can be expressed in a rather simple MPS form. In that case, $N_{\rm orb}$ is the number of spin-$1$, the physical index $m$ can take three different values $-1,0,+1$ corresponding to the three values of $S_z$. The MPS representation requires three $2 \times 2$ matrices 
\begin{eqnarray}
C^{[0]}=\matrix22{-\sqrt{\frac{1}{3}}}{0}{0}{\sqrt{\frac{1}{3}}},&C^{[+1]}=\matrix22{0}{\sqrt{\frac{2}{3}}}{0}{0}
&{\rm and}\;\; C^{[-1]}=\matrix22{0}{0}{-\sqrt{\frac{2}{3}}}{0}.\label{eq:AKLTMPS}
\end{eqnarray}

As an exercise, we can check that we indeed reproduce the ground state of the AKLT model. We focus on the ground state $\Psi^{N}_{1,1}$ in the $S_z=1$ sector. For $N=2$ spins, it decomposes into 
\begin{eqnarray}
\Psi^{N=2}_{1,1}&=&\frac{\sqrt{2}}{2}\ket{+1,0}-\frac{\sqrt{2}}{2}\ket{0,+1}\label{eq:AKLTDecompositionN2}
\end{eqnarray}
where $\ket{s_{z,1},s_{z,2}}=\ket{s_{z,1}}\otimes\ket{s_{z,2}}$ is the many-body basis where the first (resp. second) spin has its spin projection along $z$ equal to $s_{z,1}$ (resp. $s_{z,2}$). Using the matrices of Eq.~\ref{eq:AKLTMPS}, we find that picking the entry of the first row and second column, we recover (up to a normalization factor) the coefficients of the decomposition of Eq.~\ref{eq:AKLTDecompositionN2}. Choosing the entry of the matrix product (i.e. $\alpha_L$ and $\alpha_R$) is akin to select the boundary conditions. Note that this matrix element is zero for products such as $C^{[+1]}C^{[+1]}$ or $C^{[0]}C^{[0]}$, as it should be since $\Psi^{N=2}_{1,1}$ has $S_z=1$.

A similar calculation can be performed for $N=3$ spins. There the decomposition of $\Psi_{1,1}$ is
\begin{eqnarray}
\Psi^{N=3}_{1,1}&=&\frac{1}{\sqrt{7}}\left(\ket{+1,0,0}-\ket{0,+1,0}+\ket{+1,0,0}\right)-\frac{2}{\sqrt{7}}\ket{+1,-1,+1}\label{eq:AKLTDecompositionN3}
\end{eqnarray}
Once again, we can explicitly check that the MPS description leads to the correct decomposition. Note that $\Psi^{N=3}_{1,1}$ has weight neither on $\ket{-1,+1,+1}$ nor on $\ket{+1,+1,-1}$ despite having $S_z=1$. The MPS description also gives such a result.

In this example, the size of the $C^{[m]}$ matrices is equal to the number of non zero eigenvalues observed in the entanglement spectrum. As we will now show, these two quantities are related. A way to create a bipartite partition of the system is to consider $A$ being made of the indices $\{m_1,...m_{{l_A}}\}$ and $B$ built from the indices $\{m_{{l_A + 1}},...,m_{N_{\rm orb}}\}$. Following the notations of Eq.~\ref{eq:GenericPsiDecomposition}, we have $\{\ket{\mu_A}=\ket{m_1,...m_{{l_A}}}\}$ and $\{\ket{\mu_B}=\ket{m_{l_A+1},...,m_{N_{\rm orb}}}\}$. The MPS formulation of Eq.~\ref{eq:MPS} can be rewritten to make this partition apparent
\begin{eqnarray}
\ket{\Psi} &=  \sum_{\alpha=1}^{\chi}\sum_{\{m_i\}} & \left(C^{[m_1]}...C^{[m_{l_A}]}\right)_{\alpha_L,\alpha} \label{eq:MPSESFullExpression}\\
& &\left(C^{[m_{l_A+1}]}...C^{[m_{N_{\rm orb}}]}\right)_{\alpha,\alpha_R} \; \ket{m_1,...,m_{N_{\rm orb}}} \nonumber
\end{eqnarray}
Thus we obtain that 
\begin{eqnarray}
\ket{\Psi}&=&\sum_{\alpha=1}^{\chi}\ket{A: \alpha} \otimes \ket{B: \alpha}\label{eq:MPSESReducedExpression}
\end{eqnarray}
with 
\begin{eqnarray}
\ket{A: \alpha}&=&\sum_{\{m_i\}} \; \left(C^{[m_1]}...C^{[m_{l_A}]}\right)_{\alpha_L,\alpha}\ket{m_1,...m_{{l_A}}}\label{eq:MPSESLeftState}\\
\ket{B: \alpha}&=& \sum_{\{m_i\}} \;  \left(C^{[m_{l_A+1}]}...C^{[m_{N_{\rm orb}}]}\right)_{\alpha,\alpha_R}\ket{m_{{l_A + 1}},...,m_{N_{\rm orb}}}\label{eq:MPSESRightState}
\end{eqnarray}
While this decomposition looks similar to the Schmidt decomposition of Eq.~\ref{eq:SchmidtDecomposition}, the states $\ket{A: \alpha}$ and $\ket{B: \alpha}$ are neither orthonormal nor linearly independent. Two extra steps are actually required to obtain the true Schmidt decomposition, we need to extract an orthonormal complete basis from $\{\ket{A: \alpha}\}$ and $\{\ket{B: \alpha}\}$ and then to perform an SVD on the entanglement matrix. But no matter these extra steps, they can only (at worst) reduce the number of terms in the sum of Eq.~\ref{eq:MPSESReducedExpression}. Indeed, denoting $\{\ket{A: \tilde{\alpha}}\}$ (resp. $\{\ket{B: \tilde{\beta}}\}$) the orthonormalized basis extracted from $\{\ket{A: \alpha}\}$ (resp. $\{\ket{B: \alpha}\}$), we can introduce the two transformation matrices $U$ and $V$ such that 
 
\begin{eqnarray}
\ket{A: \alpha}=\sum_{\tilde{\alpha}} U_{\alpha,\tilde{\alpha}} \ket{A: \tilde{\alpha}}&{\rm and}&\ket{B: \alpha}=\sum_{\tilde{\beta}} V_{\alpha,\tilde{\beta}} \ket{B: \tilde{\beta}}\label{eq:MPSESOrthonormalBasis}
\end{eqnarray}

These basis have dimensions lower or equal to the bond dimension $\chi$. We immediately get that 

\begin{eqnarray}
\ket{\Psi}&=&\sum_{\tilde{\alpha}, \tilde{\beta}} \left(U^t V\right)_{\tilde{\alpha}, \tilde{\beta}} \ket{A: \tilde{\alpha}} \otimes \ket{B: \tilde{\beta}} \label{eq:MPSESPsiSchmidtOrthonormalBasis}
\end{eqnarray}
The entanglement matrix can be directly readout from the previous equation, leading to the ES once the SVD is performed on $U^t V$. As a consequence, if we want to write an exact MPS for $\ket{\Psi}$, the bond dimension $\chi$ cannot be lower than the number of non-zero eigenvalues of the reduced density matrix. The latter number gives the optimal size for the MPS representation of a state (as discussed in the case of the AKLT ground state). Thus any massive reduction of the system entanglement spectrum should be the sign of an efficient MPS representation of a quantum state.

\section{Observing an edge mode through the entanglement spectrum}\label{sec:EdgeandES}

In the previous section, we have shown how the ES was able to reveal the gapped edge physics of a system such as the AKLT model. We will now discuss the case of gapless edge excitation of a topological phase. Remarkably, even for a non-interacting system such as the integer quantum Hall effect allows to exemplify this unique property of the ES.

\subsection{The integer quantum Hall effect}\label{sec:IQHE}

\subsubsection{Overview and notations}\label{sec:IQHEOverview}

\begin{figure}
\centering
\includegraphics[width=0.55\linewidth]{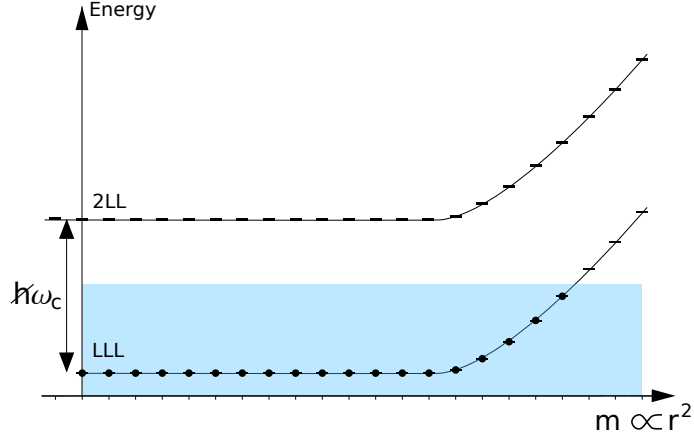}
\caption{Schematic description of the Landau levels in presence of a slowly varying confining potential. Here only the lowest Landau level (LLL) and the second Landau level (2LL) are depicted. The light blue denotes the region bellow the Fermi level.}\label{fig:landaulevelpotential}
\end{figure}

\begin{figure}
\centering
\includegraphics[width=0.9\linewidth]{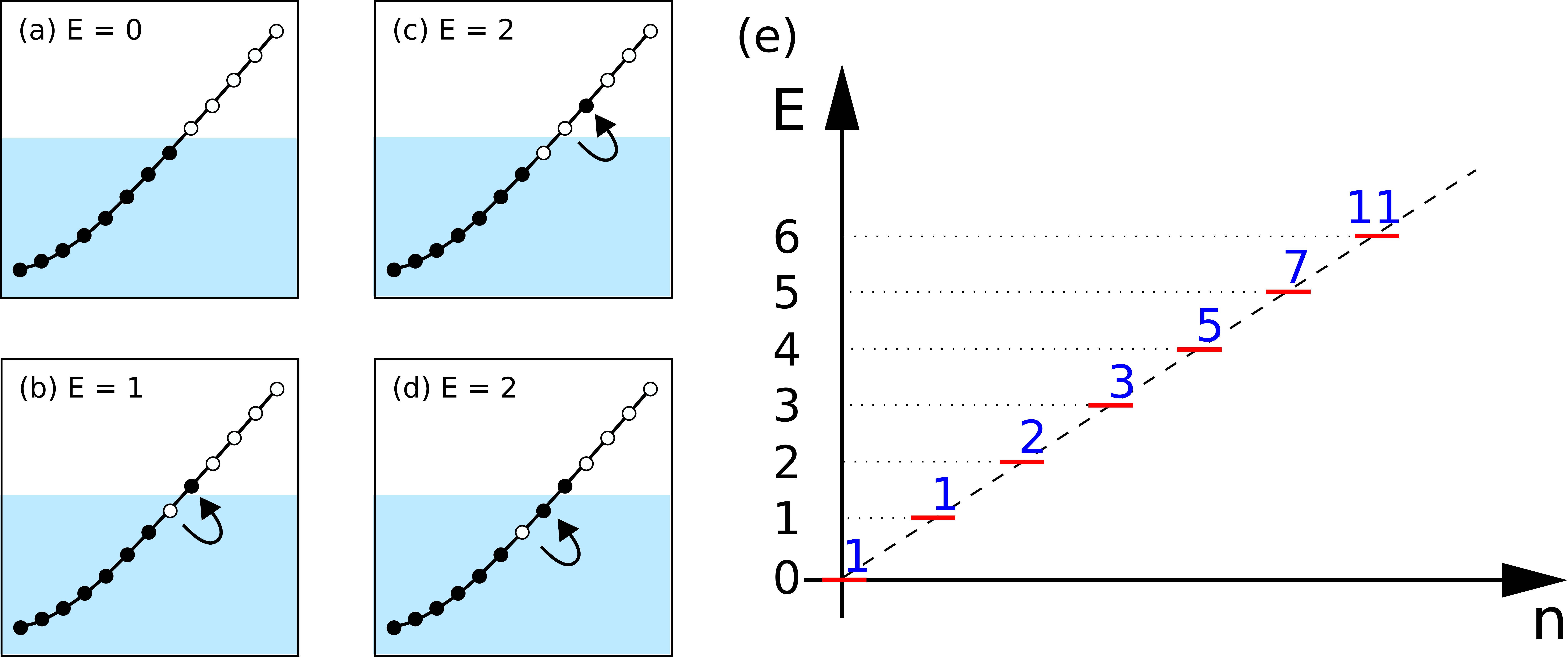}
\caption{Construction of the many-body energy spectrum of the IQHE chiral edge mode. {\it Left panel}: Starting by the filled situation (a) defining our energy of reference, we can consider the lowest energy excitation (b) by moving the topmost particle to the next empty level. (c) and (d) shows the two possible excitations with energy $E=2$. {\it Right panel}: The many-body energy $E$ of the IQHE chiral edge mode as a function of the momentum $m$. The blue number above each level gives its degeneracy.}\label{fig:edgemodeiqhe}
\end{figure}

We consider a completely filled single Landau level. For sake of simplicity and without loss of generality, we focus on the lowest Landau level (LLL). Using the symmetric gauge $\vec{A}=\left(-yB, xB,0\right)$, the one-body wave functions are given by 
\begin{equation}
\phi_m (z)= \frac{1}{\sqrt{2 \pi  2^{m} m! }} z^m e^{-\frac{1}{4l_B^2} |z|^2}.\label{eq:OrbitalOneBodyBasisDisk}
\end{equation}
where $l_B=\sqrt{\frac{h}{eB}}$ is the magnetic length that we will set to $1$ and $z=x+iy$ is the particle complex coordinate in the plane. $m\ge 0$ is an integer corresponding the angular momentum. If we have a radial confining potential with a slow variation compared to the magnetic length, the Landau levels are just bent and follow the confining potential as depicted in Fig.~\ref{fig:landaulevelpotential}. The edge mode of the IQHE corresponds to a non-interacting chiral state, with a linear dispersion relation $E=\frac{2\pi v}{\mathcal{L}} n$ where $n\ge 0$ is an integer, $v$ is the edge mode velocity and $\mathcal{L}$ is the length of the edge. From this one particle spectrum, we can build the many-body spectrum of the chiral edge as depicted in Fig.~\ref{fig:edgemodeiqhe}. For simplicity, we drop the energy scale $\frac{2\pi v}{\mathcal{L}}$. We start by filling all the levels up to the Fermi level (Fig.~\ref{fig:edgemodeiqhe}a). We define the total energy as the energy reference (i.e. $E=0$). The lowest energy excitation $E=1$ is unique and shown in Fig.~\ref{fig:edgemodeiqhe}b: it gives to a single level in the many-body spectrum at momentum $n=1$. At the energy $E=2$, we have two possible excitations Figs.~\ref{fig:edgemodeiqhe}c and~\ref{fig:edgemodeiqhe}d, leading to two levels at momentum $n=2$. Performing a similar reasoning for the other excitations, we end up with the counting per momentum $1,1,2,3,5,...$ depicted in Fig.~\ref{fig:edgemodeiqhe}e. This is the counting of the chiral $U(1)$ bosons.

The bulk of the IQHE is actually quite simple to describe. Since the system is a filled band, we are filling all the orbitals starting from the one with the lowest momentum. The number of occupied orbitals is given by the number of flux quanta $N_{\Phi}$ in the system. For $N$ electrons, the many-body wave function simply reads
\begin{equation}
\ket{\Psi_{\rm IQHE}}=\left(\prod_{m=0}^{N-1} c^\dagger_{m}\right) \ket{0}.\label{eq:IQHEwavefunction}
\end{equation}
where $c^\dagger_{m}$ is the creation operator associated to the one-body wave function $\phi_m(z)$. In first quantized notation and up to a normalization factor, Eq.~\ref{eq:IQHEwavefunction} simply writes
\begin{equation}
\Psi_{\rm IQHE}(z_1,...,z_N)=\prod_{i<j} \left(z_i-z_j\right) \exp\left(-\frac{1}{4} \sum_i \left|z_i\right|^2\right).\label{eq:IQHEwavefunctionfirstquantized}
\end{equation}

\subsubsection{Real space partition}\label{sec:RealspacePartition}

We will now discuss how to perform a partition in real space of our system. Note that this section is generic and not specific to the IQHE. Our goal is to divide the space into two separate regions $A$ and $B$. Consider the creation operator $c^\dagger$ for a given orbital. We want to split this operator into two parts $A$ and $B$ such that 

\begin{eqnarray}
c^\dagger &=& \alpha c^\dagger_A + \beta c^\dagger_B\label{eq:realspaceoneorbital}
\end{eqnarray}

where $c^\dagger_A$ (resp. $c^\dagger_B$) creates an electron living only in $A$ (resp. $B$) and $\alpha$ and $\beta$ are constants. Since $\{c,c^\dagger\}=\{c_A,c_A^\dagger\}=\{c_B,c_B^\dagger\}=1$, the two constants should satisfy $\left|\alpha\right|^2+\left|\beta\right|^2=1$. Let $\Psi(\vec{r})=\bra{\vec{r}}c^\dagger\ket{0}$ denote the wave function in first quantized notations. We can perform the following decomposition

\begin{eqnarray}
\Psi\left(\vec{r}\right)&=&\Theta_A\left(\vec{r}\right) \Psi\left(\vec{r}\right) + \left(1 - \Theta_A\left(\vec{r}\right)\right) \Psi\left(\vec{r}\right),\label{eq:wavefunctiondecompositionrealspace}\\
&=&{\mathcal N}_A \Psi_A\left(\vec{r}\right) + {\mathcal N}_B \Psi_B\left(\vec{r}\right).\nonumber
\end{eqnarray}

where $\Theta_A\left(\vec{r}\right)$ is one if $\vec{r}$ belongs to $A$ and zero otherwise, and where the two normalized (and orthogonal) wave functions $\Psi_A$ and $\Psi_B$ are given by 

\begin{eqnarray}
\Psi_A\left(\vec{r}\right)=\frac{1}{{\mathcal N}_A}\Theta_A\left(\vec{r}\right)\Psi\left(\vec{r}\right),&&{\mathcal N}_A=\int_A d\vec{r}\;|\Psi\left(\vec{r}\right)|^2,\label{eq:wavefunctionArealspace}\\
\Psi_B\left(\vec{r}\right)=\frac{1}{{\mathcal N}_B}\left(1-\Theta_A\left(\vec{r}\right)\right)\Psi\left(\vec{r}\right),&&{\mathcal N}_B=\int_B d\vec{r}\;|\Psi\left(\vec{r}\right)|^2.\label{eq:wavefunctionBrealspace}
\end{eqnarray}

Notice that ${\mathcal N}_A^2+{\mathcal N}_B^2=1$. Comparing Eq.~\ref{eq:realspaceoneorbital} and Eq.~\ref{eq:wavefunctiondecompositionrealspace}, we immediately deduce that $\alpha={\mathcal N}_A$ and $\beta={\mathcal N}_B$. The treatment that we have performed here is generic and does not depend on any specific property of $\Psi$.

\subsubsection{Back to the IQHE}\label{sec:RealspaceIQHE}

Equipped with the decomposition mentioned in the previous subsubsection, we can now perform the real space entanglement spectrum of the IQHE. While there is in principle no constraint on the choice of $A$, we would like to preserve as many quantum numbers as possible. If we use the symmetric gauge, a natural choice is to preserve the rotational symmetry. Thus we can decide to take for $A$, a disk centered at the origin and with a radius $R$. Such a cut was described in Ref.~\cite{Rodriguez-PhysRevB.80.153303} to compute the entanglement entropy of the IQHE. Each orbital $\Phi_m\left(z\right)$ can be decomposed as following following Eq.~\ref{eq:wavefunctiondecompositionrealspace}

\begin{eqnarray}
\Phi_m\left(z\right)&=&\alpha_{m} \Phi_{m,A}\left(z\right) + \beta_{m} \Psi_{m,B}\left(z\right).\label{eq:orbitaldecompositionrealspace}
\end{eqnarray}
with
\begin{eqnarray}
\alpha_{m}^2=\frac{1}{2^m m!}\int_0^R dr\;r^{2m+1}e^{-\frac{r^2}{2}}&\;\;{\rm and}\;\;&\beta_{m}^2=1-\alpha_{m}^2\label{eq:orbitalweightrealspace}
\end{eqnarray}
$\alpha_m^2$ and $\beta_{m}^2$ are incomplete gamma functions and can easily be computed numerically. Since $\Phi_{m,A}\left(z\right)$ are eigenstates of the angular momentum, they are still orthogonal (similarly for $\Psi_{m,B}\left(z\right)$).

Performing the Schmidt for such a partition is rather easy. For pedagogical reasons, we first consider the IQHE with only two particles occupying the two first orbitals $m=0$ and $m=1$. The ground state wave function $\ket{\Psi_{\rm IQHE}}$ reads
\begin{eqnarray}
\ket{\Psi_{\rm IQHE}}&=&c^\dagger_0 c^\dagger_1 \ket{0}.\label{eq:IQHEtwoparticlewavefunction}
\end{eqnarray}
Substituting the expression Eq.~\ref{eq:realspaceoneorbital} for each creation operator, we get
\begin{eqnarray}
\ket{\Psi_{\rm IQHE}}&=&\alpha_0\alpha_1 c^\dagger_{0,A} c^\dagger_{1,A} \ket{0}\nonumber\\
&&+\beta_0\beta_1 c^\dagger_{0,B} c^\dagger_{1,B} \ket{0}\label{eq:IQHEtwoparticlewavefunctionSchmidt}\\
&&+\left(\alpha_0\beta_1 c^\dagger_{0,A} c^\dagger_{1,B}\ket{0} - \alpha_1\beta_0 c^\dagger_{1,A} c^\dagger_{0,B}\ket{0}\right).\nonumber
\end{eqnarray}
The first term (resp. second term) of the right hand side of Eq.~\ref{eq:IQHEtwoparticlewavefunctionSchmidt} corresponds to the case where we have two particles in $A$ (resp. $B$). The third term is the case where we have one particle in $A$ and one in $B$ (notice the minus sign due to the fermionic statistics). When performing such a cut, we have two good quantum numbers :
\begin{itemize}
\item The total number of particles $N=N_A+N_B$ where $N_A$ (resp. $N_B$) is the number of particles in the region $A$ (resp. $B$).
\item The total angular momentum $L_{z} = L_{z,A}+L_{z,B}$ where $L_{z,A}$ (resp. $L_{z,B}$) is the total angular momentum of the particles in $A$ (resp. $B$).
\end{itemize}

The entanglement spectrum can be directly read out from the Schmidt decomposition of Eq.~\ref{eq:IQHEtwoparticlewavefunctionSchmidt} and sorted per quantum numbers: 
\begin{itemize}
\item $N_A=0$, $L_{z,A}=0$ : one level at $-\ln\left(\left|\beta_0\beta_1\right|^2\right)$.
\item $N_A=1$, $L_{z,A}=0$ : one level at $-\ln\left(\left|\alpha_0\beta_1\right|^2\right)$.
\item $N_A=1$, $L_{z,A}=1$ : one level at $-\ln\left(\left|\alpha_1\beta_1\right|^2\right)$.
\item $N_A=2$, $L_{z,A}=1$ : one level at $-\ln\left(\left|\alpha_0\alpha_1\right|^2\right)$.
\end{itemize}
As an exercise, we let the reader check that the reduced density matrix associated to this entanglement spectrum is properly normalized to one. For the generic case ($N$ electrons occupying the $N$ first orbitals), a similar calculation of the Schmidt decomposition would give
\begin{equation}
\ket{\Psi_{\rm IQHE}}=\sum_{N_A=0}^{N}\sum_{L_{z,A}}\sum_{\substack{\{m_{1},...,m_{N_A}\},\\ \sum_i m_i=L_{z,A}}}
\prod_{i=1}^{N_A}\alpha_{m_i}\prod_{j=N_A+1}^{N}\beta_{m_j} \ket{\{m_{1},...,m_{N_A}\}}\otimes  \ket{\{m_{N_A+1},...,m_{N}\}}.\label{eq:IQHENparticlewavefunctionSchmidt}
\end{equation}
The third sum runs over all the possible ways to choose $N_A$ particles among the $N$ occupied orbitals with the constraint that these $N_A$ particles have a total angular momentum $L_{z,A}$. Each valid configuration $\{m_{1},...,m_{N_A}\}$ where the $m_i$ are ordered for the smallest the largest integer leads to a state $\ket{\{m_{1},...,m_{N_A}\}}$ up to a sign (a consequence of the possible orbital reordering and the fermionic statistics, as was mentioned in Eq.~\ref{eq:IQHEtwoparticlewavefunctionSchmidt}). Once the $\{m_{1},...,m_{N_A}\}$ are selected, the occupied orbitals $\{m_{N_A+1},...,m_{N}\}$ for the particles in $B$ are automatically defined, giving the state $\ket{\{m_{N_A+1},...,m_{N}\}}$. Once again, the entanglement spectrum directly follows from such a decomposition since it is a Schmidt decomposition: The orthogonality conditions are satisfied since two different sets $\{m_{1},...,m_{N_A}\}$ give two orthogonal (and normalized) states.

\begin{figure}
\centering
\includegraphics[width=0.9\linewidth]{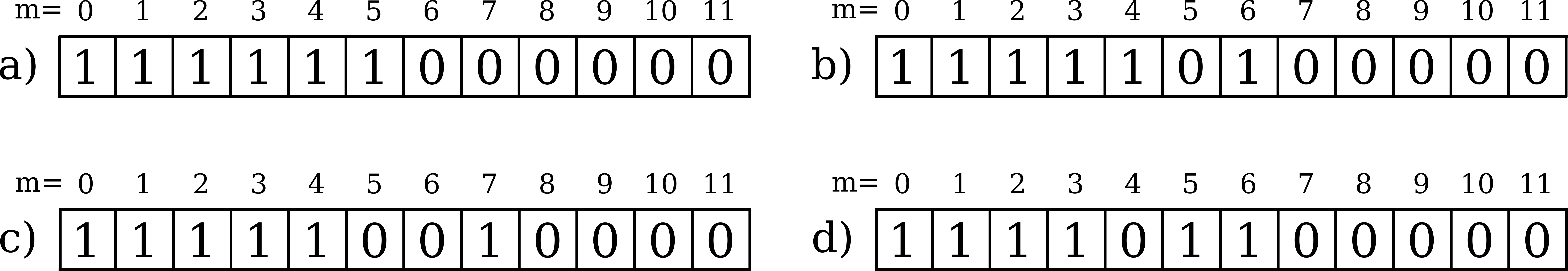}
\caption{Valid configurations that appears in the Schmidt decomposition for $N_A=N/2=6$. Each box represents an orbital with momentum $m$ (the label on top) and can be occupied (1) or empty (0). a) is the configuration with the smallest angular momentum $L_{z,A}=15$. b) is the only configuration with $L_{z,A}=16$. For $L_{z,A}=17$, we have two options shown in c) and d).}\label{fig:RSESIntegerQHECounting}
\end{figure}

If we focus on a single $N_A$ sector we can count how many entanglement energies we have per angular momentum. For concreteness, let's take $N_A=6$ and $N=12$. The system has $12$ orbitals with angular momentum going from $0$ to $11$. Each $\{m_{1},...,m_{N_A}\}$ can be represented with boxes corresponding to each orbitals as depicted in Fig.~\ref{fig:RSESIntegerQHECounting}. The smallest angular momentum $L_{z,A}=15$ is obtained by occupying the $6$ first orbitals as shown in Fig.~\ref{fig:RSESIntegerQHECounting}a. Thus there is a single level in the entanglement spectrum for $N_A=6$ and $L_{z,A}=15$. $L_{z,A}=16$, there is also a unique (and thus unique level) related to the configuration Fig.~\ref{fig:RSESIntegerQHECounting}b. Moving on to $L_{z,A}=17$, we now have two options Figs.~\ref{fig:RSESIntegerQHECounting}c and Fig.~\ref{fig:RSESIntegerQHECounting}d. We immediately see that this construction is identical to the one of the many-body spectrum for the chiral edge mode. So this latest and the entanglement spectrum in a given particle number sector have the same counting of levels.

\begin{figure}
\centering
\includegraphics[width=0.9\linewidth]{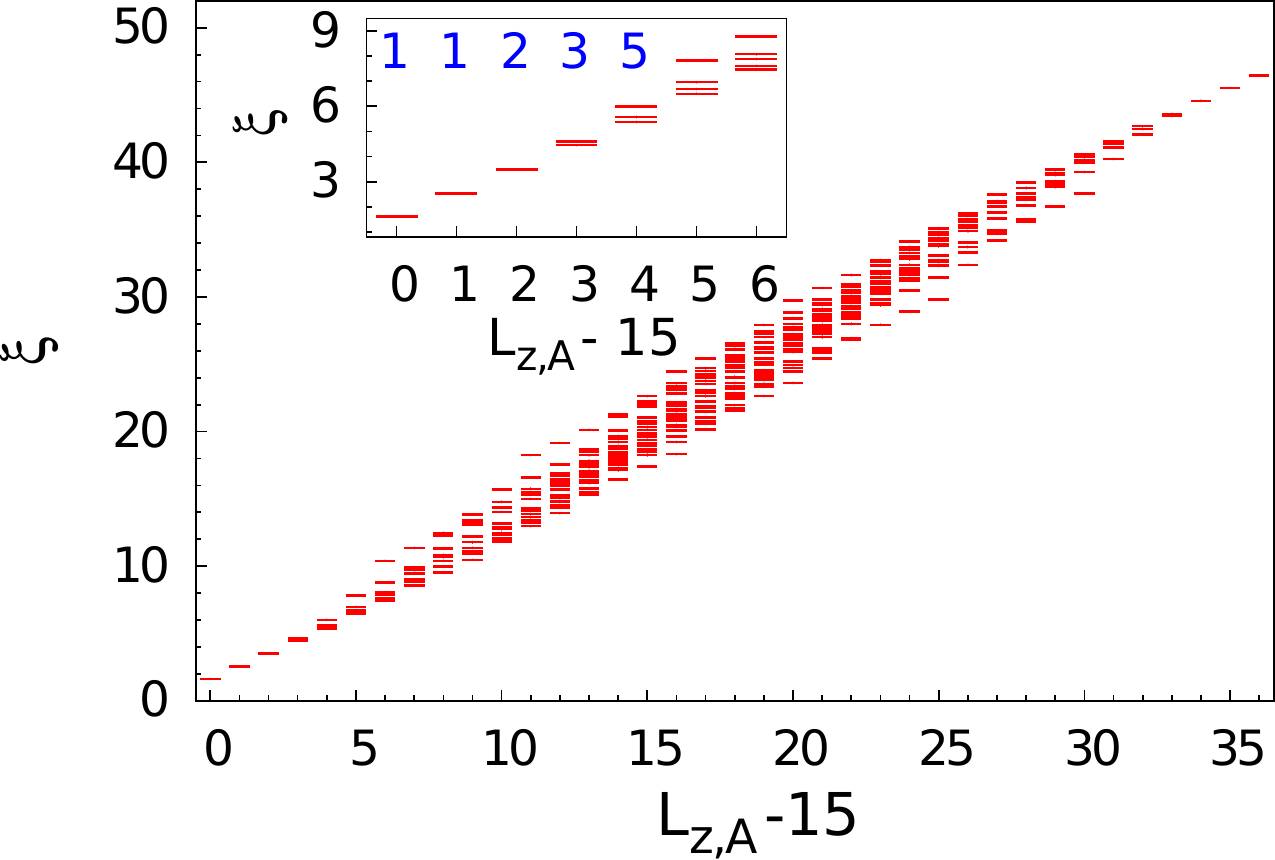}
\caption{RSES of the $\nu=1$ integer quantum Hall effect on the disk geometry for $N=12$ fermions and $N_{\Phi}=11$. We focus on the sector $N_A=6$. The inset displays the counting in the thermodynamical part (the RSES has some exact degeneracies). It matches the one of a $U(1)$ chiral boson.}\label{fig:RSESIntegerQHEDisk}
\end{figure}

Do the entanglement energies also follow a linear dispersion relation? The entanglement spectrum for $N_A=N/2=6$ is shown in Fig.~\ref{fig:RSESIntegerQHEDisk}.  The finite size effects lead to two deviations to the physical spectrum of the edge mode:
\begin{itemize}
\item They spoil the edge mode counting for large angular momentum. Indeed for a given number of flux quanta, there is a maximum angular momentum that can be reached.
\item As can be observed in Fig.~\ref{fig:RSESIntegerQHEDisk}, levels are not strictly degenerate for a given angular momentum as it should be. This property is only recovered in the thermodynamical limit.
\end{itemize}

We can wonder why we looked at a single sector of $N_A$. A key idea of Li and Haldane approach when they have introduced the entanglement spectrum, was to look at some specific block of the reduced density matrix. When we cut the system into two, we want $A$ to be a smaller droplet of the same quantum fluid up to edge excitation. In this picture, it is natural to look at a fixed number of particles. We can even be more quantitative. Consider the eigenstates of the reduced density matrix (i.e. the $\ket{A:i}$ of the Schmidt decomposition of Eq.~\ref{eq:SchmidtDecomposition}). The eigenstate related to the level with the smallest $L_{z,A}$ is a many-body where all the first $N_A$ lowest orbitals in the $\phi_{A,m}$ basis (i.e. each orbital from $m=0$ to $m=N_A-1$) are occupied. This is the densest state that one can create for $\nu=1$ IQHE with $N_A$ particles.

\subsection{Chern Insulators}\label{sec:CI}

The construction of the entanglement spectrum that we have done for the integer quantum Hall effect can be extended to any non-interacting topological insulators. For simplicity, we will focus on the first and simplest example of a topological insulator, the Chern insulator (CI) that was introduced in a theoretical work by F.D.M. Haldane in 1988 \cite{haldane-1988PhRvL..61.2015H}. It is defined by a non-zero Chern number $C$ of the occupied bands. This (first) Chern number is a topological invariant, computed over the Brillouin zone, that characterizes a given band. A key feature is that a non-zero Chern number results\cite{Thouless-PhysRevLett.49.405} in a quantized Hall conductance $\sigma_{xy} = \frac{e^2}{h} C$, similar to the quantum Hall effect, but now without the requirement of an external magnetic field.

\begin{figure}
\centering
\includegraphics[width=0.35\linewidth]{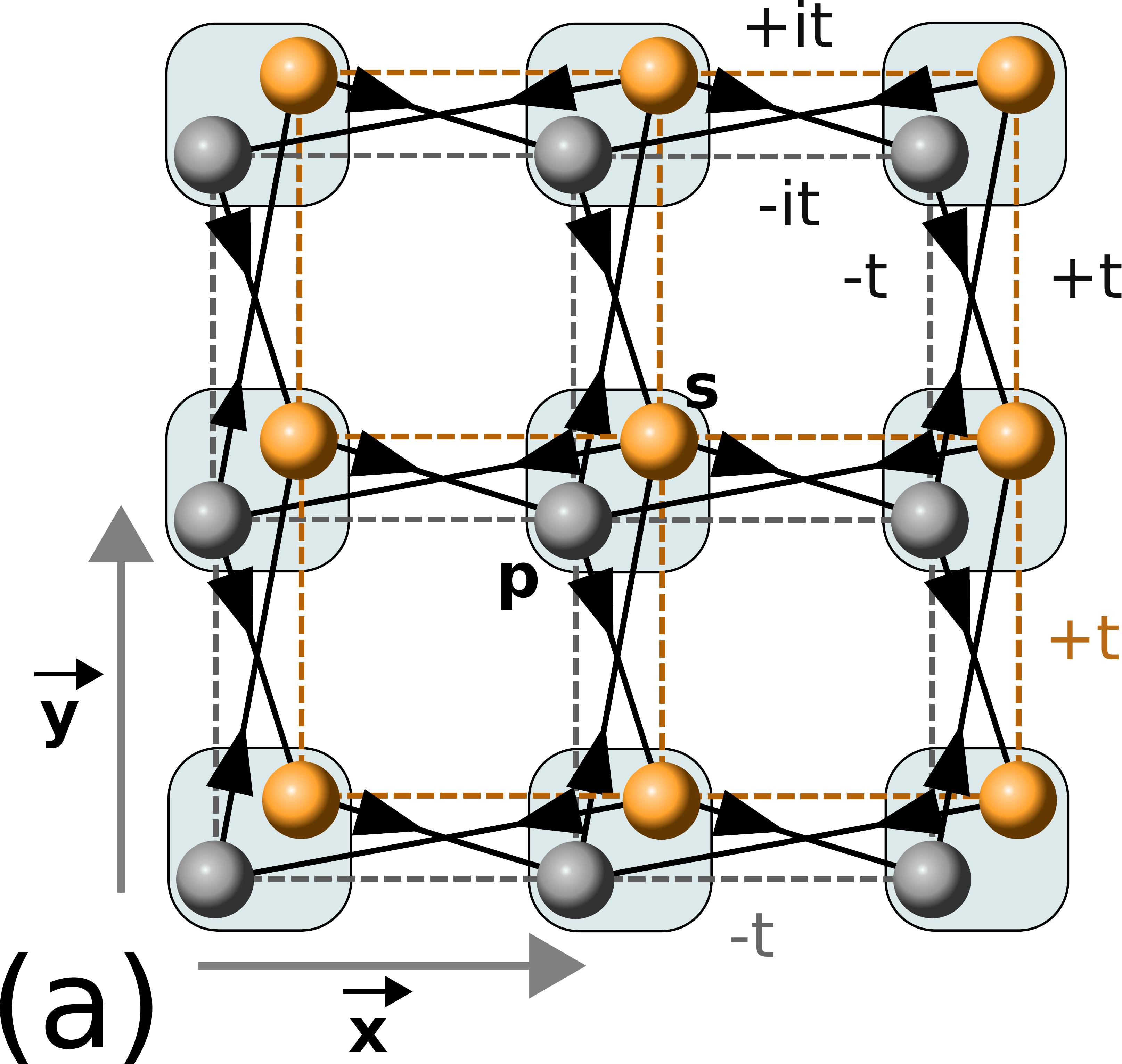}
\includegraphics[width=0.44\linewidth]{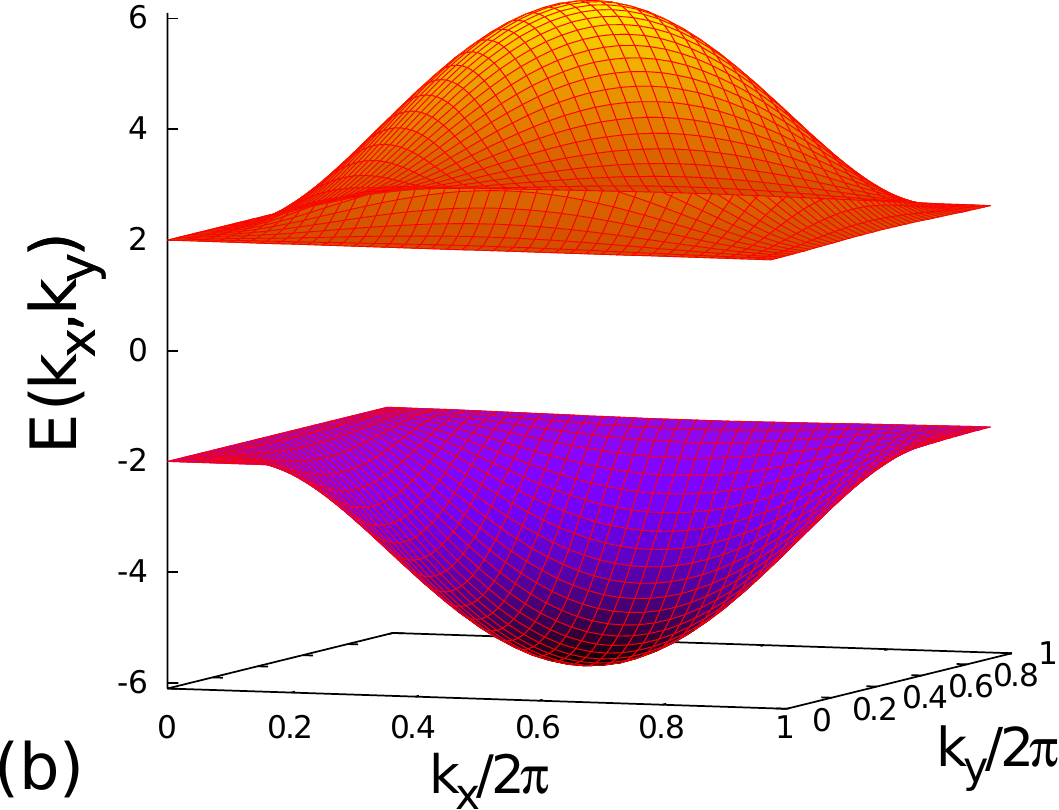}
\caption{{\it Left panel}: The two-orbital lattice model with one $s$ and one $p$ orbital per site. The choice of {\it Right panel}: The band structure for the two-orbital lattice model with a mass term set to $M=1$, plotted as a function of the momenta $k_x$ and $k_y$. Each band carries a Chern number of $C=\pm 1$.}\label{fig:twoorbitals}
\end{figure}

\begin{figure}
\centering
\includegraphics[width=0.65\linewidth]{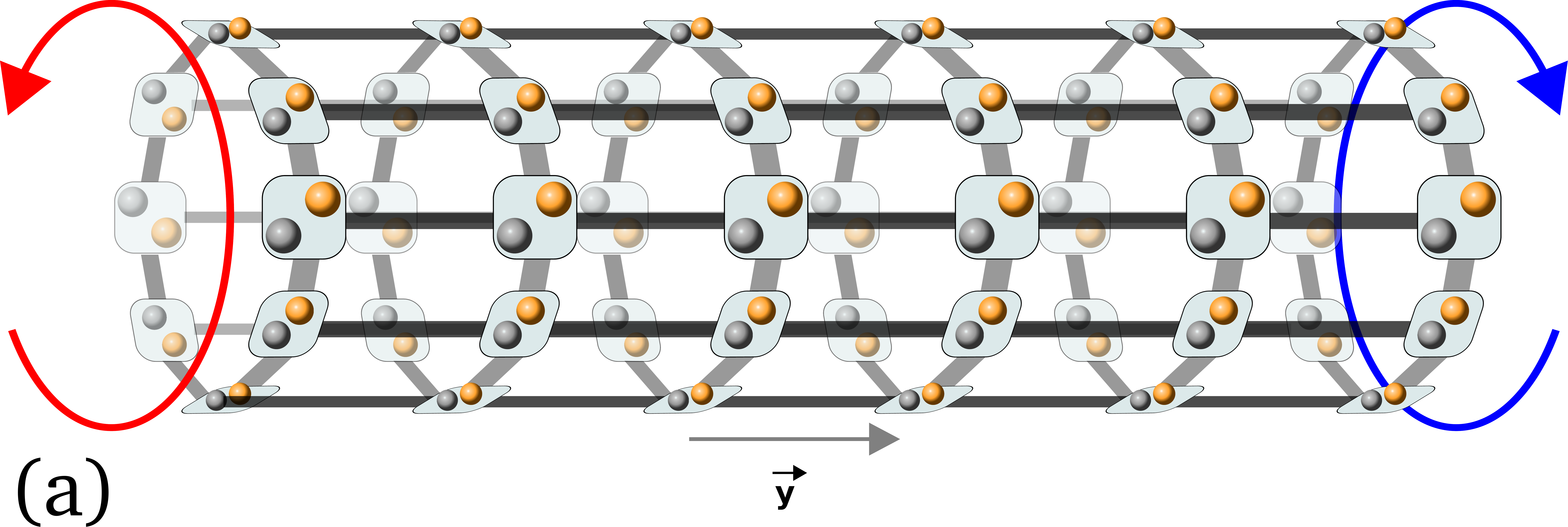}\\
\includegraphics[width=0.46\linewidth]{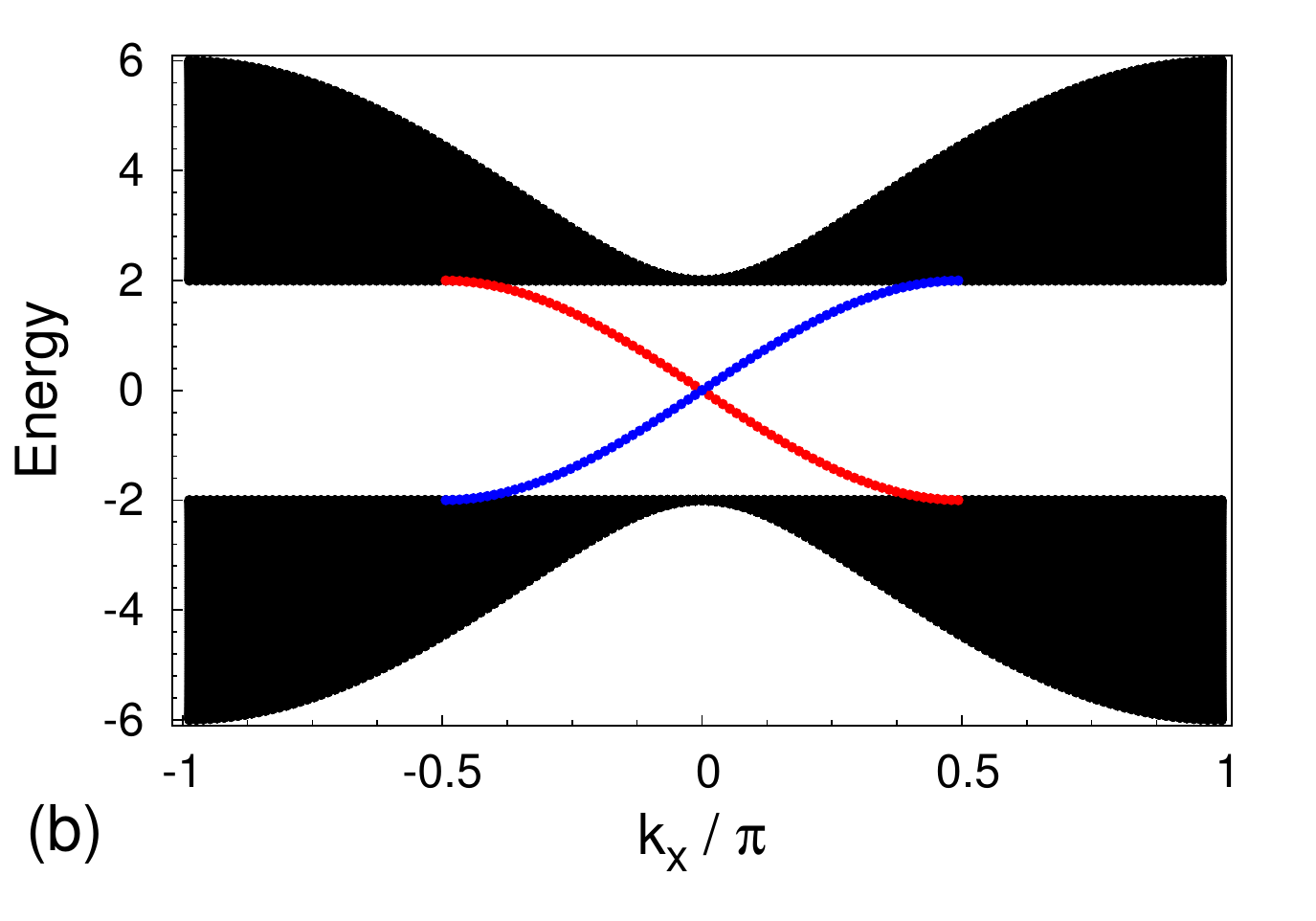}
\includegraphics[width=0.46\linewidth]{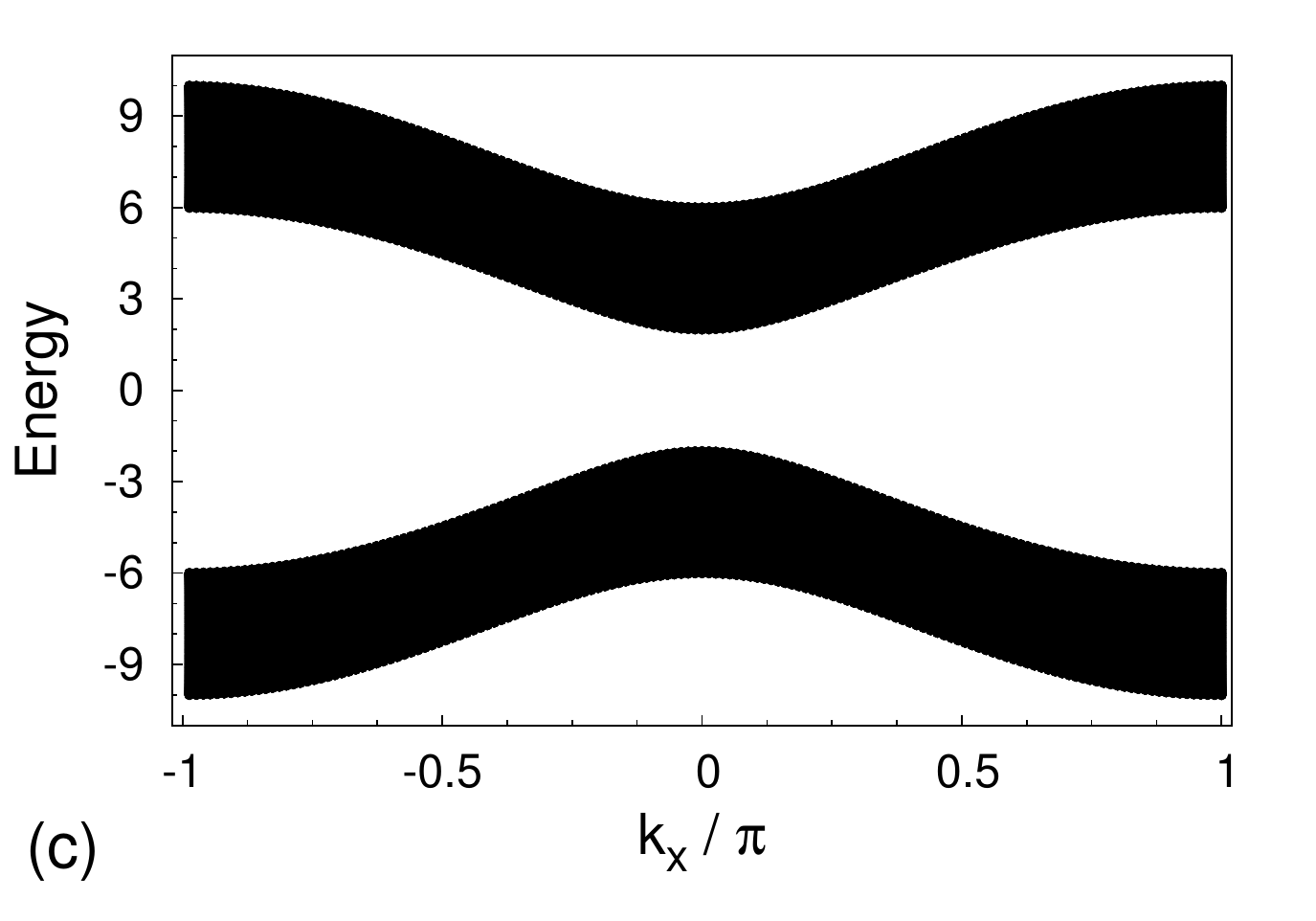}
\caption{{\it Top panel}: The two-orbital lattice model on a cylinder. At each end, there is one chiral gapless edge mode (the blue and one red arrows). {\it Middle panel}: The band structure for the two-orbital lattice model on a cylinder with a mass term set to $M=1$, plotted as a function of the momentum $k_x$ along the cylinder radius. In the gap region we clearly observe the dispersion relation of the two gapless edge modes (in blue and red), going from one band to another (in black). {\it Lower panel}: The band structure for the two-orbital lattice model on a cylinder with a mass term set to $M=3$. The system is in trivial (non-topological) phase and no gapless edge modes are observed.}\label{fig:twoorbitalcylinder}
\end{figure}

A typical example of a Chern insulator is shown in Fig.~\ref{fig:twoorbitals}a. It is based on a simple tight-binding model on a square lattice with two orbitals (one $s$ and one $p$ orbital) per site (see Ref.~\cite{Wu-2012PhysRevB.85.075116} for a more detailed description). With a suitable choice for the hopping amplitudes, the Bloch Hamiltonian for this model reads
\begin{eqnarray}
{\mathcal H}(\mathbf{k}) &=& \sum_{i=x,y} -\sin( k_i) \sigma_i - (M - \sum_{i=x,y} \cos (k_i) ) \sigma_z \label{eq:twoorbitalhamiltonian}
\end{eqnarray}

where $\sigma_{x}, \sigma_{y}$ and $\sigma_{z}$ are the three Pauli matrices and $M$ is a mass term (chemical potential between the two orbitals). The band structure is shown in Fig.~\ref{fig:twoorbitals}b, with the two bands, each carrying a Chern number equal to $\pm 1$ if $|M|<2$, and separated by a gap. If the model is put on cylinder when the system is in a topological phase (meaning that the Chern number is non-zero), the energy spectrum clearly exhibits one chiral gapless edge mode at each end of the cylinder, as shown in Fig.~\ref{fig:twoorbitalcylinder}b. Setting $|M|>2$ to drive the system to a trivial phase will suppress these gapless edge modes as in  Fig.~\ref{fig:twoorbitalcylinder}a.

\subsection{Entanglement spectrum for a CI}\label{sec:ESCI}

\begin{figure}
\centering
\includegraphics[width=0.6\linewidth]{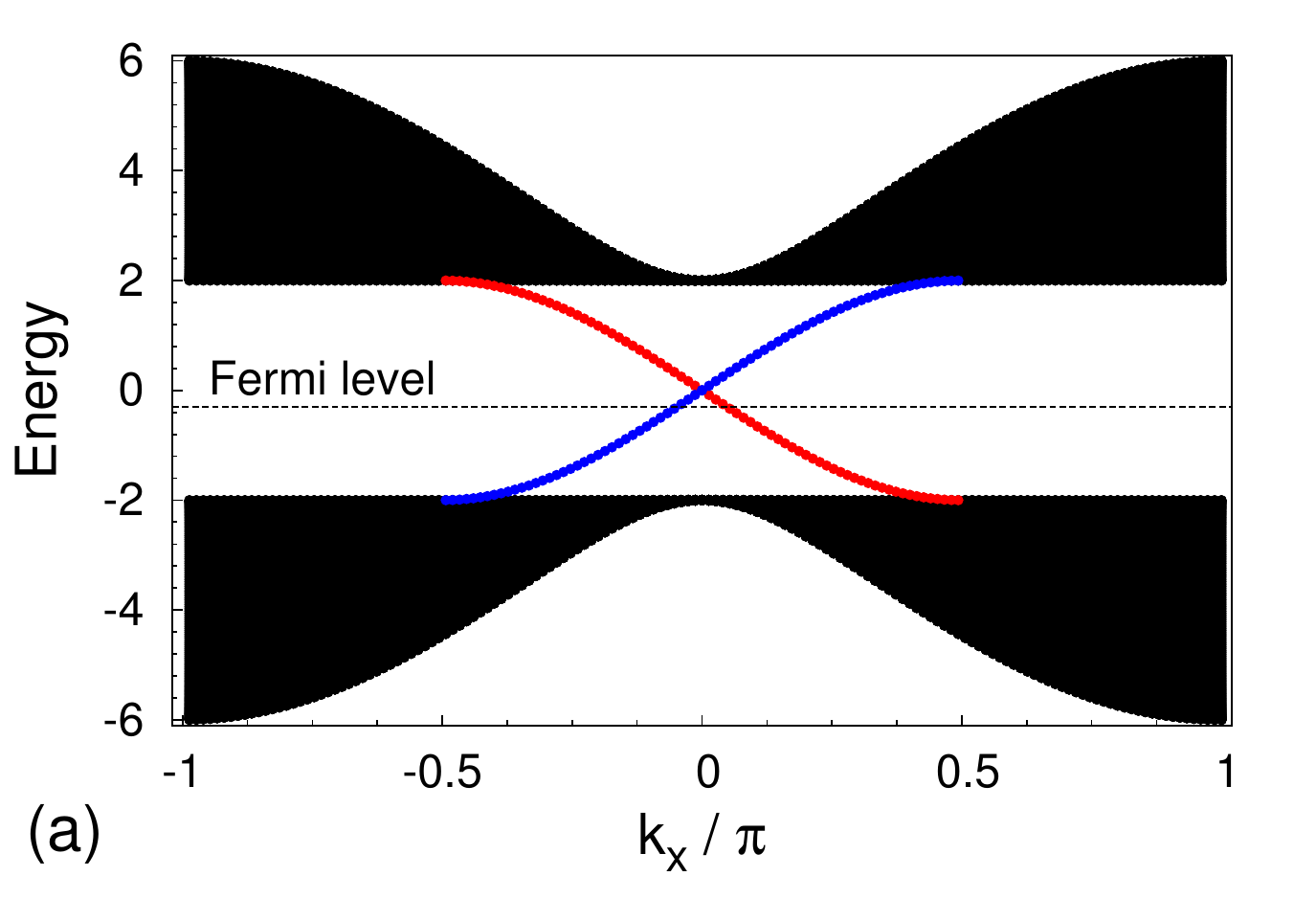}\\
\includegraphics[width=0.65\linewidth]{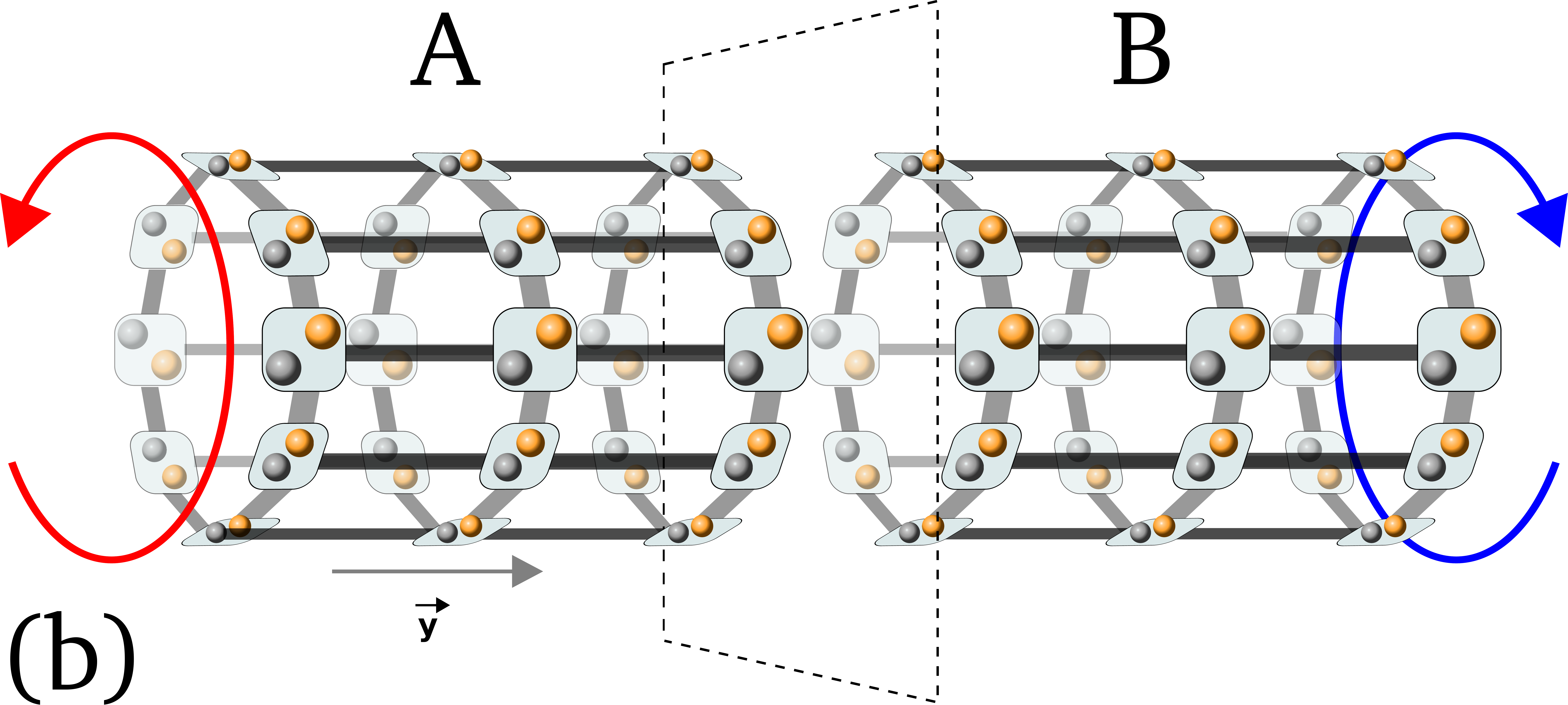}
\caption{{\it Top panel}: The band structure for the two-orbital lattice model on a cylinder with a mass term set to $M=1$ where we set the Fermi level in the bulk gap. {\it Bottom panel}: The two-orbital lattice model on a cylinder cut into part $A$ (left) and $B$ (right).}\label{fig:twoorbitalcylinderAB}
\end{figure}

By tuning the mass parameter $M$ of the two-orbital model described previously, we can drive the system from a topological insulator to a trivial insulator. In these two cases, the energy spectrum would look similar, exhibiting a bulk gap. Here we will show that the ES is able to distinguish between the trivial and the topological phase. For that purpose, we consider the case where we completely fill the lower band up to the Fermi energy $\epsilon_F$ located in the system bulk gap. The quantum state $\ket{\Psi_{\rm CI}}$ of the system is just a simple Slater determinant that can be written in second quantized notation as
\begin{eqnarray}
\ket{\Psi_{\rm CI}}&=&\prod_{k_x;\epsilon < \epsilon_F}c^\dagger_{k_x, \epsilon}\ket{0}\label{eq:CIgroundstate}
\end{eqnarray}
where $c^\dagger_{k_x, \epsilon}$ is the creation operator related to the state with momentum $k_x$ and energy $\epsilon$. We fill all the states with an energy $\epsilon$ lower than the Fermi energy $\epsilon_F$ (see Fig.~\ref{fig:twoorbitalcylinderAB}a). Any choice of $\epsilon_F$ is valid as long as it stands in the bulk gap. For sake of simplicity, the calculation is done on the cylinder geometry, motivating our choice to label the states using the momentum along the $x$ direction where we apply periodic boundary conditions. In this situation the edge mode excitations are completely frozen in a similar one choosing one of the ground state of the AKLT in Sec.~\ref{subsubsection:spinAKLT} fixes the excitations at the edge. 

We now cut the cylinder into two parts $A$ (left part) and $B$ (right part) as depicted in Fig.~\ref{fig:twoorbitalcylinderAB}b (our partition is performed far from the edges). Performing the same decomposition than Eq.~\ref{eq:orbitaldecompositionrealspace}, each creation operator can be written as a sum of two creation operators, one for each part of the system
\begin{eqnarray}
c^\dagger_{k_x, \epsilon}&=&\alpha^*_{k_x, \epsilon}c^\dagger_{k_x, \epsilon;A}+\beta^*_{k_x, \epsilon}c^\dagger_{k_x, \epsilon;B}\label{eq:CIABcreationoperator}
\end{eqnarray}
where $\alpha_{k_x, \epsilon}$ (resp. $\beta_{k_x, \epsilon}$) is the weight of the state $(k_x, \epsilon)$ on the $A$ (resp. $B$) part. These weights satisfy $\left|\alpha_{k_x, \epsilon}\right|^2+\left|\beta_{k_x, \epsilon}\right|^2=1$. Using this decomposition, $\ket{\Psi_{\rm CI}}$ can be rewritten as 
\begin{equation}
\ket{\Psi_{\rm CI}}=\sum_{N_A} \sum_{\{k_{x,A}, \epsilon_A\}}{\mathcal N}_{\{k_{x,A}, \epsilon_A\}}{\mathcal N}_{\{k_{x,B}, \epsilon_B\}}\ket{\{k_{x,A}, \epsilon_A\}}\otimes\ket{\{k_{x,B}, \epsilon_B\}}\label{eq:CIgroundstateAB}
\end{equation}
Here $N_A$ is the number of particles in the part $A$ and the sum over $\{k_{x,A}, \epsilon_A\}$ corresponds to all the possible way to choose $N_A$ states among the original states used to build the Slater determinant of the full band. Note that once we have fixed the choice for the states we consider to be in $A$, the state for $B$ is completely determined and unique, labeled here by $\{k_{x,B}, \epsilon_B\}$. This equation is the exact analogue of Eq.~\ref{eq:IQHENparticlewavefunctionSchmidt} for the IQHE. The other quantities in Eq.~\ref{eq:CIgroundstateAB} are 
\begin{eqnarray}
{\mathcal N}_{\{k_{x,A}, \epsilon_A\}}&=&\prod_{(k_x,\epsilon)\in \{k_{x,A}, \epsilon_A\}} \alpha^*_{k_x, \epsilon}\label{eq:CINAfactor}\\
\ket{\{k_{x,A}, \epsilon_A\}}&=&\prod_{(k_x,\epsilon)\in \{k_{x,A}, \epsilon_A\}}c^\dagger_{k_x,\epsilon;A}\ket{0}\label{eq:CIAstate}
\end{eqnarray}
and similarly for the quantity with a $B$ index. 

By construction, we have just realized the Schmidt decomposition of $\ket{\Psi_{\rm CI}}$. So the ES can be readout directly from Eq.~\ref{eq:CIgroundstateAB}. Indeed, the spectrum of the reduced density matrix is just given by $\left\{\left|{\mathcal N}_{\{k_{x,A}, \epsilon_A\}}{\mathcal N}_{\{k_{x,B}, \epsilon_B\}}\right|^2\right\}$. Unfortunately, this spectrum still requires a factorial effort to be computed due to combinatorial factor of choosing $N_A$ states among the occupied ones, as discussed in Sec.~\ref{sec:RealspaceIQHE}. But it was pointed out in Refs.~\cite{Fidkowski-PhysRevLett.104.130502} and~\cite{Rodriguez-PhysRevB.80.153303} that this many-body ES can be deduced from a one-body ES in a similar way one constructs the many-body energy spectrum of non-interacting particles from the one-body spectrum. For such a system, the reduced density matrix $\rho_A$ can be rewritten as
\begin{eqnarray}
\rho_A&=\sum_\alpha e^{-\xi_\alpha} \ket{A : \alpha}\bra{A : \alpha}&=e^{-\hat{H}_{\rm ent}}\label{eq:CIrhoAEntHam}
\end{eqnarray}
where the entanglement hamiltonian $\hat{H}_{\rm ent}$ is a one-body hamiltonian
\begin{eqnarray}
\hat{H}_{\rm ent}=\sum_{k_x,i,j}h(k_x)_{i,j}\; c^\dagger_{k_x,i}c_{k_x,j}\label{eq:CIEntHamdefinition}
\end{eqnarray}
Here the indices $i$ and $j$ denote any site (or orbital) that belongs to the $A$ part. Note that one can always define an entanglement hamiltonian using Eq.~\ref{eq:CIrhoAEntHam}. The possibility to express it as a one-body hamiltonian is a specific feature of $\ket{\Psi_{\rm CI}}$ being a product state (i.e. a single Slater determinant) or more generally a Gaussian state\cite{Bravyi:2005:LRF:2011637.2011640}.

To compute the spectrum of $h(k_x)$, we use its relation to the propagator 
\begin{eqnarray}
G^A_{i,j}(k_x)&=&\bra{\Psi_{\rm CI}}c^\dagger_{k_x,i}c_{k_x,j}\ket{\Psi_{\rm CI}}\label{eq:CIpropagator}
\end{eqnarray}
Indeed, theses two matrices, $G^A(k_x)$ and $h(k_x)$, are related by the following equation (see Ref.~\cite{Alexandradinata-PhysRevB.84.195103} for a detailed derivation)
\begin{eqnarray}
{G^A}^t&=&\frac{1}{1+e^{h(k_x)}}\label{eq:CIpropagatorentham}
\end{eqnarray}
Note that this propagator can be written of a sum of projectors onto the the $A$ part. As a consequence, its eigenvalues are between 0 and 1.

\begin{figure}
\centering
\includegraphics[width=0.48\linewidth]{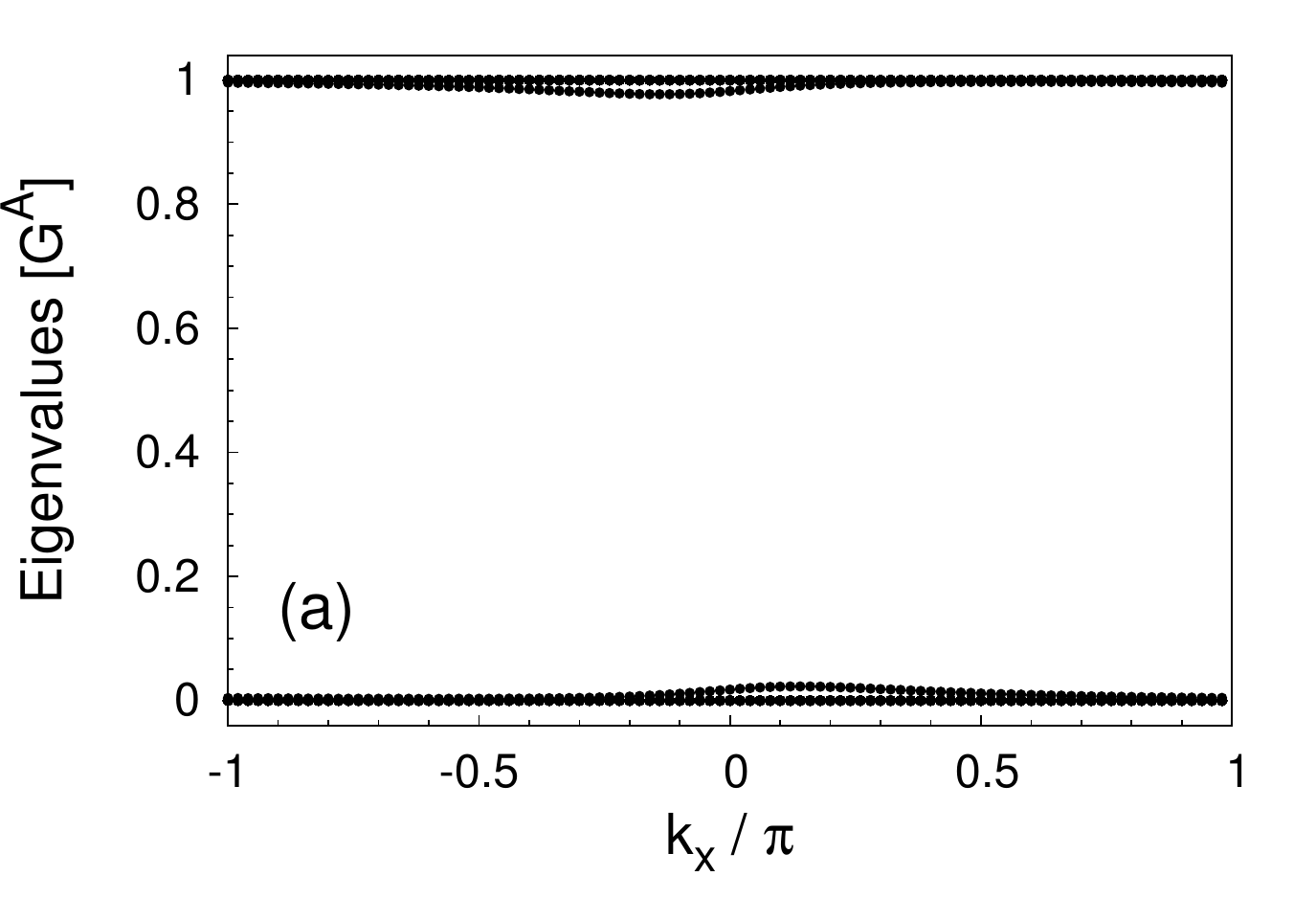}
\includegraphics[width=0.48\linewidth]{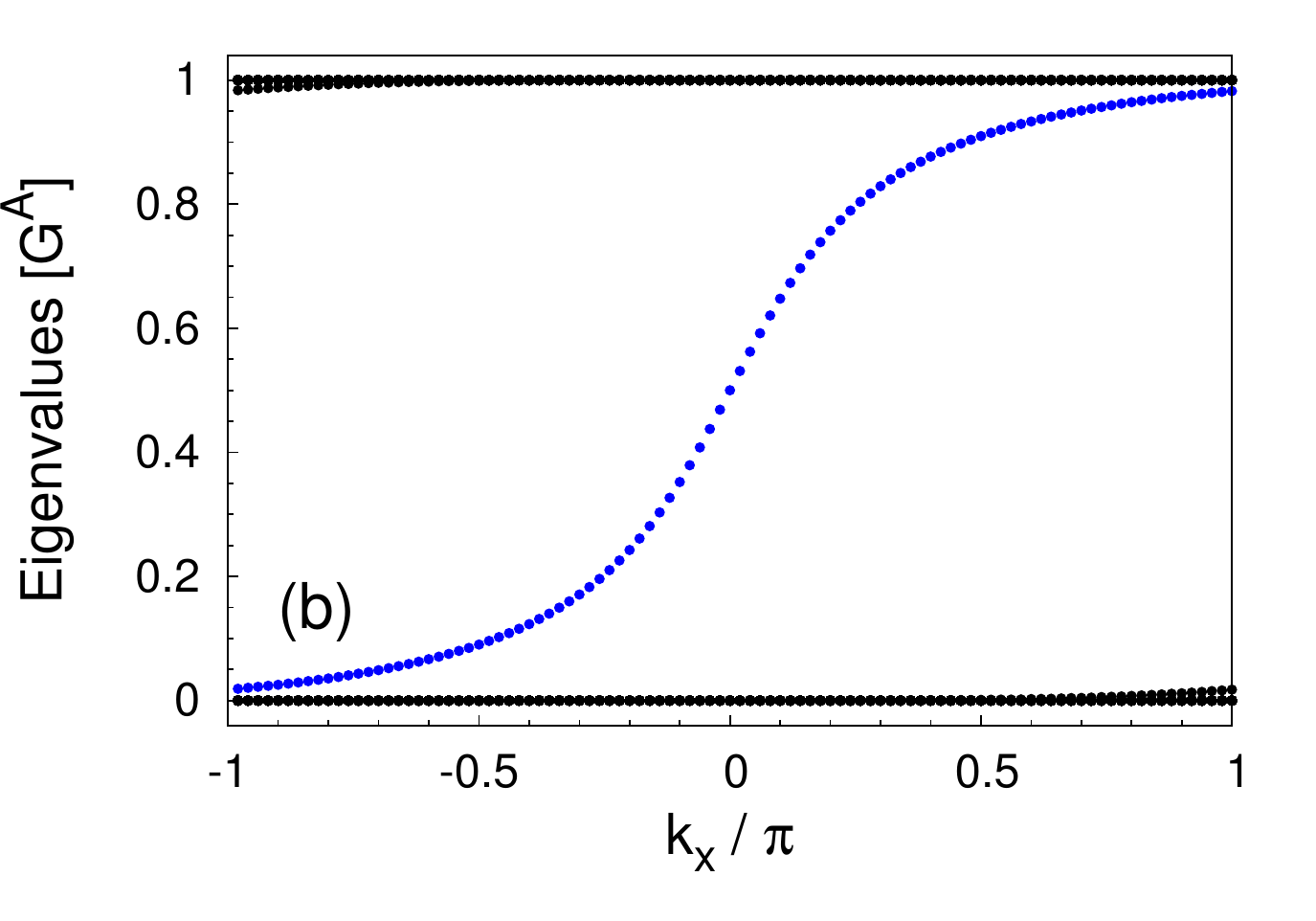}
\caption{{\it Top panel}: Eigenvalues of the propagator of Eq.~\ref{eq:CIpropagator} for the two-orbital lattice with a mass term set to $M=3$ (trivial phase). The partition is exactly half of the cylinder. There is a clear gap between eigenvalues localized around 0 and eigenvalues localized around 1. {\it Lower panel}: Eigenvalues of the propagator of Eq.~\ref{eq:CIpropagator} for the two-orbital lattice with a mass term set to $M=1$ (topological phase). The partition is exactly half of the cylinder. The blue dots that mimic the chiral edge mode, interpolate between the two bands. This is similar to the true edge mode shown in Fig.~\ref{fig:twoorbitalcylinder}b.}\label{fig:twoorbitalonebodyentspec}
\end{figure}

While the many-body ES exhibits the chiral edge mode of the CI (this ES is actually similar to the one that we will describe in Sec.\ref{sec:RSES} for interacting systems), the spectrum $G^A(k_x)$ (or $h(k_x)$) allows to unveil the same information from a one-body calculation. In Fig.~\ref{fig:twoorbitalonebodyentspec}, we show the spectrum of $G^A(k_x)$ for both the topological (Fig.~\ref{fig:twoorbitalonebodyentspec}a) and the trivial phase (Fig.~\ref{fig:twoorbitalonebodyentspec}b). Similar to its energy spectrum, the one-body ES of the trivial phases exhibits two bands (one located around 0, the other one located around 1) separated by a gap. On the other hand, the topological case clearly a chiral mode connecting the two bands. Like the AKLT model discussed in Sec.~\ref{subsubsection:spinAKLT}, the partition has introduced an artificial edge and the ES mimics the true edge spectrum of system.

Compared to the spin chain discussed in the Sec.~\ref{subsubsection:spinAKLT}, we see that plotting the entanglement energies as a function of conserved quantum number (here the momentum along the $x$ direction) is really helpful to directly observe the edge mode. But it is more a matter of convenience than a requirement. Indeed, using the ES to diagnose topological order can be done without this additional information. In the presence of disorder, $k_x$ would not be a good quantum number. Still by looking at the level statistics in the ES\cite{Prodan-PhysRevLett.105.115501}, we can differentiate a trivial phase from a topological phase. The CI case is one example of a two dimensional system where the ES can be computed analytically. Actually, this derivation holds true for any of the non-interacting topological insulators. Unfortunately such an analytical derivation is not feasible, in general, for strongly interacting systems as we will now discuss in the case the fractional quantum Hall systems.

\section{Fractional quantum Hall effect and entanglement spectra}\label{sec:FQHEandES}

In this section, we review the different aspects of entanglement spectra applied to the fractional quantum Hall effect. We provide a short (and partial) introduction to this topic. We discuss the different partitions that have been proposed and their relation. In particular, we show how much information about the excitations can be extracted from the ground state by using the entanglement spectra.

\subsection{Fractional quantum Hall effect: Overview and notations}\label{sec:FQHEOverview}

\begin{figure}
\centering
\includegraphics[width=0.30\linewidth]{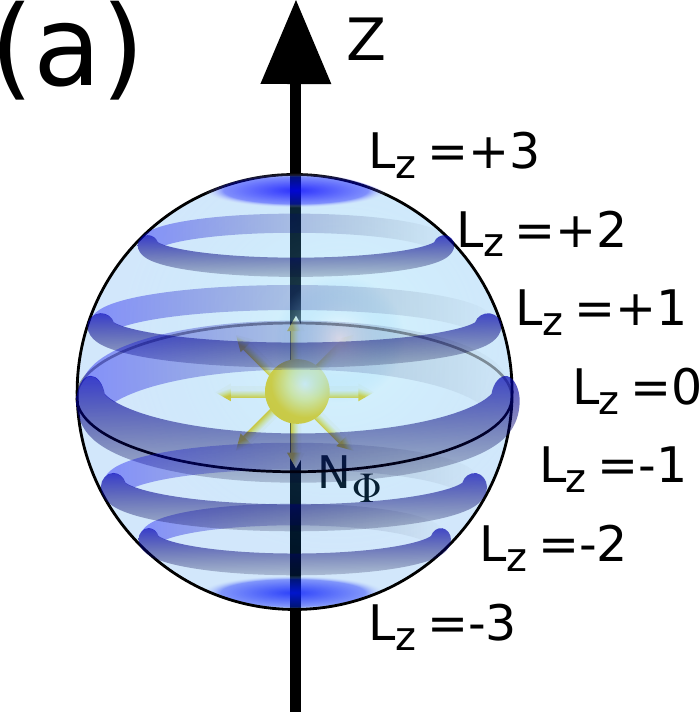}
\includegraphics[width=0.55\linewidth]{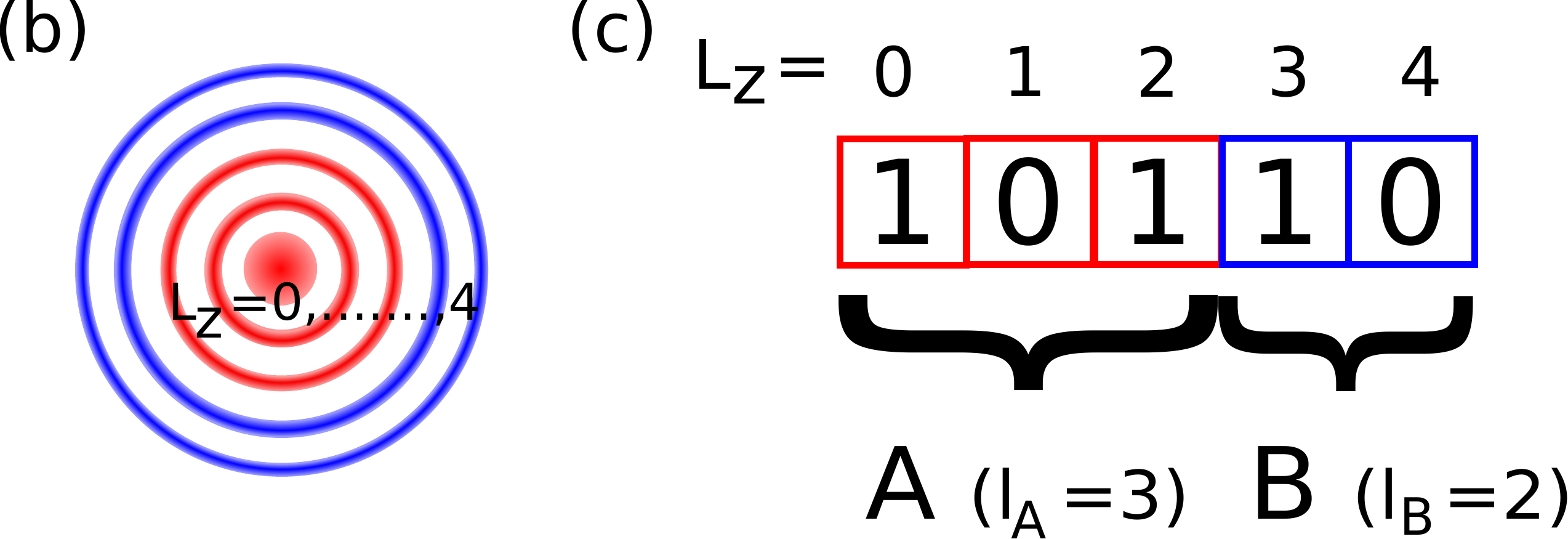}
\caption{Schematic representation of the orbital basis (a) on the sphere geometry for $N_{\Phi}=6$ and (b) on the disk geometry . (c) We show a typical $n$-body state of the occupation basis having three particles in orbital with angular momentum $L_z=0,2$ and $3$. When we perform an orbital partition into the $l_A$ leftmost orbitals (denoted in red here), the $A$ part in real space is roughly the domain that contains the three red orbitals.}\label{fig:spherediskorbitals}
\end{figure}

In this lecture notes, we restrict to the case of spinless particles occupying the lowest Landau level. The natural geometry to consider is the plane (or disk). For technical reasons, other geometries having periodic boundary conditions (in one or two directions) such as the cylinder\cite{rezayi-PhysRevB.50.17199}, the torus\cite{Haldane:1985p2786} or the sphere\cite{haldane-PhysRevLett.51.605}, are more convenient when it comes to finite size (numerical) studies. In the following, we will mostly focus on the genus zero surfaces and in particular the disk and the sphere. We note $N$ the number of particles in the system and $N_{\Phi}$ the number of flux quanta. The filling factor is defined (in the thermodynamical limit) as $\nu=N / N_{\Phi}$. A convenient choice for the one-body basis on the plane (using the symmetric gauge as discussed in Sec.~\ref{sec:IQHEOverview}) and on the sphere leads to the following set of wave functions:
\begin{equation}
\phi_m ({\bf r})= \left\{
\begin{array}{lcc}
\frac{1}{\sqrt{2 \pi  2^{m} m! }} z^m e^{-\frac{1}{4} |z|^2} & & \text{ plane} \\
& & \\
\sqrt{\frac{(N_{\Phi}+1)! }{4 \pi m! (N_{\Phi} - m)! } } u^{m} v^{N_{\Phi} - m} & & \text{ sphere}
\end{array}
\right.\label{eq:OrbitalOneBodyBasis}
\end{equation}
On the plane (or disk) $z=x+iy$ is the particle coordinate, $L_z=m$ is the angular momentum (where $m \ge 0$ is an integer). On the sphere, $u=\cos(\theta/2) e^{i \varphi/2}$ and $v=\sin(\theta/2) e^{-i \varphi/2}$ are the spinor coordinates with the polar coordinates $(\theta,\varphi)$, and $L_z=N_{\Phi}/2 - m$ is the angular momentum along $z$ where $m=0,1,\ldots,N_{\Phi}$. $N_{\Phi}$ is the number of flux quanta that pierce through the sphere. On such a closed geometry, both the radius ($\propto \sqrt{N_{\Phi}}$) of the sphere and the number of orbitals ($N_{\Phi}+1$) are fixed by the strength of the magnetic monopole at its center. Figs~\ref{fig:spherediskorbitals}a and \ref{fig:spherediskorbitals}b schematically describe these orbitals for both geometries.

On the plane geometry, a general quantum Hall wave function for $N$ particles in the lowest Landau level can be expressed as 
\begin{eqnarray}
\Psi \left(z_1,...,z_N\right)&= &P\left(z_1,...,z_N\right) e^{-\frac{1}{4}\sum_i |z_i|^2}\label{eq:GenericWavefunction}
\end{eqnarray}
where $P$ is a polynomial in the $N$ complex variables associated with the particle positions $z_1,...,z_N$. If we restrict to fermionic wave function, this polynomial has to be anti-symmetric. Note that any wave function written on the disk can also be obtained on the sphere using the stereographic projection by identifying $z \equiv u/v$ (up to some global factor). To simplify our equations, we will drop the gaussian factor in any wave functions. So when discussing model wave function on the plane or the sphere geometry, it is sufficient to provide $P$ and we can drop all the other factors. We can decompose this wave function in the occupation basis, using the orbitals of Eq.~\ref{eq:OrbitalOneBodyBasis}
\begin{eqnarray}
\Psi\left(z_1,...,z_N\right) &=& \sum_{\{\lambda\}} c_{\lambda} \mathcal{M}_{\lambda}\left(z_1,...,z_N\right) \label{eq:GenericWavefunctionDecomposition}
\end{eqnarray}
$\mathcal{M}_{\lambda}$ is the normalized Slater determinant that has its orbital occupation given by the configuration $\lambda$ (such a configuration is shown in Fig.~\ref{fig:spherediskorbitals}c). The functions $\mathcal{M}_{\lambda} $ form a set of orthonormal free many-body states. When the wave function is obtained through numerical simulation, such a decomposition is directly accessible: One diagonalizes an Hamiltonian by expressing it in a convenient basis which is generally the occupation basis for the FQHE. For model wave functions such as the Laughlin\cite{Laughlin:1983p301} or Moore-Read\cite{Moore:1991p165} states, one can use an efficient recursive algorithm\cite{bernevig-09prl206801} that provides the corresponding decomposition. 

The archetype of fractional quantum Hall (FQH) model wave function is the celebrated Laughlin state\cite{Laughlin:1983p301}
\begin{eqnarray}
\Psi_{\rm Lgh}\left(z_1,...,z_N\right) &=& \prod_{i<j} \left(z_i-z_j\right)^m\label{eq:Laughlin}
\end{eqnarray}
$m$ is the only variational parameter. Actually, $m$ is related to the filling factor $\nu=1/m$. On the sphere geometry, Eq.~\ref{eq:Laughlin} implies the relation $N_{\Phi}=m (N-1)$. For a fermionic wave function, $m$ has to be odd. $m=1$ corresponds to the completely filled lowest Landau level as pointed out in Eq.~\ref{eq:IQHEwavefunctionfirstquantized}. It is a single Slater determinant (Vandermonde determinant) and thus is a product state in the occupation basis. At $\nu=1/3$, the Laughlin wave function is a very accurate approximation of the FQH ground state obtained through any realistic simulation. Being an intrinsic topological phase in its full glory (as opposed to the integer quantum Hall effect), the Laughlin wave function is degenerate when placed on a higher genus surface. For example, it is $m$-fold degenerate on the torus geometry.

\begin{figure}
\centering
\includegraphics[width=0.95\linewidth]{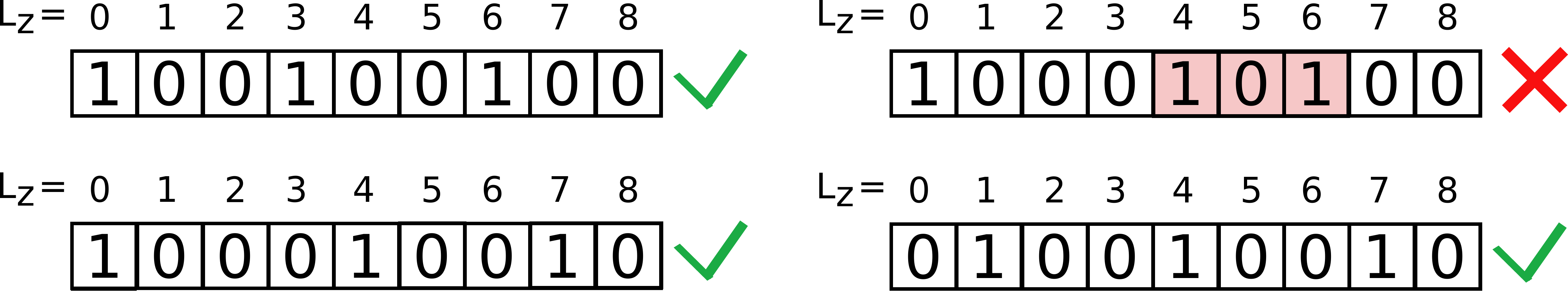}
\caption{An example of the Haldane's exclusion principle. We consider a system with $N=3$ particles in $9$ orbitals on the disk with momenta going from $L_z=0$ to $L_z=8$. Among the four configurations described here, only three of them satisfy the Haldane's exclusion principle of the $\nu=1/3$ Laughlin: No more than one particle in three consecutive orbitals. The violation of this principle in the top right configuration is shown in light red. For each compatible partition, one can easily compute the corresponding total $L_z$ value (for example $9$ for the left topmost configuration).}\label{fig:exclusionprinciple}
\end{figure}

Bulk excitations can be nucleated by removing (for quasielectron) or inserting (for quasihole) fluxes. Each excitation carries a fractional charge ($\pm e/m$) and obeys fractional statistics. For one quasihole located at the position $\eta$, we can write the corresponding wave function
\begin{eqnarray}
\Psi_{\rm Lgh, 1qh}\left(z_1,...,z_N;\eta\right) &=& \prod_{i} \left(z_i-\eta\right) \Psi_{\rm Lgh}\left(z_1,...,z_N\right)\label{eq:LaughlinOneQH}
\end{eqnarray}
Changing the quasihole position $\eta$ spans a subspace described by a basis of $N+1$ quasihole states, each of them having a well and unique defined angular momentum. More generally, the number of quasihole states for given values of $N$ and $N_{\Phi}$ is a signature of the phase and acts as a fingerprint that can be tracked in numerical simulations. This counting of states can be obtained by the Haldane's exclusion principle\cite{Haldane-PhysRevLett.67.937} (or Haldane statistics). For the $\nu=1/m$ Laughlin, this number is identical to the number of configurations with $N$ particles and $N_{\Phi}+1$ orbitals where there is no more than $1$ particle in $m$ consecutive orbitals. Fig.~\ref{fig:exclusionprinciple} give some simple examples of compatible configurations. Note that both the Laughlin wave function and its quasihole excitations are the only zero energy states of a local two-body model interaction\cite{haldane-PhysRevLett.51.605}, Eq.~\ref{eq:LaughlinOneQH} being among them the unique densest zero energy state.

\begin{figure}
\centering
\includegraphics[width=0.75\linewidth]{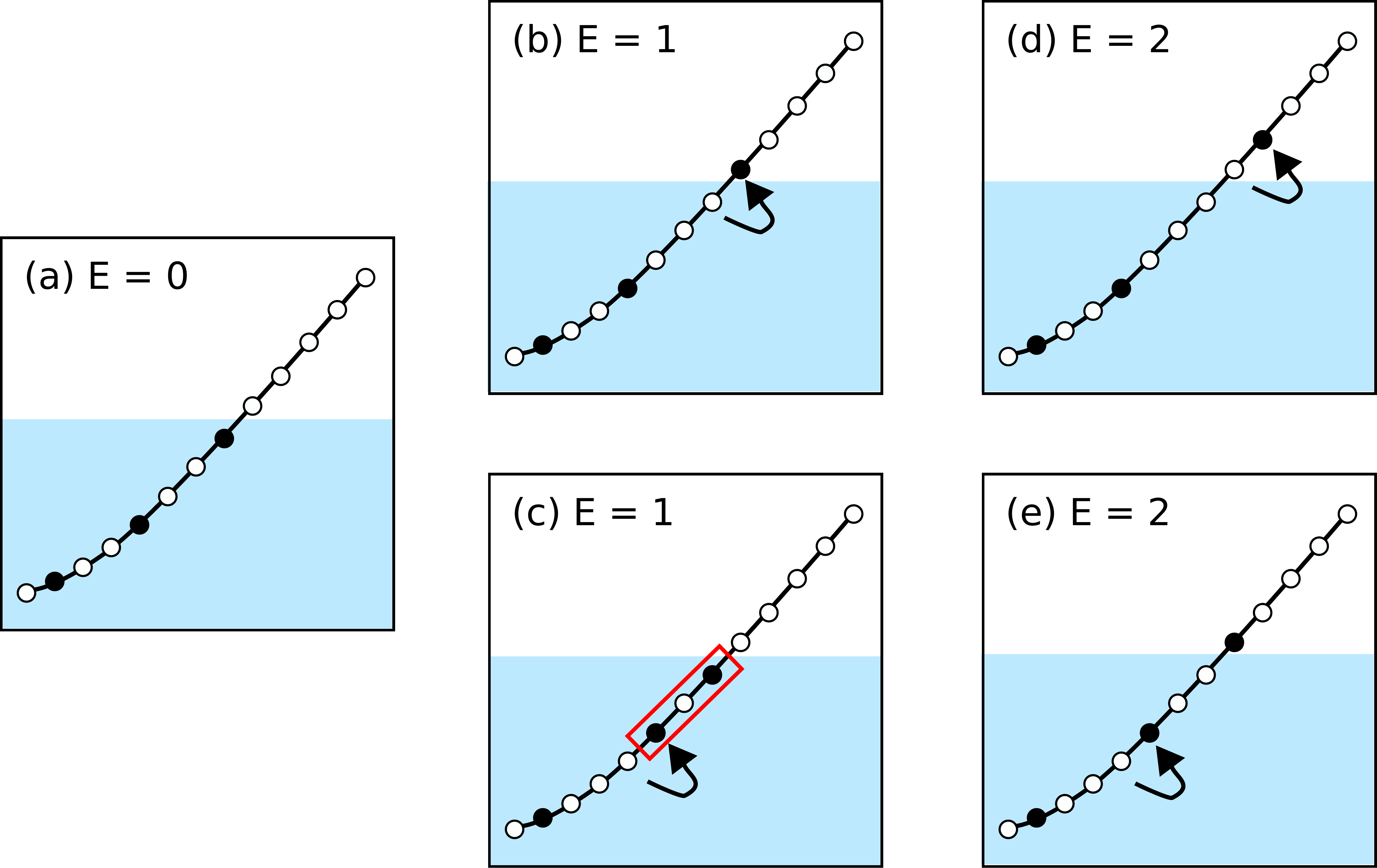}
\caption{A description of the chiral $U(1)$ edge mode counting at $\nu=1/3$. (a) The ground state with energy $E=0$. It obeys the Haldane statistics (no more than one particle in three consecutive orbitals). The light blue denotes the region bellow the Fermi level. (b) The lowest energy excitation ($E=1$) that satisfies the Haldane statistics. (c) An example of an excitation at the same energy but violating (in red) the Haldane statistics. (d) and (e) The two possible excitations at $E=2$.}\label{fig:u1counting}
\end{figure}

The edge excitations of the Laughlin state are described by a chiral $U(1)$ boson. For an edge of length ${\mathcal L}$, the dispersion relation is given by $E \simeq \frac{2 \pi v}{{\mathcal L}} n$ where $n$ is an integer, $v$ is the edge mode velocity. The degeneracy of each energy level can be deduced from the picture described in Fig.~\ref{fig:u1counting}. Using the Haldane statistics for the Laughlin $\nu=1/m$ state, starting from the ground state, we obtain the sequence $1,1,2,3,...$ irrespective of $m$, identical to the counting of the IQHE discussed in Sec.~\ref{sec:IQHEOverview}. Like in the case of the quasihole state, this counting is a fingerprint of the edge excitations.

Finally, we give another example of model wave functions: The Moore-Read state\cite{Moore:1991p165}. This model is considered to be the prototype model wave function to explain the appearance of a Hall conductance plateau at filling factor $\nu=5/2$ (i.e. in the second Landau level). It can be written as 
\begin{eqnarray}
\Psi_{\rm MR}\left(z_1,...,z_N\right) &=& {\rm Pf} \left( \frac{1}{z_i-z_j} \right) \prod_{i<j} \left(z_i-z_j\right)^2\label{eq:MooreRead}
\end{eqnarray}
The Moore-Read state possesses two kind of excitations: Abelian excitations with a charge $\pm e/2$ and non-Abelian excitations carrying a charge $\pm e/4$. In a similar way to the Laughlin case, the number of quasihole states can be derived from the Haldane's exclusion principle (in that case, no more than $2$ particles in $4$ consecutive orbitals). The Moore-Read state has two edge modes: A charge edge mode similar to the one of the Laughlin state and a neutral Majorana fermion edge mode. Note that a natural way to build the Moore-Read state is based on the conformal field theory (CFT)\cite{Moore:1991p165}, rewriting Eq.~\ref{eq:MooreRead} as a correlator (using the CFT of the Ising model in that present case). 

\subsection{Orbital entanglement spectrum}\label{sec:OES}

\begin{figure}
\centering
\includegraphics[width=0.80\linewidth]{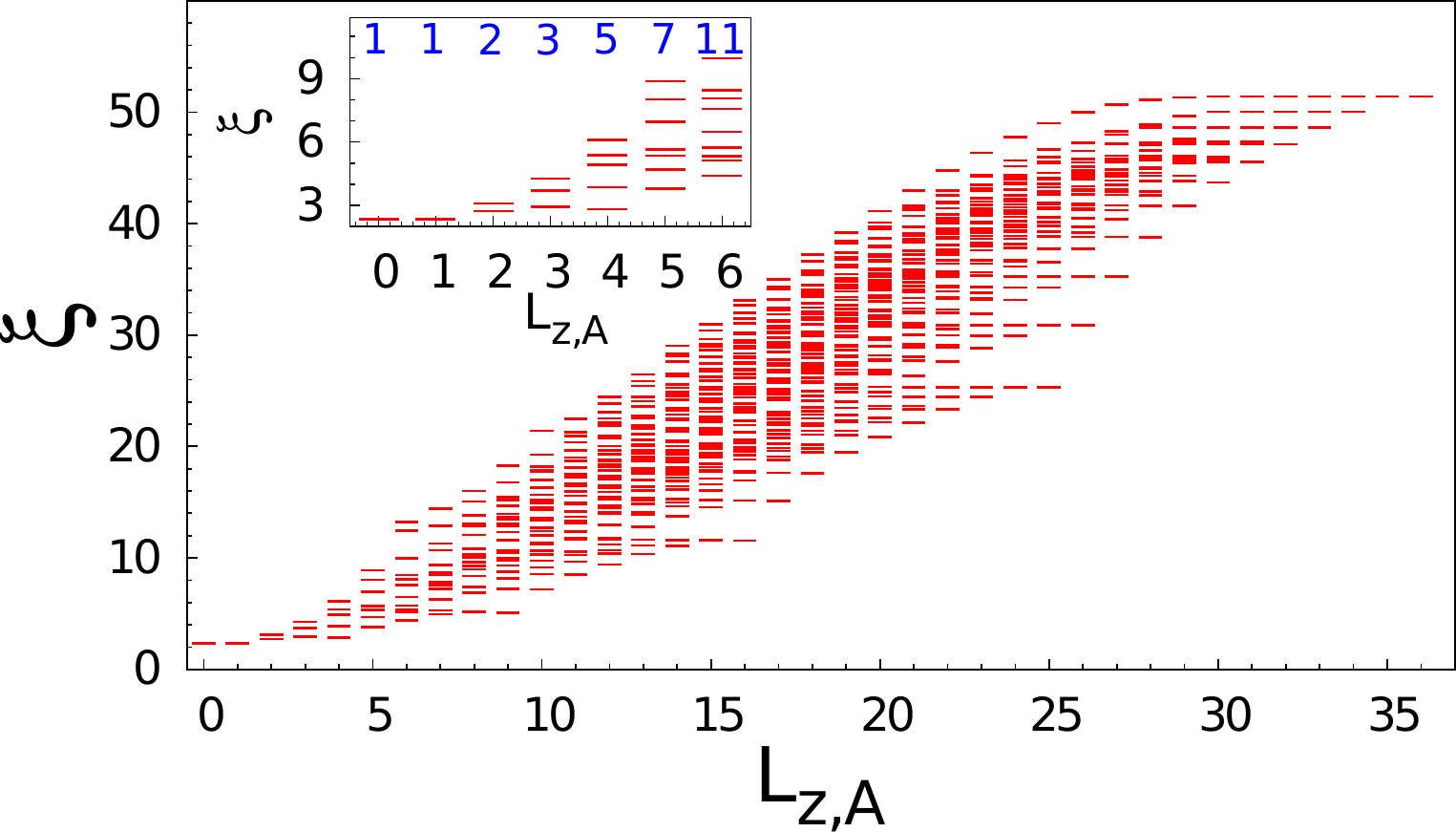}
\caption{OES for the $\nu=1/3$ Laughlin with $N=12$ fermions on the sphere geometry, keeping $l_A=17$ orbitals and looking at the fixed number of particles sector $N_A=6$. The inset provides a zoom on the entanglement spectrum related to the $U(1)$ edge mode counting of the Laughin state. As expected this counting is $1,1,2,3,5,7,11$. For this cut, the deviation of the OES counting to the edge mode counting (due to finite size effects) starts at $L_{z,A}=7$: The OES gives 13 levels while the $U(1)$ counting is 15.}\label{fig:oeslaughlin}
\end{figure}

Li and Haldane~\cite{li-08prl010504} proposed to compute the ES of a FQH state using a partition in orbital basis. We call this type of ES the orbital entanglement spectrum (OES). As already pointed out in Ref.~\cite{Haque-PhysRevLett.98.060401} where the authors tried to extract the topological entanglement entropy of the Laughlin state from the wave function, a cut in the orbital basis roughly mimics a cut in real space. The OES is defined by the number of consecutive orbitals that are considered. This number will be denoted $l_A$, the number of orbitals for the $B$ part being $l_B$ satisfying $l_A+l_B=N_{\Phi} + 1$ on the sphere geometry. When we compute the OES for a FQH state on the geometry such as the sphere or the disk, one can use two good quantum numbers to label the blocks of the reduced density matrix: $N_A$ the number of particles in $A$ and $L_{z,A}$ the projection of the total angular momentum of the particles in $A$. The OES is generally represented as the entanglement energies $\xi$ as a function of $L_{z,A}$ for a fixed value of $N_A$. A typical example is available in Fig.~\ref{fig:oeslaughlin} for the $\nu=1/3$ Laughlin state. Note that for sake of simplicity and as opposed to many of the original publications, the non-trivial part of the OES is located at the left hand side of the plot. We also shift the origin of $L_{z,A}$ such that the leftmost entanglement level state has $L_{z,A}=0$.

The OES of the Laughlin state is highly specific: Any random state with the same symmetry would produce much more entanglement energy levels, i.e. it would have much less zero eigenvalues in the reduced density matrix. Actually, not only additional entanglement energy levels would be present in sector of $L_{z,A}$ where there is no level for the Laughlin state, but the total number of level would be exponentially larger. Thus such a model state induces large constraints on the reduced density matrix. In Fig.~\ref{fig:oeslaughlin}, we observe that the counting of entanglement energies starting from the left matches the sequence $1,1,2,3,5,7,11$. This is the expected counting sequence for a chiral $U(1)$ boson edge mode as discussed in Sec.~\ref{sec:FQHEOverview}. Beyond a given $L_{z,A}$, the OES counting starts being lower than the $U(1)$. Knowing we are dealing with a finite size system (both in the number of orbitals and particles), there is a maximum value of $L_{z,A}$ that can be reached and there is single state with $l_A$ orbitals and $N_A$ fermions that can reach it. Thus it is clear than both countings should differ at some point since the $U(1)$ counting keeps growing. We name the thermodynamical region the part of the OES where no size effect affects the edge mode counting. This region increases with the system size. In a simplified picture, we can think about the unique state at $L_{z,A}=0$ of the OES as a Laughlin liquid droplet for $N_A$ particles. Slightly increasing $L_{z,A}$ corresponds to generate edge mode excitations. A more rigorous derivation of this schematic point of view will be described in Sec.~\ref{sec:PES}.

From this observation, Li and Haldane conjectured that in the thermodynamical limit the OES should be identical to the energy spectrum of the edge mode of the model state. This statement goes beyond the counting argument which is in itself a signature of the edge physics. To corroborate this idea, one can look at the evolution of the entanglement energies when we increase the system size\cite{Thomale-PhysRevLett.104.180502}. These energies should mimic the dispersion relation of the gapless edge mode $\frac{2 \pi v}{{\mathcal L}} L_{z,A} $ where ${\mathcal L}$ is the cut perimeter. Despite some indication that this description is correct, the finite size calculations are unable to make a definitive conclusion. A more accurate approach will be discussed in Sec.~\ref{sec:RSES}, and will provide more convincing evidences of this conjecture.

\begin{figure}
\centering
\includegraphics[width=0.45\linewidth]{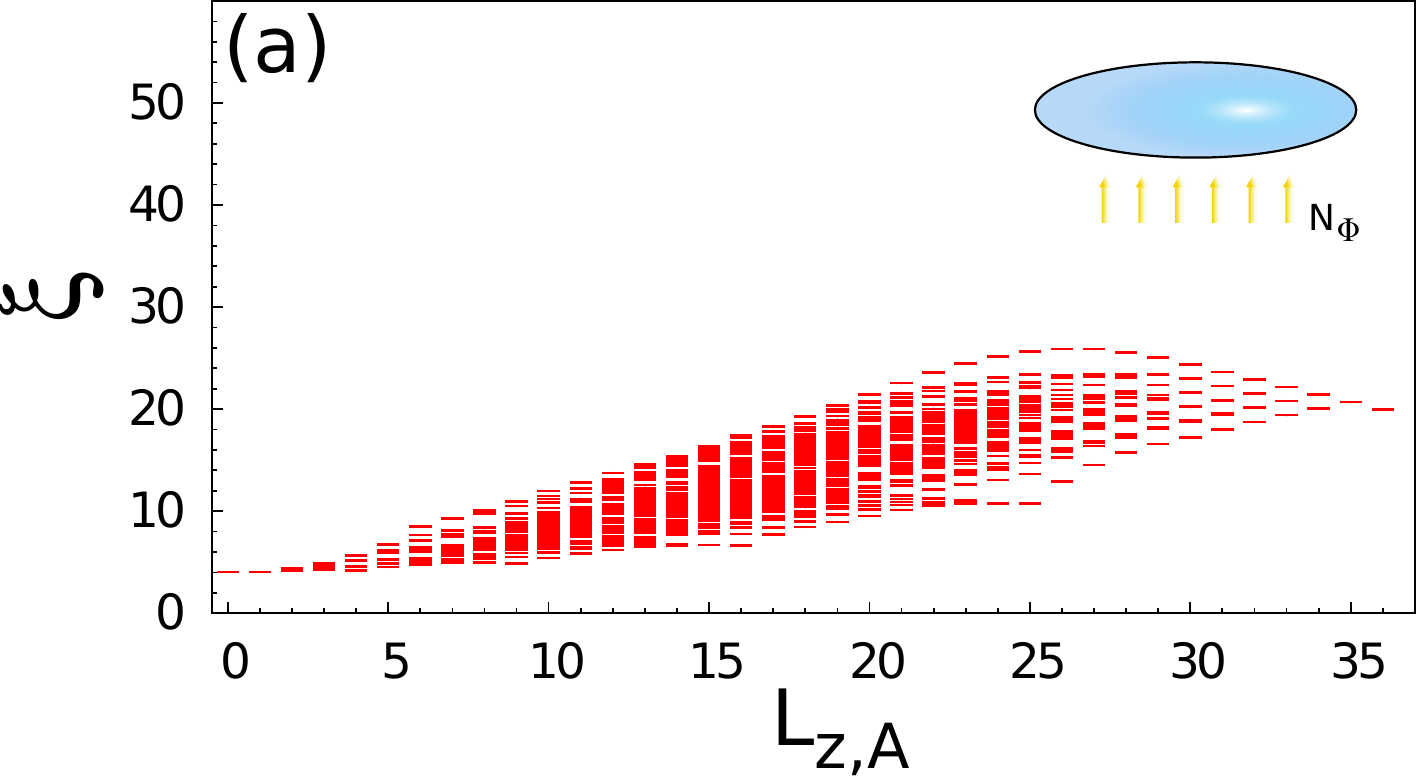}
\includegraphics[width=0.45\linewidth]{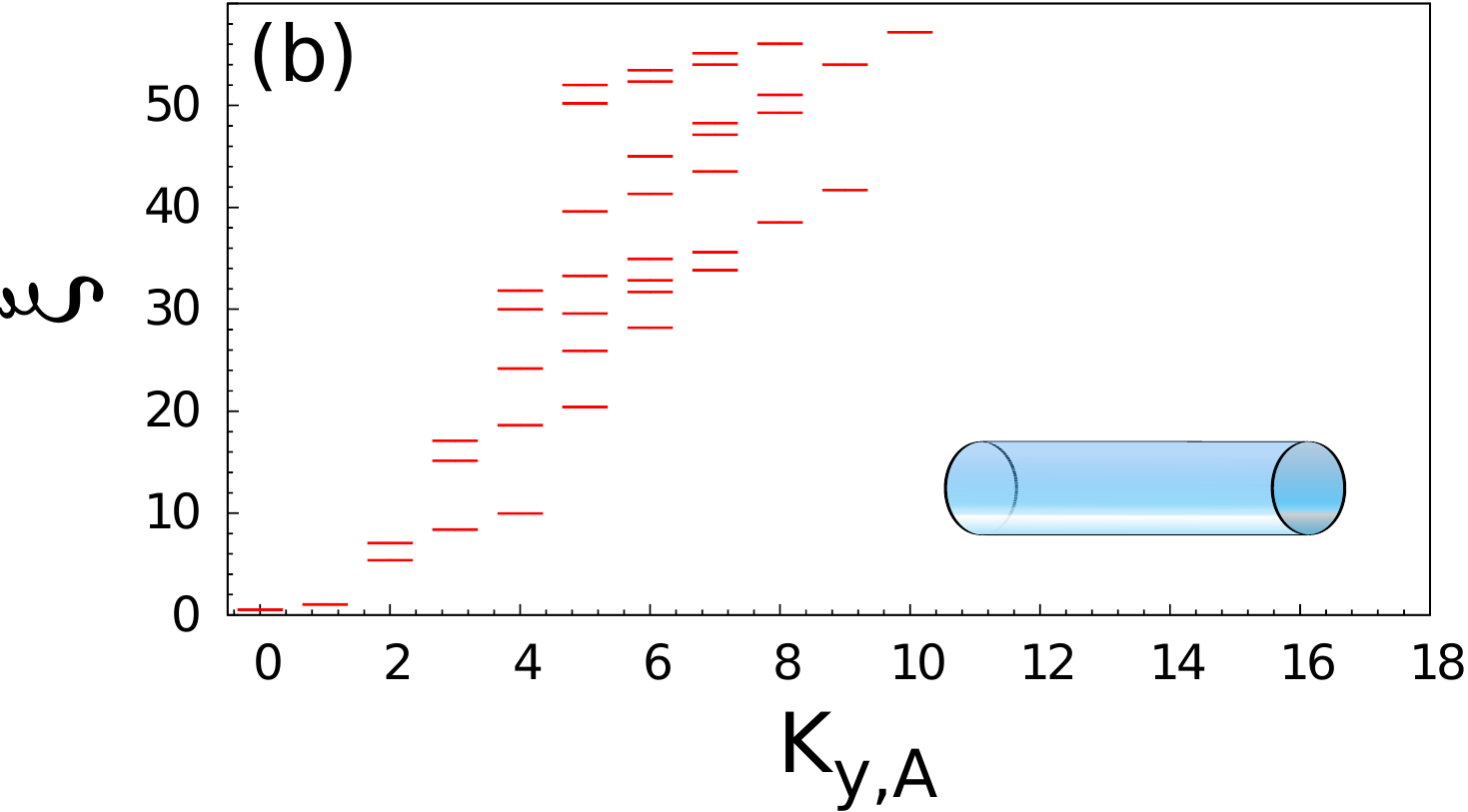}\\

\includegraphics[width=0.45\linewidth]{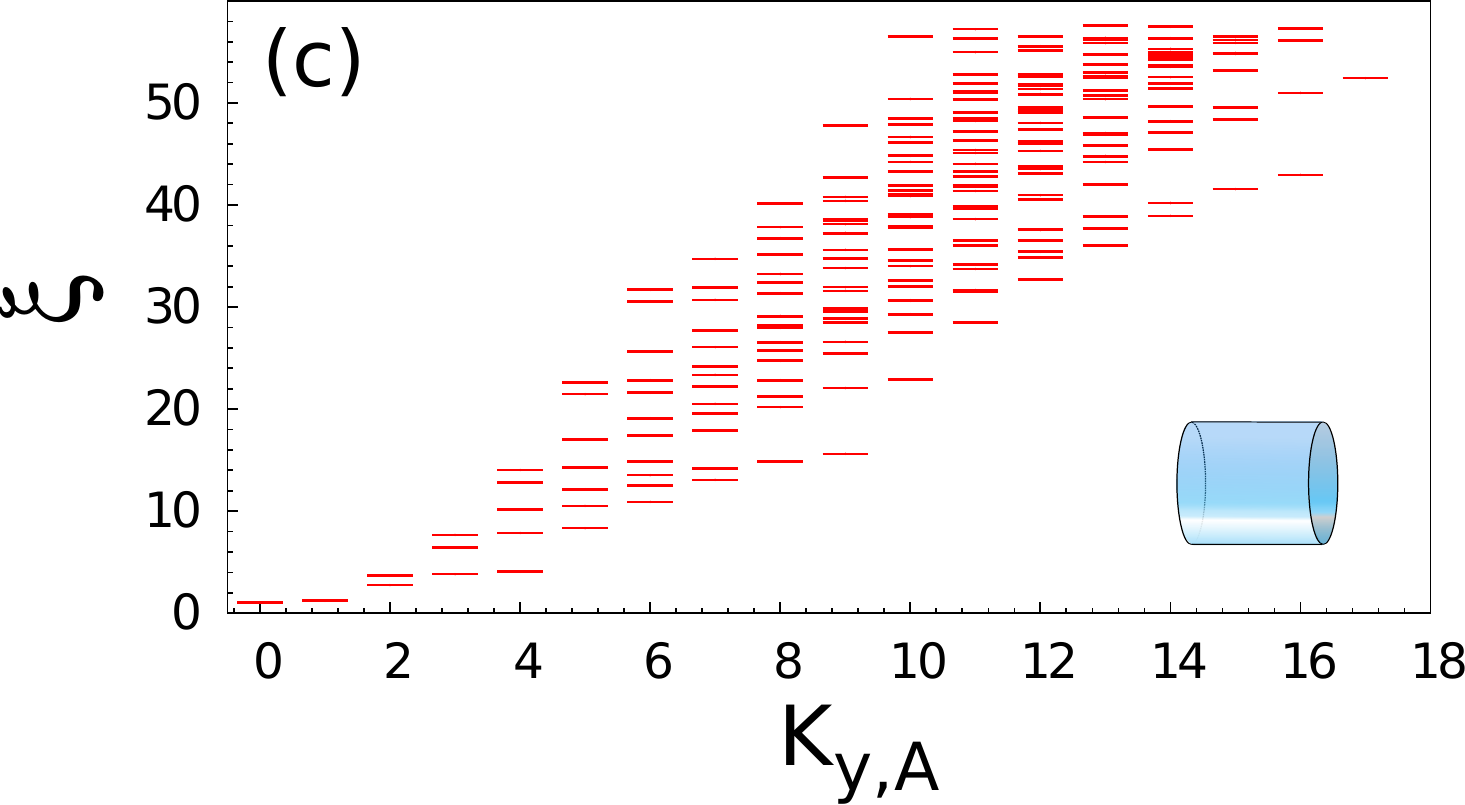}
\includegraphics[width=0.45\linewidth]{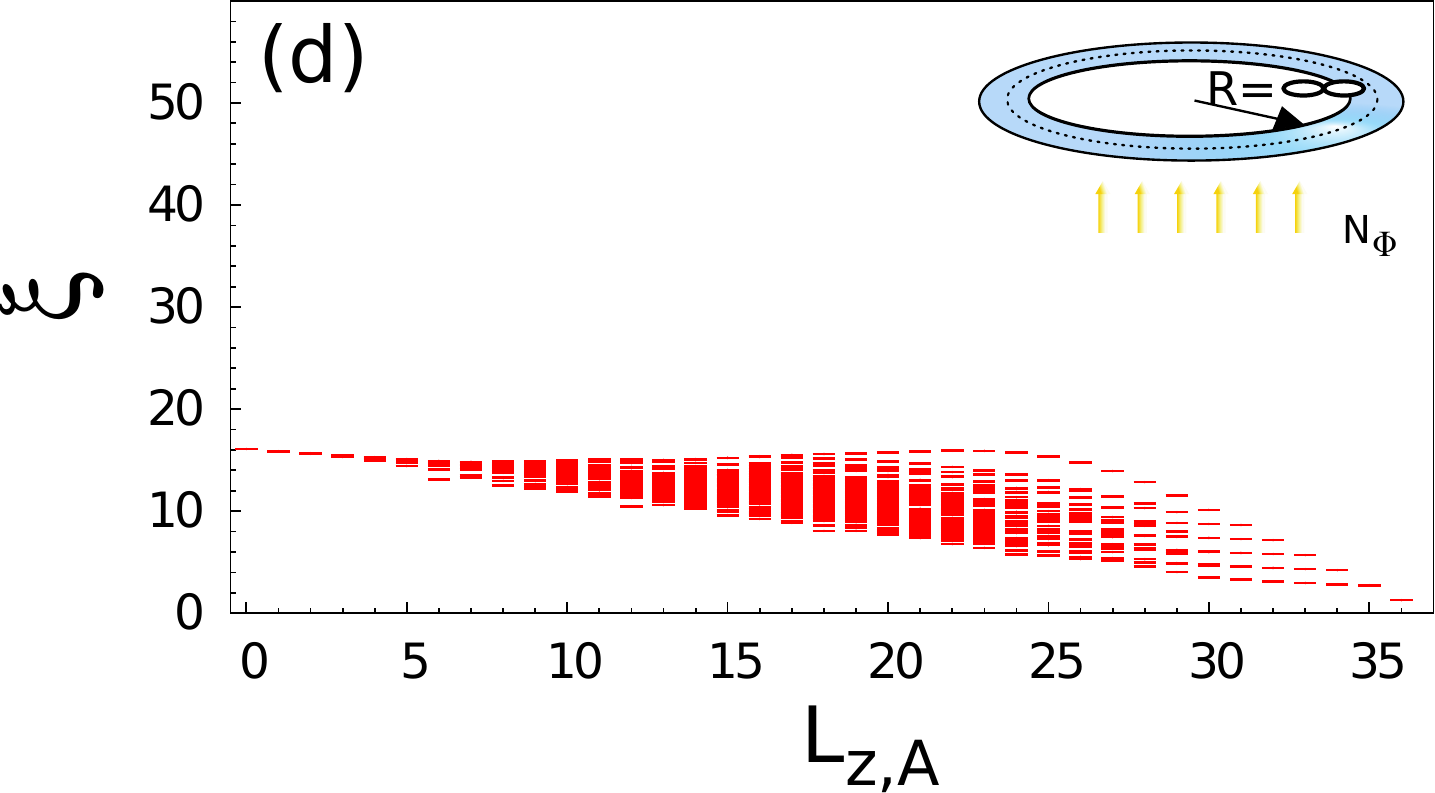}
\caption{{\it From left to right, top to bottom:} Orbital entanglement spectrum for the $\nu=1/3$ Laughlin with $N=12$ fermions, keeping $l_A=17$ orbitals and looking at the fixed number of particles sector $N_A=6$, on different geometries: The disk geometry (a), a thin cylinder with a perimeter of $L=10 l_B$ (b), a thicker cylinder with a perimeter of $L=15 l_B$ (c) and the thin annulus limit (d). Note that for the cylinder geometry, we use the momentum along the cylinder perimeter $K_{y,A}$ instead of the angular momentum $L_{z,A}$.}\label{fig:oesgeometries}
\end{figure}

A similar calculation can be performed on other geometries. Figs~\ref{fig:oesgeometries}a-d shows the OES of the $\nu=1/3$ Laughlin state for the disk, the cylinder and the thin annulus (or conformal limit\cite{Thomale-PhysRevLett.104.180502}). While the shape of the OES depends on the geometry, the counting remains identical as long as one considers genus zero surfaces. The OES on the two different cylinder in Figs~\ref{fig:oesgeometries}c and \ref{fig:oesgeometries}d are a clear consequence of the area law. While the OES is an approximation of a real space cut, its shape depends on the length of the cut. On the cylinder, this length is the cylinder circumference (or perimeter), and does not vary with the number of flux quanta $N_{\Phi}$ (the usual hemisphere cut for the sphere would give a length proportional to $\sqrt{N_{\Phi}}$). A smaller perimeter and thus a smaller entanglement entropy, results in an OES with a steeper slope.

Moving to a higher genus surface like the torus leads to a slightly different picture\cite{lauchli-PhysRevLett.104.156404}. The usual orbital basis on the torus is translational invariant along one direction of this geometry. Performing the bipartite partition gives rise to two artificial and spatially separated edges. The OES mimics the physics of two counter-propagating edge modes. A consequence of this interplay between these two modes is the absence of zero eigenvalue in the reduced density matrix. This is a major difference with the OES for the genus zero geometries.

Beyond the thermodynamical region of the OES, finite size effects start to appear. There, the spectrum also has a non-trivial structure compared to a generic wave function. For most of the model wave functions, there is no quantitative understanding of this non-thermodynamical part of the OES. In the case of the $\nu=1/m$ Laughlin state, it was shown\cite{Hermanns-PhysRevB.84.121309} that the counting of this region can actually be deduced from a generalized exclusion principle that depends on $m$. Actually, it is a nice example where finite size effects allow to get more information than the thermodynamical limit: While all $\nu=1/m$ Laughlin states have the same edge theory, a chiral $U(1)$ boson, the compactification radius of the bosons depends on $m$ ($\sqrt{m}$ in that case). The thermodynamical region gives access to the $U(1)$ counting whereas the finite size effects encode the value $m$.

\begin{figure}
\centering
\includegraphics[width=0.49\linewidth]{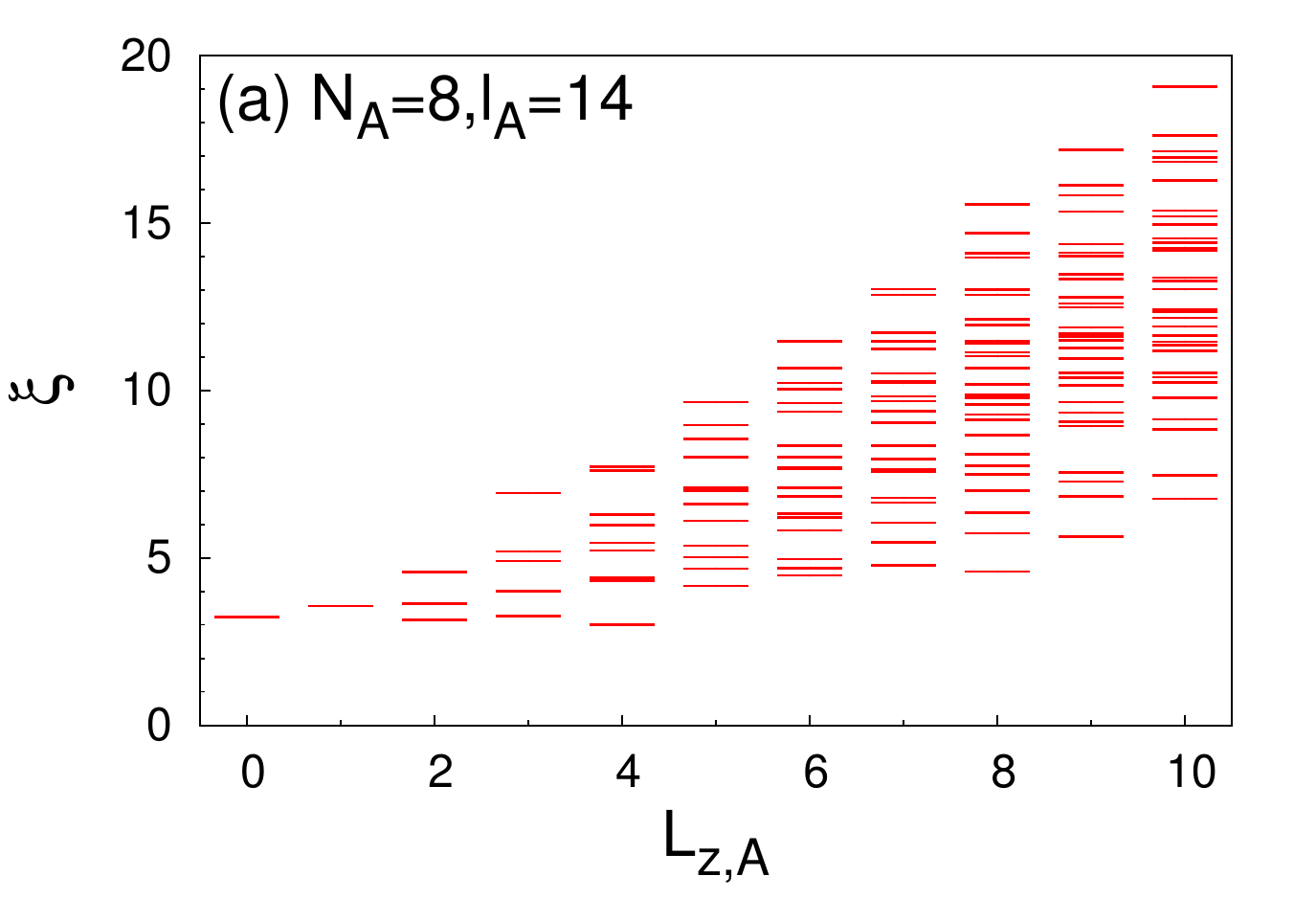}
\includegraphics[width=0.49\linewidth]{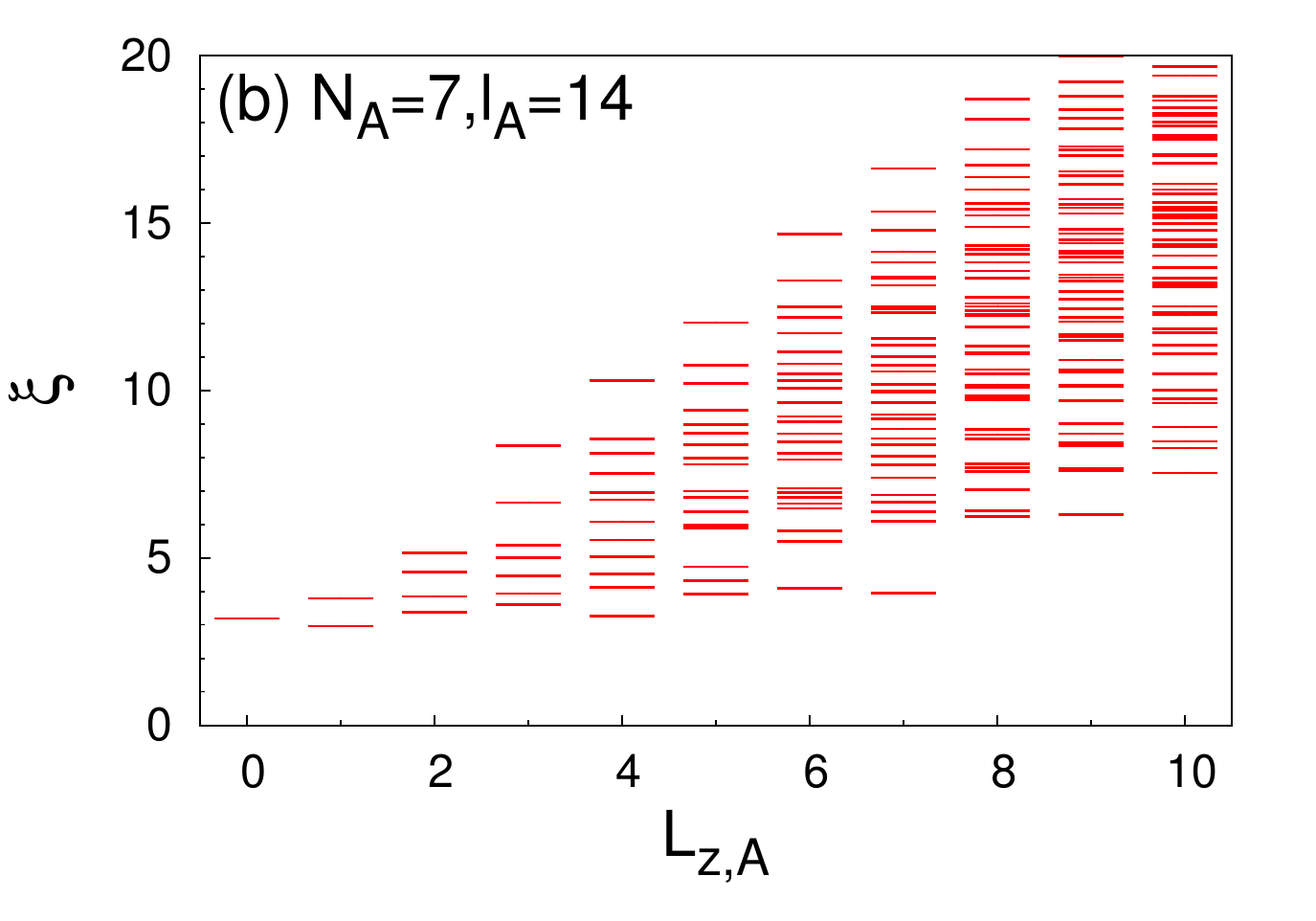}
\caption{OES for the $\nu=2+1/2$ Moore-Read state with $N=16$ fermions keeping $l_A=14$ orbitals. (a) Setting $N_A=8$ we observe the counting $1,1,3,5,...$ which is related to the identity sector. (b) Setting $N_A=7$, another counting emerges $1,2,4,7,...$ corresponding to the $\psi$ sector. In both cases and for this system size, the counting starts to deviate from the CFT one starting from $L_z=4$.}\label{fig:oesmooreread}
\end{figure}

More complex model wave functions exhibit a richer OES structure. We focus on the Moore-Read state. Figs.~\ref{fig:oesmooreread}a and \ref{fig:oesmooreread}b give the OES for this state using the same partition (here $l_A=14$) but looking at two different blocks of the reduced density matrix, namely $N_A=8$ for Fig.~\ref{fig:oesmooreread}a and $N_A=7$ for Fig.~\ref{fig:oesmooreread}b. For these two cases, the counting in the thermodynamical region is different, reflecting the two sectors of the CFT (namely the identity and the $\psi$ sectors) used to build this state. A surprising result here is that the state itself is built only using one of the two sectors, whereas the OES exhibits both.

\begin{figure}
\centering
\includegraphics[width=0.48\linewidth]{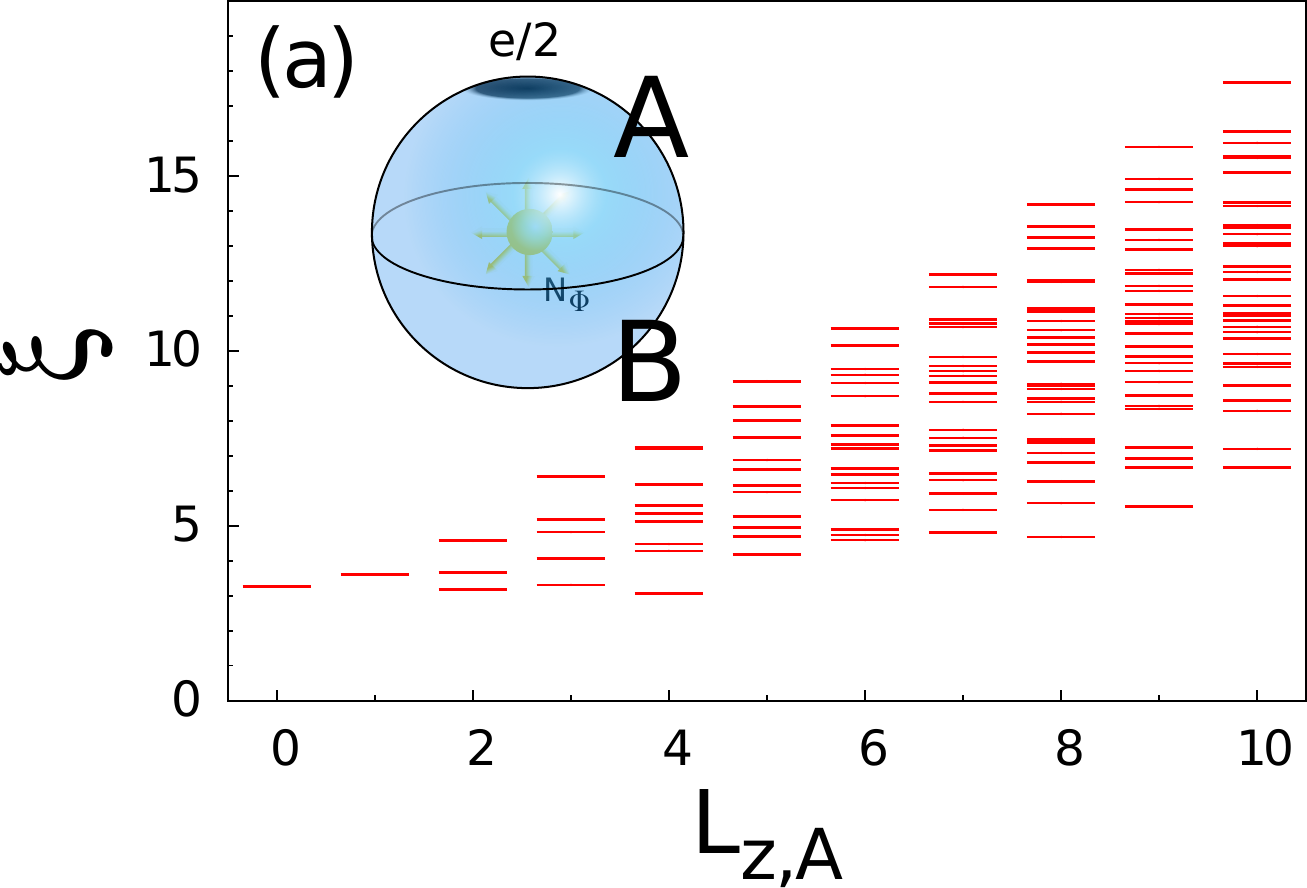}
\includegraphics[width=0.48\linewidth]{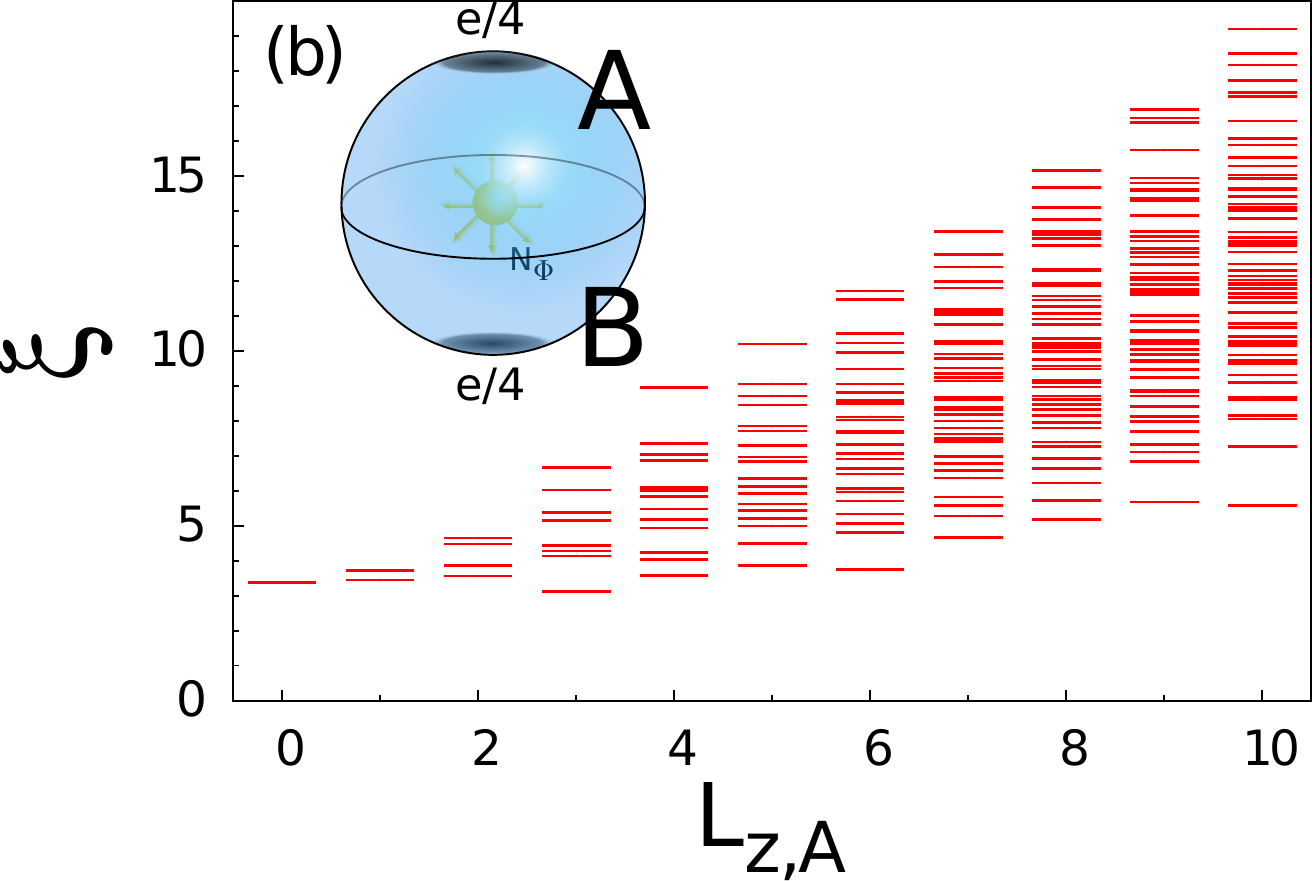}
\caption{Orbital entanglement spectrum for the $\nu=2+1/2$ Moore-Read state with quasiholes for $N=16$ fermions and $N_{\Phi}=30$ flux quanta, keeping $N_A=8$ particles and $l_A=15$ orbitals. We show two different situations: (a) one Abelian excitation of charge $e/2$ is located at the north pole in the hemisphere $A$, (b) two non-Abelian excitations, each carrying a charge $e/4$, one at each pole. Both situations clearly exhibit a different counting, namely here $1,1,3,5,...$ for (a) and $1,2,4,8,...$ (non-Abelian sector).}\label{fig:oesmoorereadExcitations}
\end{figure}

For the time being, we have only looked at the OES for the ground state (i.e. in the absence of excitations) of model wave functions. But the OES in presence of pinned excitations is also quite insightful\cite{Papic-PhysRevLett.106.056801}. Fig.~\ref{fig:oesmoorereadExcitations} shows the OES for the Moore-Read state in presence of pinned quasihole excitations. In order to preserve the rotation symmetry along the $z$ axis of the sphere, the excitations are located at the poles. We consider two situations: One Abelian excitation of charge $e/2$ is located at the north pole in the hemisphere $A$ (Fig.~\ref{fig:oesmoorereadExcitations}a) and two non-Abelian excitations, each carrying a charge $e/4$, one at each pole (Fig.~\ref{fig:oesmoorereadExcitations}b). In both cases, the system has the same number of particles and flux quanta, and the OES is computed for the same parameters for $l_A$ and $N_A$. Still, the OES clearly exhibits a different counting: For Fig.~\ref{fig:oesmoorereadExcitations}a we recover the counting of the vacuum sector (like Fig.~\ref{fig:oesmooreread}a) and for Fig.~\ref{fig:oesmoorereadExcitations}b we recover the counting of the non-Abelian sector sector (like Fig.~\ref{fig:oesmooreread}b). Thus the OES can be used as a probe to check the parity of the number of non-Abelian excitations in a region of the system. Note that the OES of the Laughlin state is not modified by the presence of pinned quasihole excitations.

As a final note about the OES for model wave functions, we should stress once again that the rank of the reduced density matrix being exponentially smaller than any random state with the same symmetry is a major property. We have discussed in Sec.~\ref{sec:MPSES} that such a feature is a signal that an efficient MPS formula might exist. In the case of the FQHE, the edge modes are gapless. The ES reflecting the edge physics, we expect this MPS to be infinite, as opposed to the AKLT example. Indeed, recent developments\cite{zaletel-PhysRevB.86.245305,Estienne-PhysRevB.87.161112,Estienne-2013arXiv1311.2936E} have shown that an (infinite) MPS formulation was available for a large class of model wave functions, with a well controlled truncation parameter that allows numerical calculations.

\subsection{OES beyond model wave functions}\label{sec:OESRealistic}

While the OES has already allowed to get some insight on the information encoded within the ground state of a topological phase, we would like to use it as a probe to detect topological order. For that purpose, we need to move away from model state. When dealing with more realistic description of FQH systems, several assumptions are made. In general, we suppose that there is no Landau level mixing, and in many cases we also assume that electrons are spin polarized. For low filling factor (such as $\nu=1/3$) these hypothesis are quite accurate. Moreover, the disorder is neglected. In this scenario and for $\nu < 1$, the effective Hamiltonian reads
\begin{eqnarray}
{\mathcal H}&=&{\mathcal P}_{\rm LLL}\sum_{i<j} V\left(\vec{r}_i - \vec{r}_j\right) {\mathcal P}_{\rm LLL}\label{eq:ExactHamiltonian}
\end{eqnarray}
${\mathcal P}_{\rm LLL}$ is the projector onto the lowest Landau level. The two particle interaction $V$ has to be thought as the effective interaction, including effects such as screening, finite confinement of the electron gas,... In a crude approach, it is generally assumed that this interaction is just the 3-dimensional Coulomb interaction, $V(\vec{r})=\frac{1}{r}$. The ground state of this Hamiltonian can be computed for a small number of particles and flux quanta using exact diagonalization techniques such as the Lan\'czos algorithm.

\begin{figure}
\centering
\includegraphics[width=0.70\linewidth]{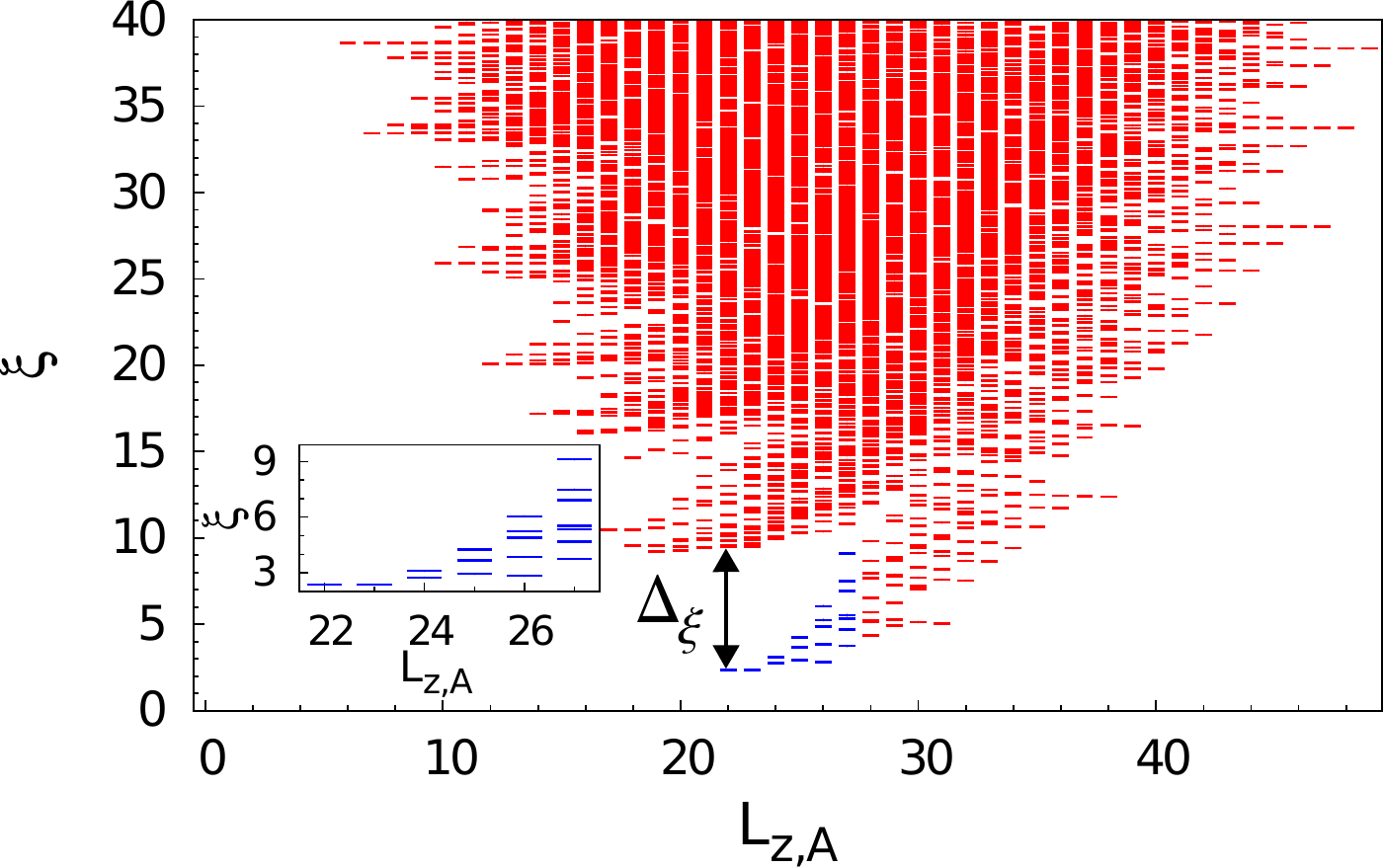}
\caption{OES for the ground state of the Coulomb interaction with $N=12$ fermions and $N_{\Phi}=33$ on the sphere geometry, keeping $l_A=17$ orbitals and looking at the fixed number of particles sector $N_A=6$. We use the same system size and parameters than the OES of the $\nu=1/3$ Laughlin in Fig.~\ref{fig:oeslaughlin}. The levels in blue are those that are related to the edge mode of the Laughlin state. $\Delta_\xi$ denotes the entanglement gap between the edge mode counting and the non-universal part of the spectrum. The inset provides a zoom on the entanglement spectrum related to the $U(1)$ edge mode counting of the Laughlin state.}\label{fig:oescoulombsphere}
\end{figure}

In Fig.~\ref{fig:oescoulombsphere}, we have computed the OES for the ground state of the projected Coulomb interaction $\ket{\Psi_{\rm exact}}$, using exact diagonalization. The overlap between this state and the $\nu=1/3$ Laughlin state $\ket{\Psi_{\rm Lgh}}$ is $|\braket{\Psi_{\rm exact}}{\Psi_{\rm Lgh}}|^2=0.9819$. In the low entanglement energy part of the spectrum, we clearly distinguish a structure similar to the one of the Laughlin state in Fig.~\ref{fig:oeslaughlin} that we have related to the edge mode excitations. As opposed to the example of the spin-1 Heisenberg model discussed in Sec.~\ref{subsubsection:spinAKLT}, the entanglement gap does not extend along all momentum sectors. But the edge mode counting is clearly separated from the higher entanglement energy levels. The low energy part related to the edge physics of the mode state is called the universal part of the ES. The higher energy part is dubbed the non-universal part of the ES. In this example, the idea of looking at the entanglement spectrum per momentum sector is crucial: Without resolving the OES as a function of $L_{z,A}$, the entanglement gap would not be visible.

\begin{figure}
\centering
\includegraphics[width=0.80\linewidth]{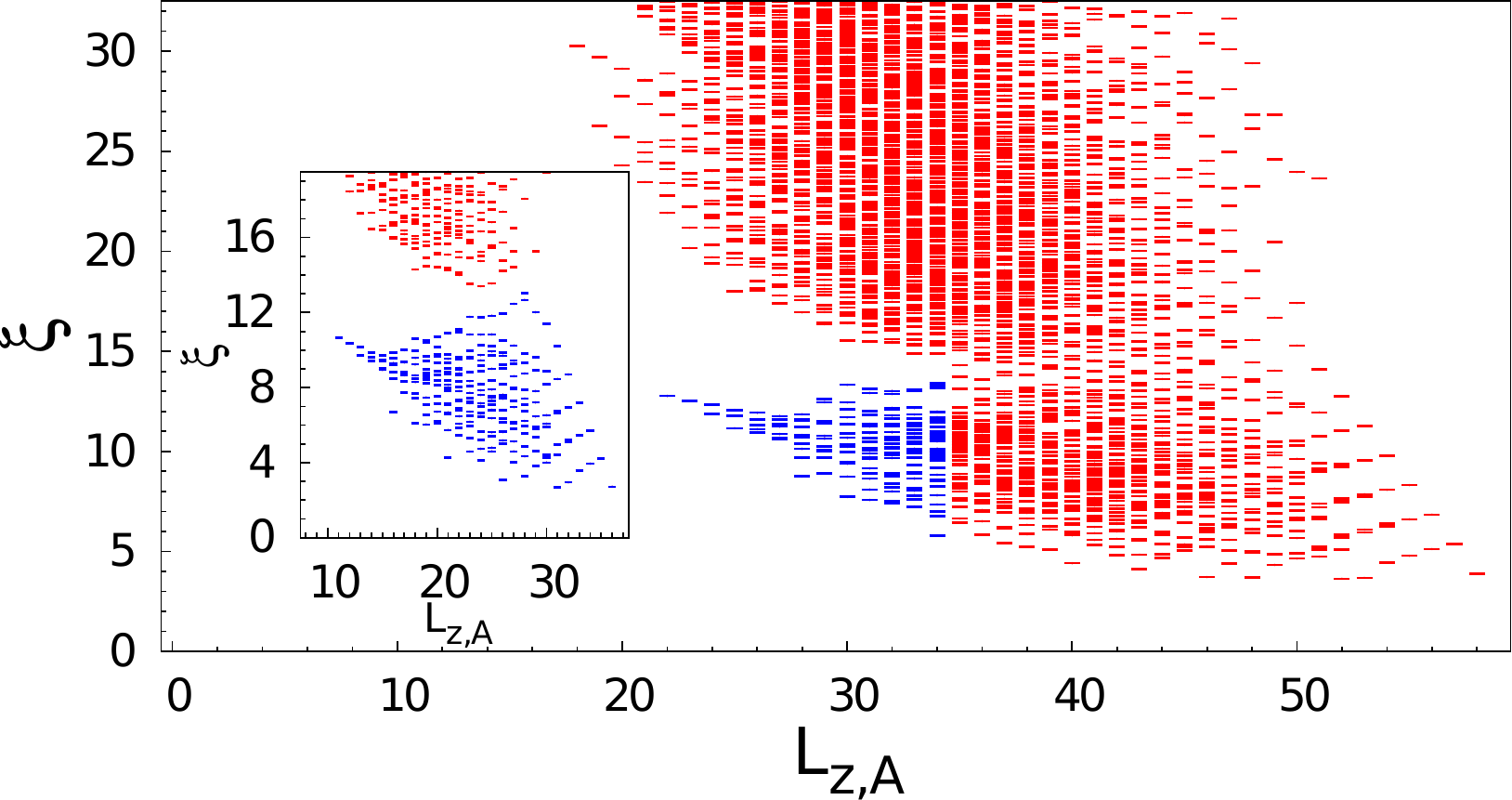}
\caption{OES for the ground state of the Coulomb interaction on the thin annulus geometry, with the same system size and parameters than in Fig.~\ref{fig:oescoulombsphere}. The levels in blue have an identical counting than the one of the Laughlin state and is separated from the non-universal part of the spectrum. The inset shows the OES for the same kind of system but with a lower number of particles ($N=10$). In that case, the structure associated with the Laughlin state (in blue) clearly detaches from the non-universal part.}\label{fig:oescoulombconformal}
\end{figure}

The fact that the entanglement gap $\Delta_{\xi}$ does not spread over the full spectrum could appear as a failure of the OES to find the universality class. First we should focus on the part of the spectrum that has reached the thermodynamical limit, i.e. in the Li and Haldane picture the region that should match the edge physics. From that perspective, what should be relevant is the presence of the entanglement gap in this region that grows when we increase the system size. In the article that has introduced the ES\cite{li-08prl010504}, convincing numerical results were provided that $\Delta_{\xi}$ does not collapse when the system size is increased. Moreover, the extension of region where there is an entanglement gap tightly depends on the geometry in finite size calculations. For example, performing the OES of the same state but on a thin annulus (also called the conformal limit\cite{Thomale-PhysRevLett.104.180502}) leads to a modified picture as shown in Fig.~\ref{fig:oescoulombconformal}. In some cases, one clearly separates the full universal part (the one of some model state) from the non-universal of the ES. As discussed in Ref.~\cite{Thomale-PhysRevLett.104.180502}, this can happen even when the exact state has a moderate overlap with the model state. In some examples, one can even adiabatically go from the model state to the exact state without closing the entanglement gap.

\begin{figure}
\centering
\includegraphics[width=0.49\linewidth]{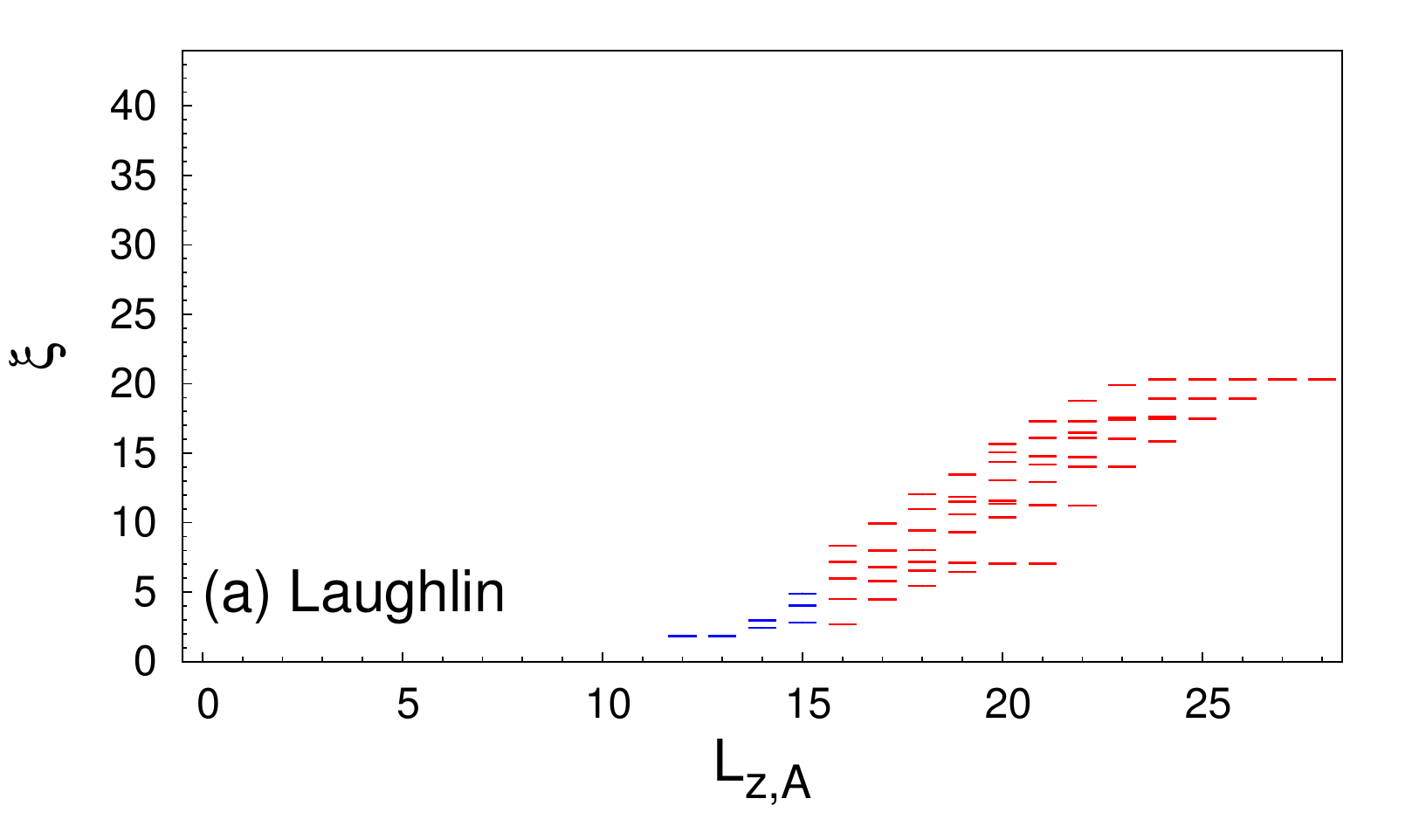}
\includegraphics[width=0.49\linewidth]{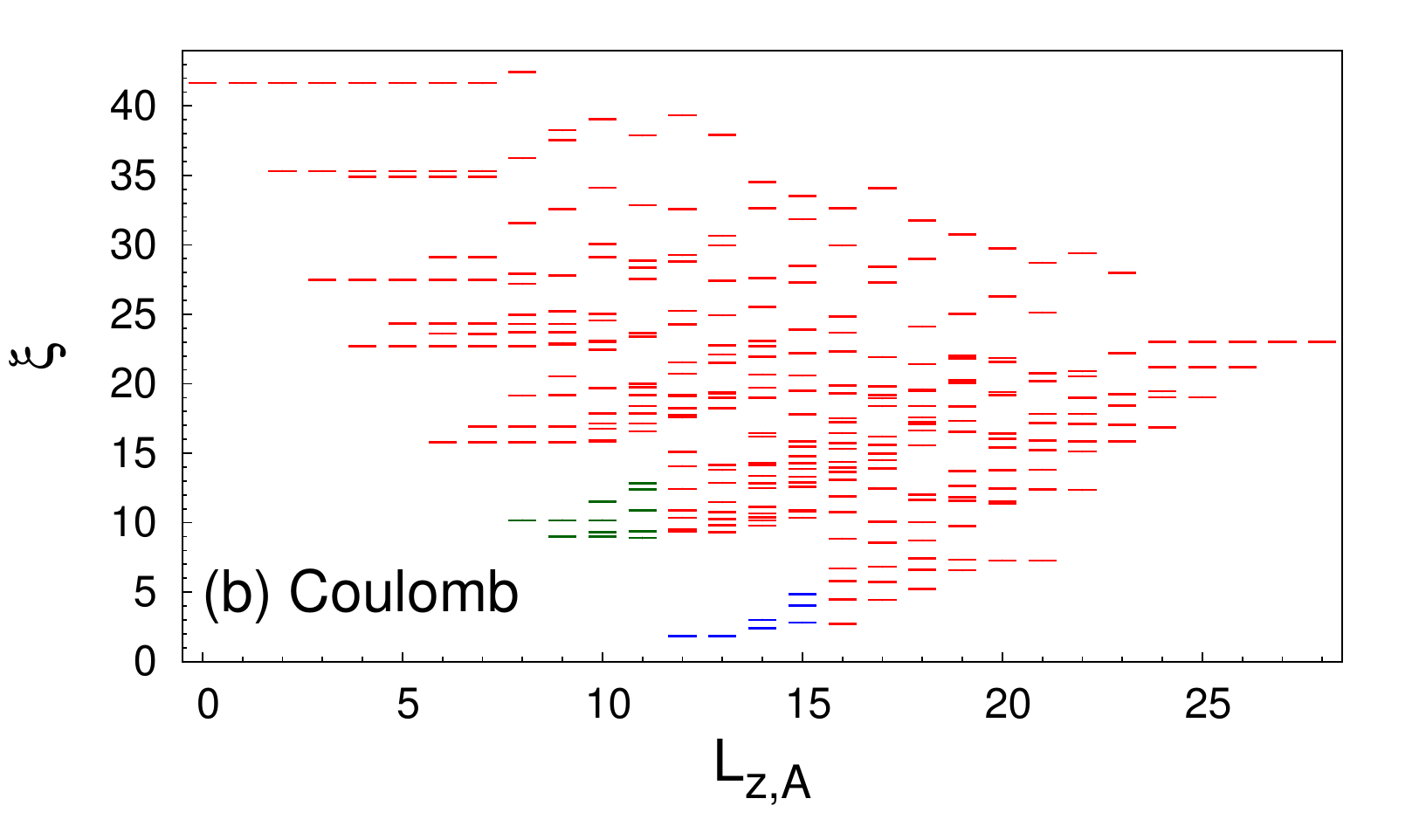}
\caption{OES for $N=8$ fermions and $N_{\Phi}=21$ flux quanta on the sphere geometry, setting $N_A=4$ and $l_A=11$. {\it Left panel:} OES of the $\nu=1/3$ Laughlin state. {\it Right panel:} OES of the ground state of the Coulomb ground state. We clearly observe three branches, the lowest (in blue) being related to the $\nu=1/3$ Laughlin state. The second lowest branch is denoted in green.} \label{fig:oeshierarchy1}
\end{figure}

\begin{figure}
\centering
\includegraphics[width=0.49\linewidth]{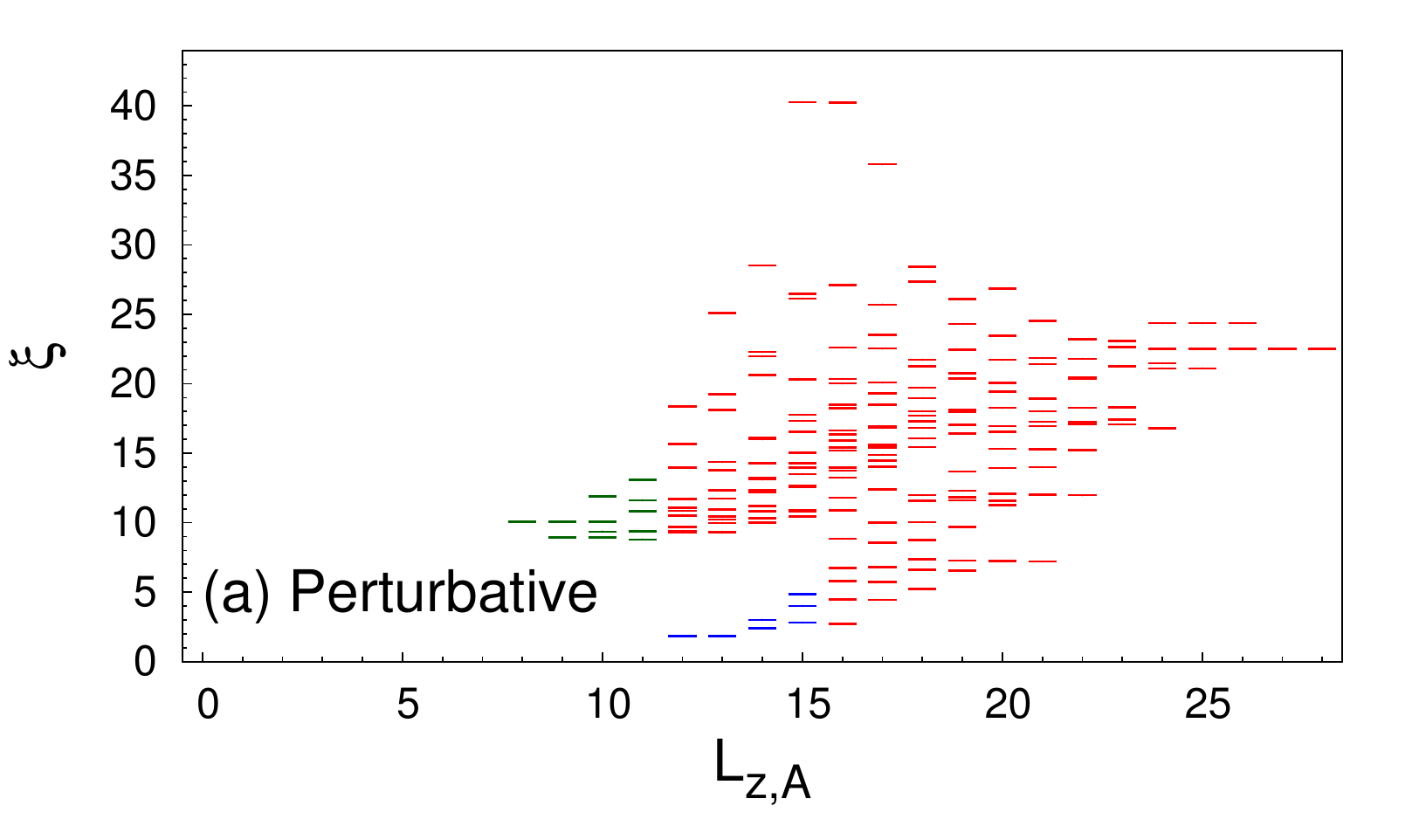}
\includegraphics[width=0.49\linewidth]{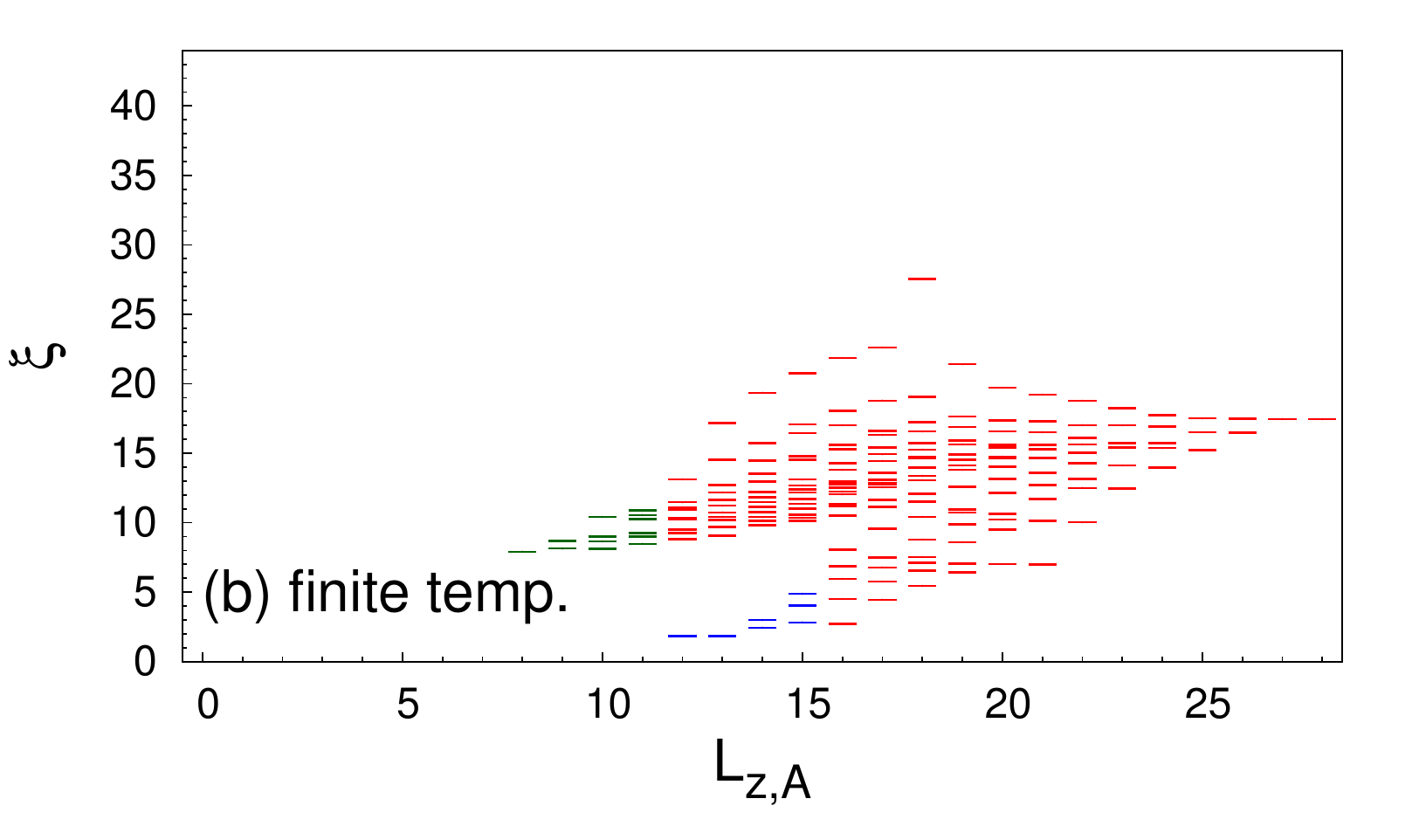}
\caption{{\it Left panel:} OES from the linear combination of the $\nu=1/3$ Laughlin state and the first neutral excitation having the same symmetry than the Laughlin state. For convenience, the linear combination is optimized to maximize the overlap with the Coulomb ground state. A fine tuning is not required to see that this technique reproduces the two lowest branches, indicating that the second branch (in green) is related to neutral excitations. {\it Right panel:} OES from the finite temperature calculation as defined in Eq.~\ref{eq:ThermalOES}. We truncate the energy spectrum to only include the lowest energy neutral excitations (the magneto-roton mode). The temperature is set to $\beta=7$ to mimic the OES of the Coulomb state. Once again, we clearly deduce that the second low entanglement energy branch is related to the lowest energy neutral excitations. For both figures, we use the same system sizes than Fig.~\ref{fig:oeshierarchy1}.}\label{fig:oeshierarchy2}
\end{figure}

Until now, we have mostly focused on the universal part of the ES. Looking at the Fig.~\ref{fig:oeshierarchy1}, we observe that the non-universal part exhibits several branches. Indeed, these branches can be related to the neutral excitations (the excitations that do not involve to change the number of particles or the number of flux quanta) of the system\cite{Sterdyniak-NewJourPhys-13-10-105001}. For the FQHE, these neutral excitations are quasihole-quasielectron excitons. Two approaches are available to test this idea. One can build an approximation of the exact ground state based on the model state and the lowest energy neutral excitation that has the same symmetry than both the model and the exact states (see Fig.~\ref{fig:oeshierarchy2}a). This .The other option is to consider the model state but in finite temperature where the full density matrix is given by 
\begin{eqnarray}
\rho=\frac{1}{\sum_n e^{-\beta E_n}}\sum_n e^{-\beta E_n} \ket{\Psi_n}\bra{\Psi_n}\label{eq:ThermalOES}
\end{eqnarray}
where the $\ket{\Psi_n}$ and $E_n$ are respectively the eigenstates and the eigenvalues of the Hamiltonian that produces the model state (see Fig.~\ref{fig:oeshierarchy2}b). In both cases, we see that the resulting OES correctly captures the non-universal part. This exercise also appears to support the idea that the entanglement spectrum of the ground state of a realistic Hamiltonian contains information not only about the universality class of the ground state but also about its excitations.

\subsection{Particle entanglement spectrum}\label{sec:PES}

The concepts of entanglement entropy and entanglement spectrum are not specifically related to a partition in real space. Indeed, the OES is strictly speaking a partition in momentum space which in the specific case of the FQHE can be roughly related to a spatial cut. Partitioning a system in different ways can unveil different type of information, as it was shown in the case of quantum spin chains\cite{Thomale-PhysRevLett.105.116805}. Among the possible partition, a simple one is based on removing particles from the system, realizing a particle partition. In the context of the entanglement entropy for FQHE, such a partition was introduced in Refs.~\cite{Haque-PhysRevLett.98.060401} and \cite{Zozulya-PhysRevB.76.125310}. The related entanglement spectrum, named the particle entanglement spectrum (PES),was introduced later in Ref.~\cite{Sterdyniak-PhysRevLett.106.100405}. As opposed to the OES, the geometry (i.e. the number of orbitals) is preserved, the particles are divided into two groups $A$ and $B$, holding respectively $N_A$ and $N_B$ particles. In first quantized notation and for a generic wave function $\Psi \left(x_1,...,x_{N}\right)$ for $N=N_A+N_B$ particles, the reduced density matrix is given by

\begin{eqnarray}
\rho_A(x_1,...,x_{N_A};{x'}_1,...,{x'}_{N_A})=\int dx_{N_A+1}...dx_{N} &&  \Psi^*(x_1,...,x_{N_A},x_{N_A+1},...,x_{N})\label{eq:defPES}\\
&&\times \Psi({x'}_1,...,{x'}_{N_A},x_{N_A+1},...,x_{N})\nonumber
\end{eqnarray}

As a first example, one can look at the completely filled lowest Landau level, i.e. the $\nu=1$ integer quantum Hall effect. The ground state on the sphere geometry for $N=N_{\Phi}+1$ fermions is given by
\begin{eqnarray}
\ket{\Psi_{\nu=1}}&=&\ket{-\frac{N_\Phi}{2},...,\frac{N_\Phi}{2}}\label{eq:groundstatenu1}
\end{eqnarray}
This state is a product state in the orbital basis, leading to a trivial OES with a single non-zero eigenvalue. For the PES, the picture is different: The counting is given by the number of ways one can choose $N_A$ particles among the $N$ particles of the system (it is the same counting that the RSES for the IQHE in finite size discussed in Sec.~\ref{sec:RealspaceIQHE}) . This case clearly stresses that different partitions probe different properties of the same system.

\begin{figure}
\centering
\includegraphics[width=0.45\linewidth]{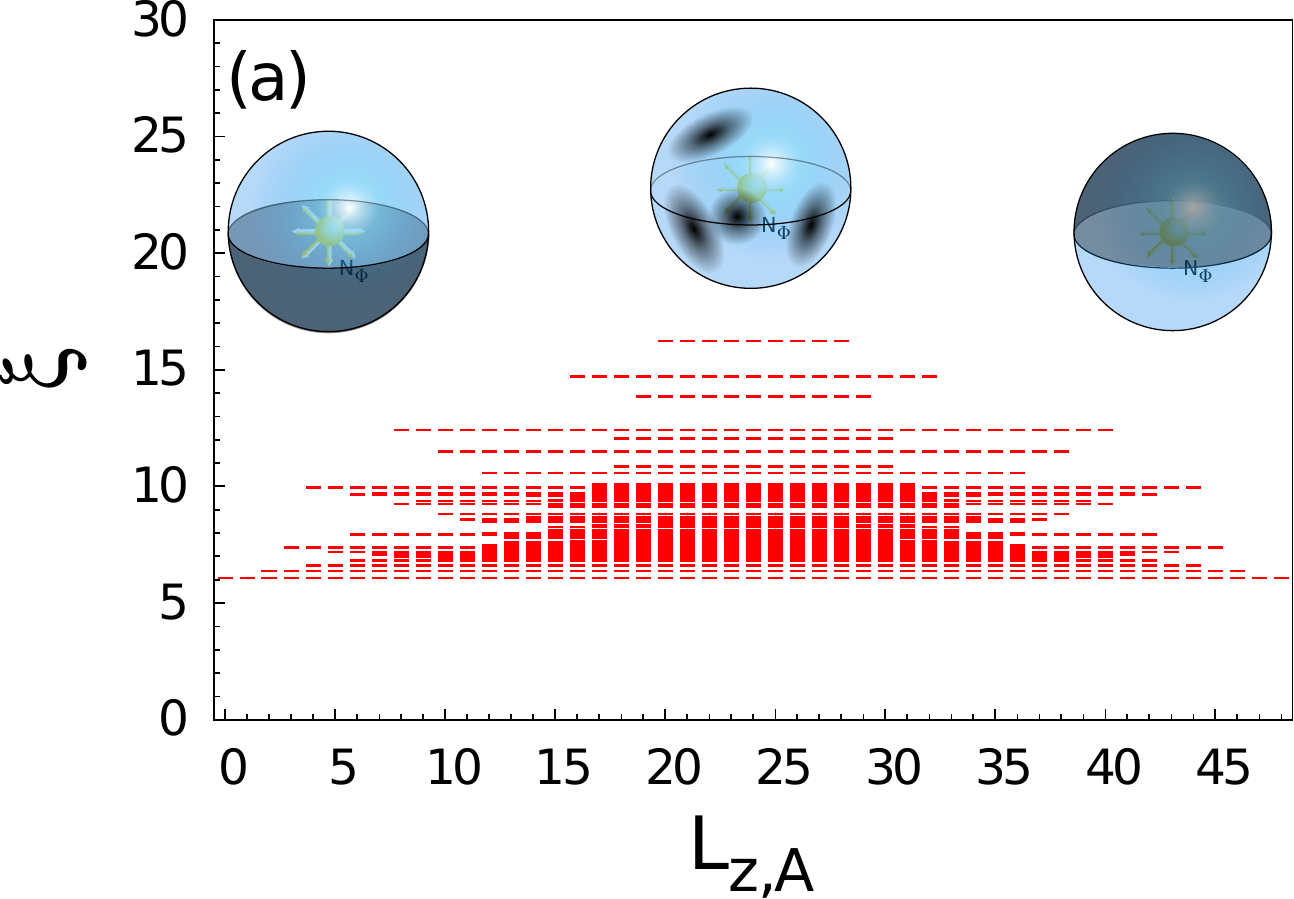}
\includegraphics[width=0.45\linewidth]{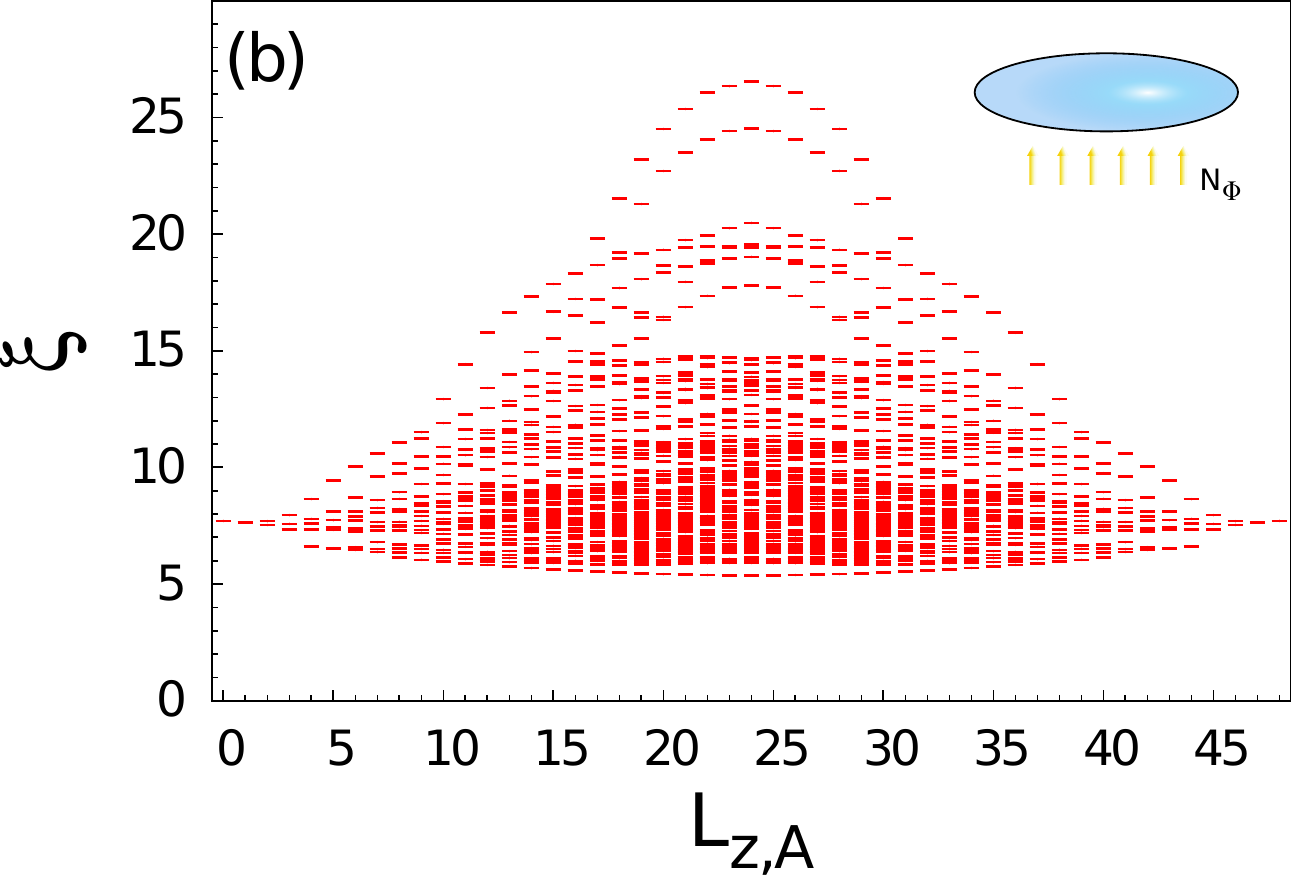}
\caption{PES for the $\nu=1/3$ Laughlin state with $N=8$ fermions, $N_{\Phi}=21$ and $N_A=4$ on the sphere geometry (a) and on the disk geometry (b). In both cases, the counting per momentum is given by the number of quasihole states of the Laughlin state for $N_A$ particles and $N_{\Phi}=21$. The high degeneracy observed for the PES on the sphere is a consequence of the total angular momentum $L_A^2$ being a good quantum number when the PES is performed on a state with a total angular momentum equal to zero (such as the Laughlin ground state). For the sphere case, we schematically describe the types of quasihole states that correspond to the leftmost, center and rightmost levels.}\label{fig:peslaughlin}
\end{figure}

We now turn to the cases of interacting states, focusing on the Laughlin $\nu=1/3$ state. Fig.~\ref{fig:peslaughlin}a and b give the PES on the sphere and the disk geometry. As in the case of the OES, the counting is non-trivial (i.e. the number of non-zero eigenvalues is much lower than the naive dimension of the reduced density matrix) and does not depend on the geometry. What was empirically found in Ref.~\cite{Sterdyniak-PhysRevLett.106.100405} is that the counting matches (per momentum sector) the number of quasihole states of the same state with $N_A$ and $N_{\Phi}$ flux quanta (the particle partition does not affect $N_{\Phi}$). This statement was checked for a large series of model wave functions. When these model wave functions are unique zero energy states of some local model Hamiltonian, one can prove that the counting is bounded by the number of quasihole states. Indeed, any eigenstate of $\rho_A$ corresponding to a non-zero eigenvalue, has to be a quasihole state (meaning a zero energy state of the model Hamiltonian). Until now, there is no mathematical proof in the generic case that this bound has to be saturated. Note that the PES for $\nu=1$ we have previously discussed above, can also be understood as the quasihole excitations of the integer quantum Hall state.

If we admit that the conjecture about the bound saturation is valid, then we completely understand the counting of the PES including any finite-size effect (as opposed to the OES). Both entanglement spectra, the OES and the PES, are actually related in the thermodynamical region\cite{Chandran-PhysRevB.84.205136}. In Fig.~\ref{fig:peslaughlin}a, we give a schematic description of the quasihole states in each part of the PES. The leftmost angular momentum sector ($L_{z,A}=0$) corresponds to the case where all quasiholes are located in the south hemisphere which is then completely depleted. We are left with a Laughlin droplet occupying the north hemisphere. Slightly moving away from $L_{z,A}=0$ is equivalent to slight deformations of the droplet, i.e. the edge excitations. Indeed, the counting starting from $L_{z,A}=0$ is $1,1,2,3,...$ as expected from the Laughlin edge mode. Ref.~\cite{Chandran-PhysRevB.84.205136} proved that the entanglement matrices (as defined in Sec.~\ref{subsubsection:definitions}) associated with the thermodynamical region in both the PES and the OES must have the same rank. Using this bulk-edge (or PES-OES) correspondence, the proof of the Li-Haldane conjecture is reduced (at least for the class of model states that have been considered) to the proof of the bound saturation.

\begin{figure}
\centering
\includegraphics[width=0.45\linewidth]{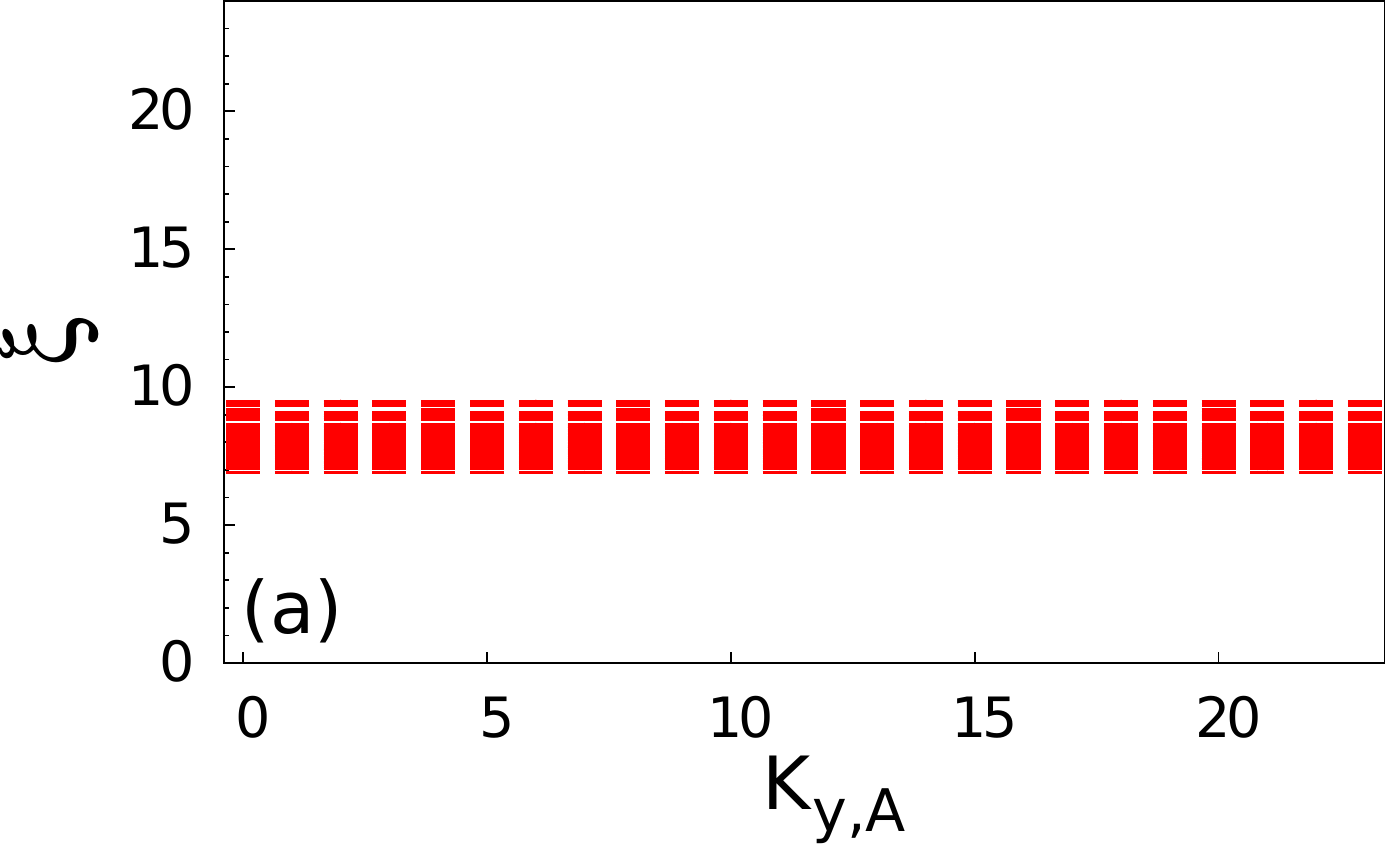}
\includegraphics[width=0.45\linewidth]{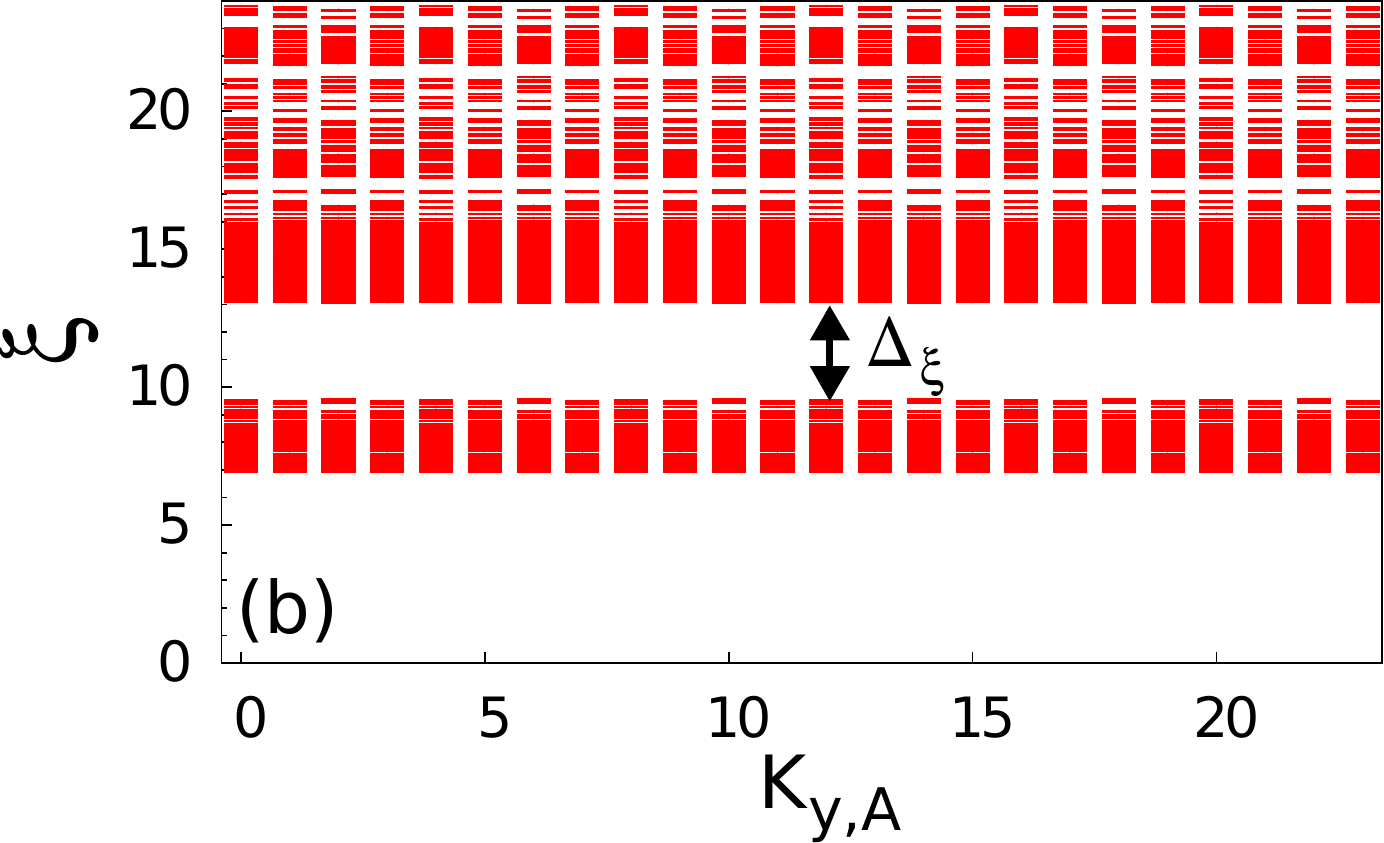}
\caption{PES on the torus geometry for $N=8$ fermions and $N_{\Phi}=24$, keeping $N_A=4$ particles. (a) The $\nu=1/3$ Laughlin state. The counting per momentum sector is exactly given by the number of quasihole states with $N_A=4$ fermions and $N_{\Phi}=24$. (b) The Coulomb ground state. We observe a clear entanglement gap $\Delta_\xi$ between a low entanglement energy structure having the same counting than the PES of the Laughlin state and a higher entanglement energy part.}\label{fig:pestorus13}
\end{figure}

The PES can also be computed on the torus geometry. The FQH phases being topological phases, the degeneracy of their ground state changes with the genus of the surface they live on. For example on the torus, the Laughlin $\nu=1/m$ state is $m$-fold degenerate and the Moore-Read state is $6$-fold degenerate. Thus, multiple choices for the density matrix are available. For the PES, we use the incoherent density matrix where one sums up all sectors:
\begin{eqnarray}
\rho&=&\frac{1}{d}\sum_{i=0}^{d} \ket{\Psi_i}\bra{\Psi_i}\label{eq:densitymatrixtorus}
\end{eqnarray}
where $\{\ket{\Psi_i}\}$ with $i=1,...,d$ forms an orthogonal basis of the degenerate ground state manifold ($d$ being the total degeneracy). As defined, this density matrix commutes with the magnetic translation operators and does not depend on a particular basis choice. The PES calculations are performed using the translation symmetry along one direction (here $y$), and the eigenvalues of $\rho_A$ can be labeled by the corresponding $K_{y,A}$ momentum. Fig.~\ref{fig:pestorus13}a shows the PES for the Laughlin state on the torus. The properties are identical to those of the PES on the sphere: The counting matches the one of the quasihole states and the corresponding eigenstates of $\rho_A$ span the subspace of the quasihole states. This is a clear difference with the OES on the torus where the counting is trivial as discussed in Sec.~\ref{sec:OES}. For the ground state of the Coulomb interaction at $\nu=1/3$, the PES is quite interesting: As observed in Fig.~\ref{fig:pestorus13}b, there is a clear entanglement gap separating a low entanglement energy structure having the same counting than the PES of the Laughlin and a higher entanglement energy part. From the different examples that have been studied, the PES behaves nicely on the torus geometry. This property will be used as a powerful tool to probe the physics of fractional Chern insulators in Sec.~\ref{sec:FCI}.

\subsection{Real space entanglement spectrum}\label{sec:RSES}

When we have described the OES in Sec.~\ref{sec:OES}, we have argued that this type of partition was an approximation of a partition in real space, thanks to the specific properties of the orbital basis. The OES appears as a fuzzy cut and not a sharp cut. Several articles\cite{Sterdyniak-PhysRevB.85.125308,Dubail:2012p2980,Rodriguez-PhysRevLett.108.256806} have addressed the question of the real space entanglement spectrum (RSES) using a sharp real space partition for FQH states. It is an natural extension from the non-interacting IQHE case described in Sec.~\ref{sec:IQHE}. If one chooses a cut that preserves the rotation along $z$ for the sphere or the disk, then $L_{z,A}$ is still a good quantum number. This makes the connection with the other entanglement spectra easier. As in the case of the OES, $N_A$ is also a good quantum number. The generic principle of the RSES was described in Sec.~\ref{sec:ESCI}. Following Eq.~\ref{eq:CIABcreationoperator}, we can split the creation operation associated to the orbital $m$ into
\begin{eqnarray}
c^\dagger_{m}&=&\alpha_{m}c^\dagger_{m;A}+\beta_{m}c^\dagger_{m;B}\label{eq:FQHEABcreationoperator}
\end{eqnarray}
where $\alpha^2_{m}$ (resp. $\beta^2_{m}$) is the weight of the orbital $m$ in the $A$ (resp. $B$) part. While on the disk  $\alpha^2_{m}$ and $\beta^2_{m}$ are related to incomplete gamma functions as shown in Eq.~\ref{eq:orbitalweightrealspace}, these coefficients on the sphere can be expressed as incomplete beta functions. A key property of the RSES is that a block of its entanglement matrix $M^{\rm RSES}_{N_A}$ with a fixed $N_A$ can be related to the entanglement matrix $M^{\rm PES}_{N_A}$ of the PES for $N_A$ particles. Using Eq.~\ref{eq:FQHEABcreationoperator} we can deduce the relation $M^{\rm RSES}_{N_A} = S M^{\rm PES}_{N_A} Q$ where $S$ and $Q$ are diagonal matrices with non-zero diagonal matrix elements. These elements are purely one-body geometrical factors $\alpha_{m}$ and $\beta_{m}$ coming from the space partition. As a consequence, the two matrices $M^{\rm PES}_{N_A}$ and $M^{\rm RSES}_{N_A}$ have the same rank and thus the two entanglement spectra have the same counting. It should be notice that if we take the weights $\alpha_m=\beta_m=1/\sqrt{2}$, we recover exactly the PES. As discussed in Sec.~\ref{sec:PES}, the OES for the $\nu=1$ state (i.e. the integer quantum Hall effect) is trivial, where $\rho_A$ has a single non-zero eigenvalue.  Having the same counting that the PES, the RSES thus strongly differs from the OES as shown in Fig.\ref{fig:RSESIntegerQHECounting}. 

\begin{figure}
\centering
\includegraphics[width=0.48\linewidth]{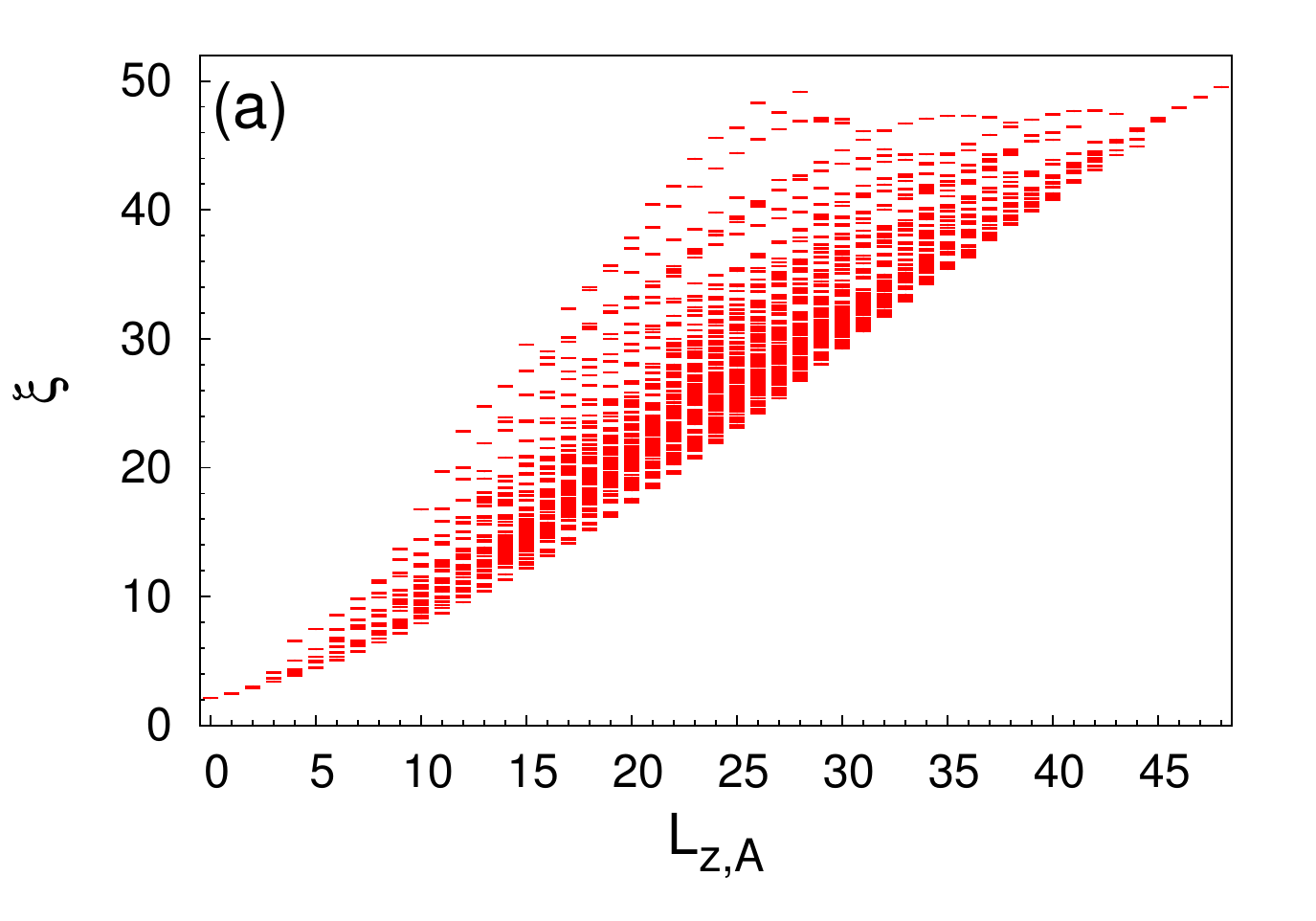}
\includegraphics[width=0.42\linewidth]{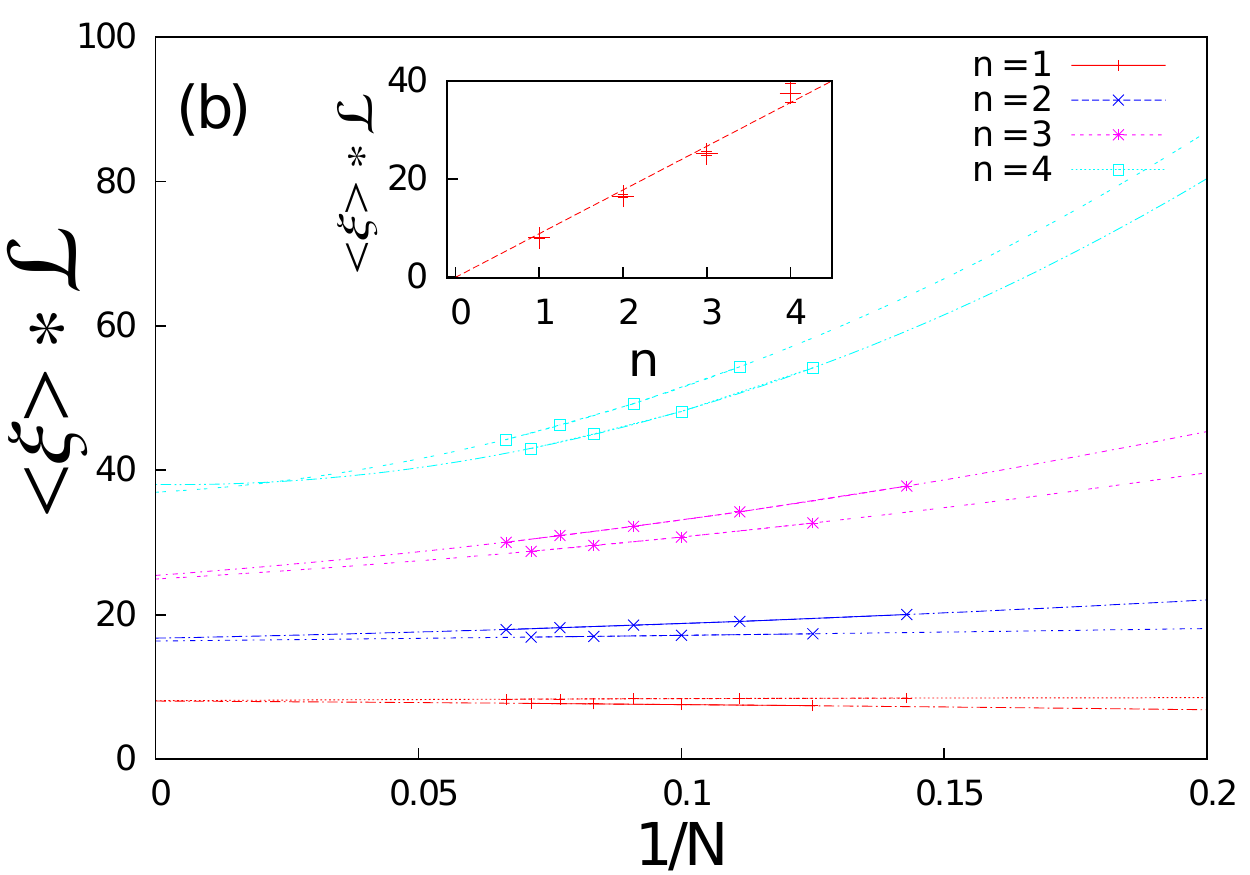}
\caption{{\it Left panel: } RSES of the $\nu=1/3$ Laughlin state on the sphere geometry for $N=8$ fermions and $N_{\Phi}=21$. We have used the (sharp) hemisphere cut and display the RSES in the sector $N_A=4$. The counting is identical to the one of Fig.~\ref{fig:peslaughlin}a. {\it Right panel: } The average entanglement energy $< \xi >$ times the perimeter of the cut ${\mathcal L}$ per momentum sector (here $n=L_z^A$) extrapolated at the thermodynamic limit as a function of $1/N$ .The even-odd effect is just a consequence of $N_A$ being the integer part of $N/2$. The velocity of the edge mode is $v = 1.41(5)$ (see inset). Such value would be compatible with a rescaling of the $\nu=1$ edge mode velocity with a factor $1/\sqrt{3}$.}\label{fig:rsesLaughlin}
\end{figure}

In Fig.~\ref{fig:rsesLaughlin}a, we show the RSES of the $\nu=1/3$ Laughlin state on the sphere when $A$ is made of the north hemisphere, having a sharp cut at the equator. As expected the counting per momentum sector is identical to the one of the PES. The shape of the spectrum itself is reminiscent of the OES, being due to the geometrical cut. Beyond the counting, one could ask if the entanglement energies of RSES mimics the dispersion relation of the edge mode, in a better way than the OES. In both cases, the spread between the smallest and the largest entanglement energy in a given angular momentum sector seems to converge to zero. Fig.~\ref{fig:rsesLaughlin}b gives the extrapolation of the average entanglement energy per angular momentum sector to the large number of particles limit. If in this limit the RSES was equivalent to the edge mode dispersion relation, we should expect these energies to be of the form $\frac{2 \pi v}{{\mathcal L}} n$ where $n$ is an integer, $v$ is the edge mode velocity and ${\mathcal L}$ is the cut perimeter. The finite size calculation has a roughly good agreement with this picture. Most of these properties have been recently confirmed in much larger system sizes using the MPS description of the Laughlin state\cite{zaletel-PhysRevB.86.245305}. These results underline once again that the ES of the system ground state contains the description of the edge excitations, reinforcing the bulk-edge correspondence.

\section{Entanglement spectrum as a tool: Probing the Fractional Chern Insulators}\label{sec:FCI}

As a practical application of the entanglement spectroscopy, we discuss the physics of Chern insulators in the strong interacting regime. We emphasize that the ES can conveniently replace the overlap calculations when those are not available. We show that the entanglement spectroscopy can discriminate between two phases where simple energetic analysis fails.

\subsection{From Chern Insulators to Fractional Chern Insulators}

In the context of quantum Hall effect, strong interactions are known to give rise to the exotic physics of the FQHE. Current work suggests that, in an analog way to the FQHE, introducing strong interactions coupled with fractional filling of the topological insulator bands can give rise to novel and remarkable topological phases of matter. The first class of topological insulators that was studied in the strongly interaction regime was the Chern insulator described in Sec.~\ref{sec:CI}. With the addition of strong interactions and fractionally filled bands these systems are known as fractional Chern insulators (FCI). At the beginning of 2011, several papers present evidence from numerical simulations \cite{sheng-natcommun.2.389,neupert-PhysRevLett.106.236804,regnault-PhysRevX.1.021014} which demonstrated that FCIs could be implemented in principle for model systems (see Refs.~\cite{Parameswaran-2012PhRvB..85x1308P} and~\cite{BERGHOLTZ-IntJournal} for reviews). 

\begin{figure}
\centering
\includegraphics[width=0.40\linewidth]{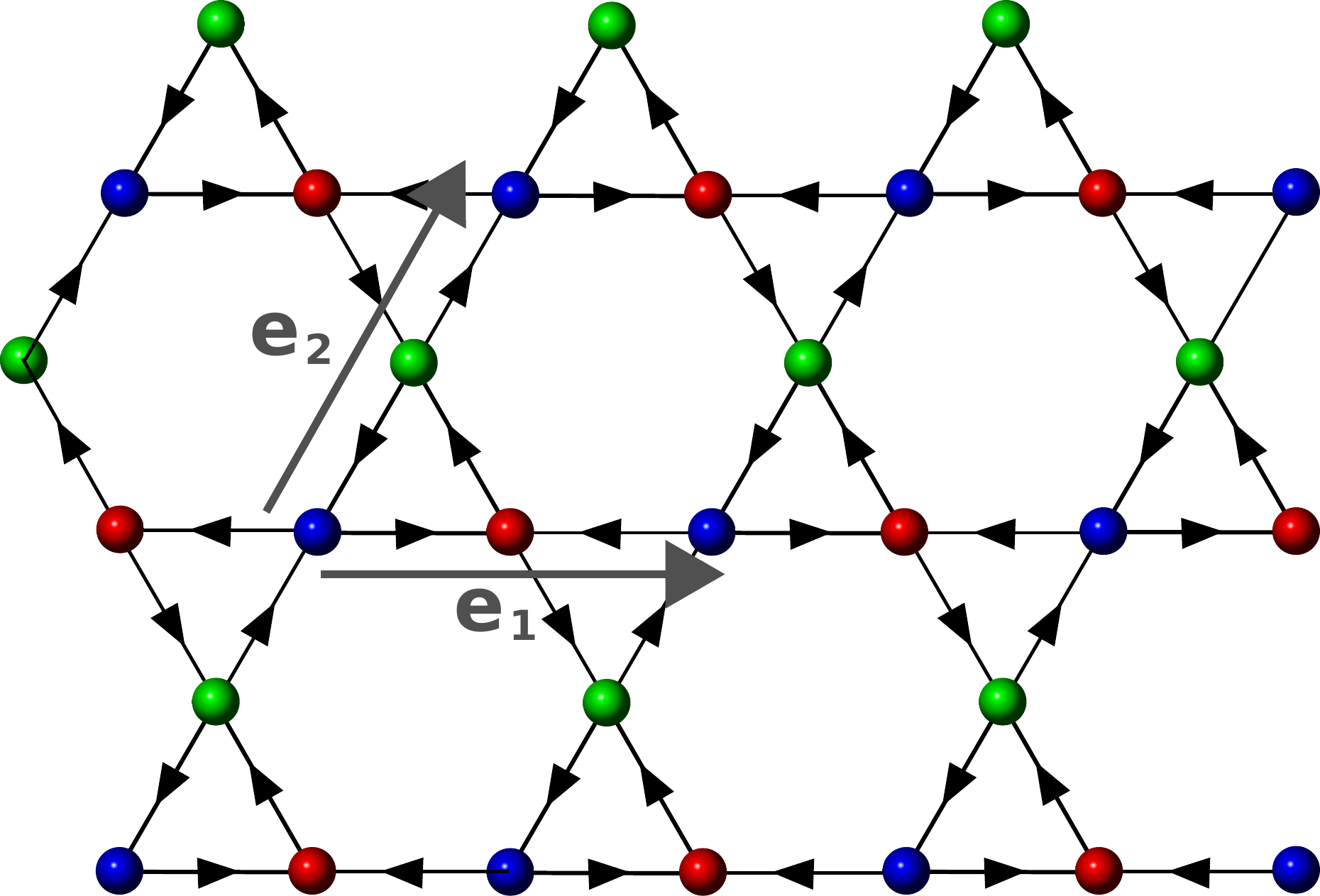}\hspace{1cm}
\includegraphics[width=0.45\linewidth]{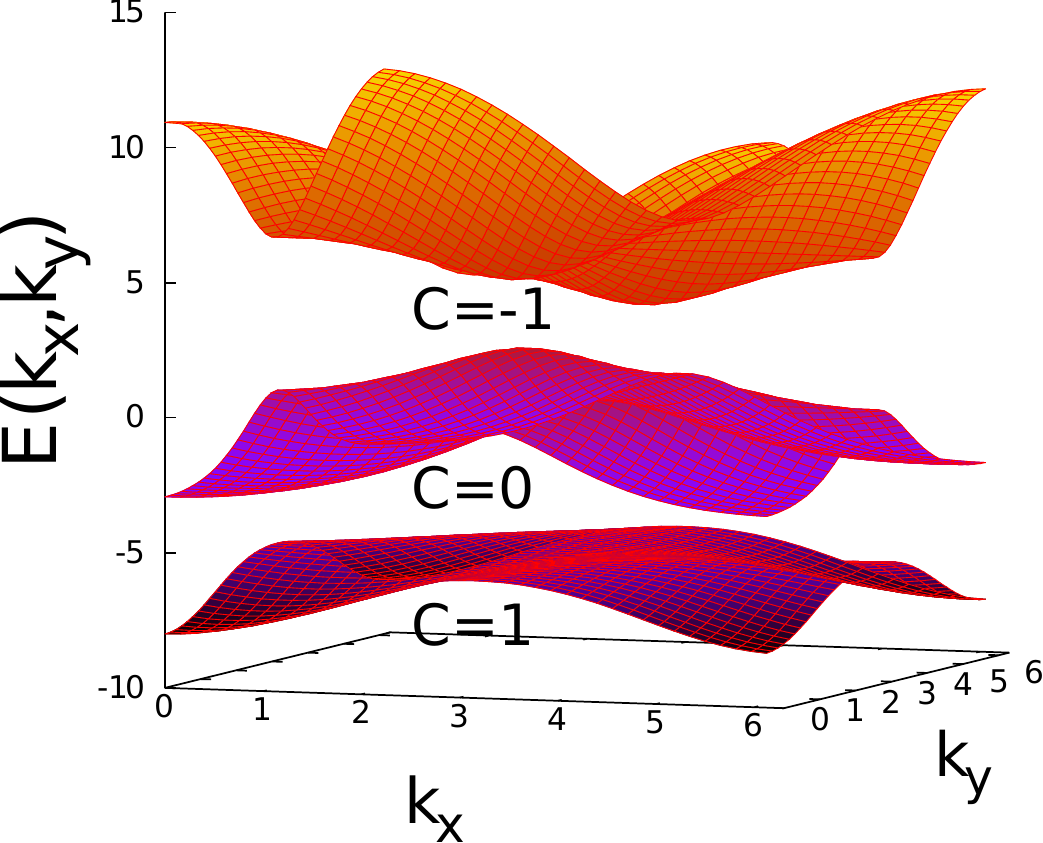}
\caption{{\it Left panel}: The Kagome lattice model as discussed in Ref. \cite{tang-PhysRevLett.106.236802} with three sites per unit cell. In the simplest case, this model has only a single complex nearest neighbor hopping term. {\it Right panel}: The band structure for the Kagome lattice model with a hopping term of $\exp\left({i\frac{\pi}{4}}\right)$. The topmost and the lowest band have a Chern number of $C=\pm 1$, while the middle band is trivial (i.e. a Chern number $C=0$).}\label{fig:Kagome}
\end{figure}

The emergence of a FQH-like phase in FCI strongly depends on the underlying one-body model\cite{Wu-2012PhysRevB.85.075116}. In this section, we consider a slightly more complex model than the two-orbital model discussed in Sec.~\ref{sec:CI} (it was shown that this model does not exhibit any FQH state in the strongly interacting regime). It is based on the Kagome lattice\cite{tang-PhysRevLett.106.236802} (see Fig.~\ref{fig:Kagome}), a triangular lattice with three sites per unit cells, with a complex hopping term $t \exp (i \varphi)$ between neighboring sites. The Bloch Hamiltonian for this model is given by
\begin{eqnarray}
{\mathcal H}(\mathbf{k}) &=& - \left[ \begin{array}{c c c}
				0            &  e^{i\varphi}(1 + e^{-ik_x})      &  e^{-i\varphi}(1 + e^{-ik_y})       \\
				 & 0                   &  e^{i\varphi}(1 + e^{i(k_x -k_y)}) \\
				{\rm h.c.} &  & 0
				\end{array}
			  \right]\label{eq:kagomeBloch}
\end{eqnarray}
where $k_x=\mathbf{k}\cdot \boldsymbol e_1$ and $k_y=\mathbf{k}\cdot \boldsymbol e_2$, $\boldsymbol  e_1$ and $\boldsymbol  e_2$ are the lattice translation vectors as described in Fig.~\ref{fig:Kagome}. The magnitude of the hopping term is set to 1. The dispersion relation is displayed in Fig.~\ref{fig:Kagome}, showing the three bands with two of them carrying a non-zero Chern number. This model with short range repulsion was shown to host Laughlin-like phases both for fermions\cite{Wu-2012PhysRevB.85.075116} and bosons\cite{Liu-PhysRevB.87.205136}.

The simplest way to look at FCI is to work in the flat-band limit\cite{regnault-PhysRevX.1.021014}: We focus on the interaction and the topological properties of the band structure, whereas the effect of band dispersion and band mixing are discarded. This allows to mimic the usual hypothesis of the FQHE calculations as described in Sec.~\ref{sec:OESRealistic}. We start from the original Bloch Hamiltonian ${\mathcal H}(\mathbf{k})=\sum_n E_n(\mathbf{k})P_n(\mathbf{k})$ where $E_n(\mathbf{k})$ and $P_n(\mathbf{k})$ are the dispersion and the projector onto the $n$-th band, respectively. Then we focus on the $i-{\rm th}$ band (the lowest band for the case of the Kagome model). We can conveniently consider an equivalent system with the same one-body wave functions but with perfectly flat bands ${\mathcal H}_{\rm FB}(\mathbf{k})= \sum_n n P_n(\mathbf{k})$. From the energy perspective, this is the same situation than a single Landau level.

\begin{figure}
\centering
\includegraphics[width=0.49\linewidth]{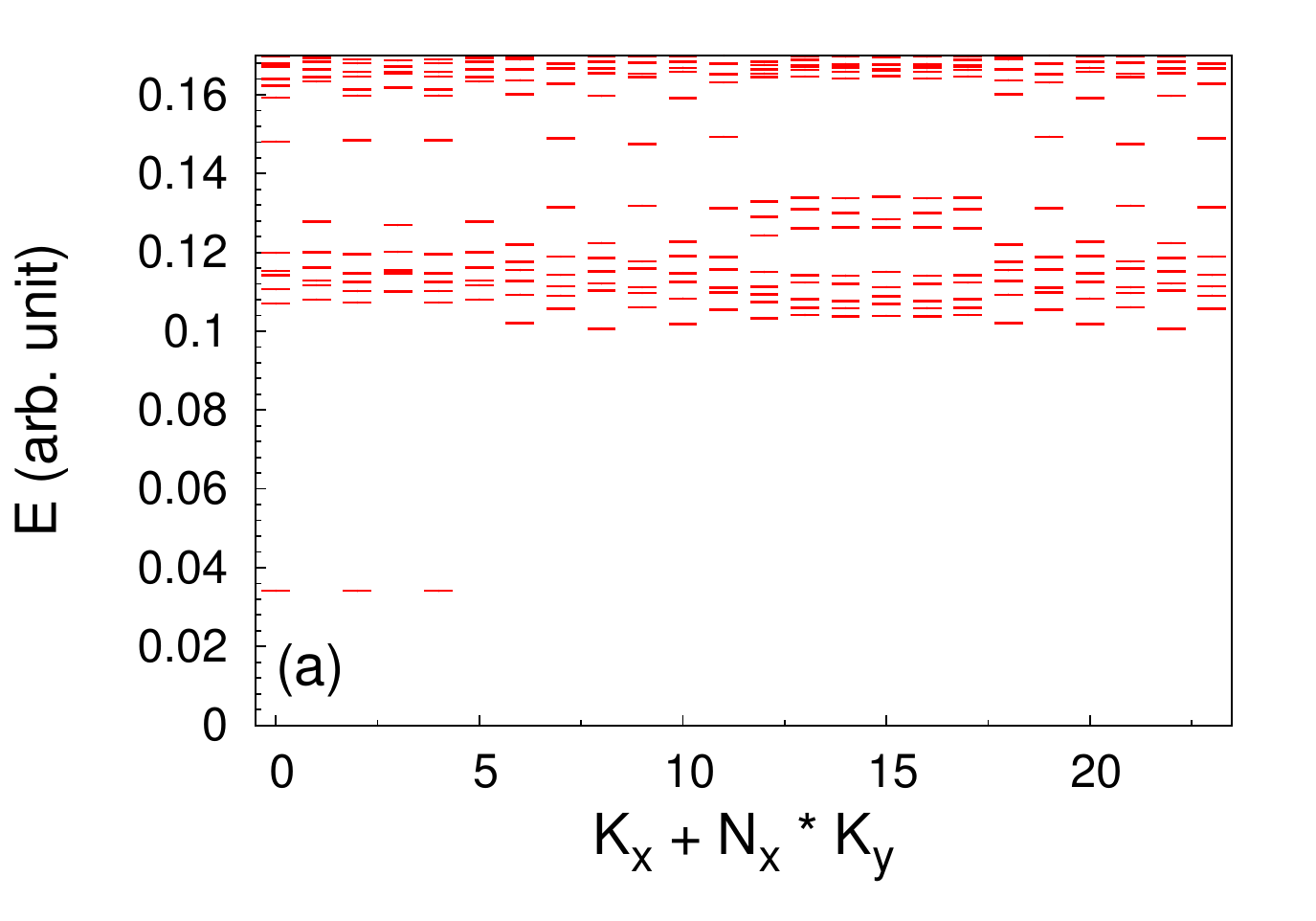}
\includegraphics[width=0.49\linewidth]{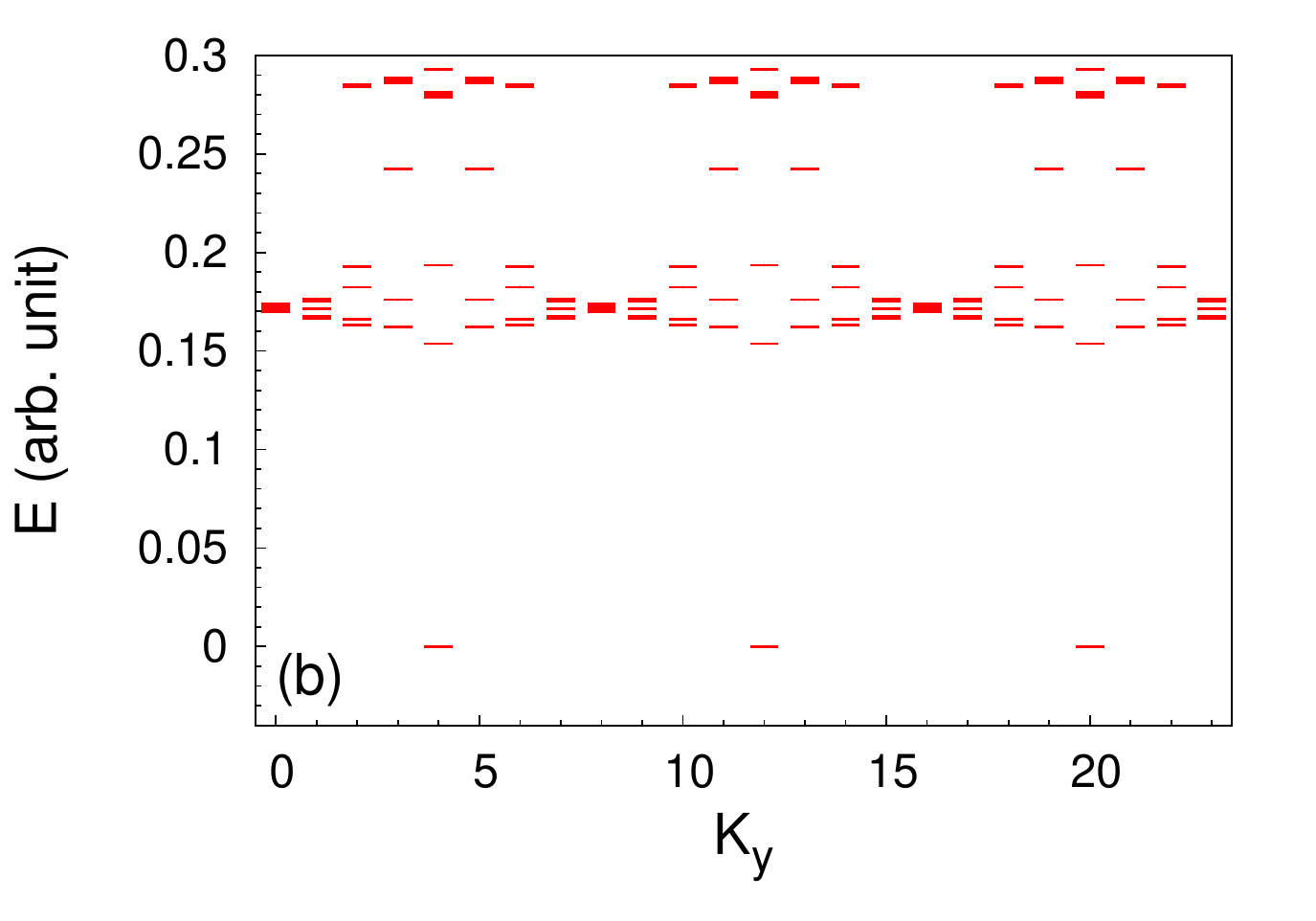}
\caption{{\it Left panel:} Low energy spectrum for $N=8$ fermions on a $N_x \times N_y= 6 \times 4$ unit cell Kagome lattice with periodic boundary conditions. $K_x$ and $K_y$ denote the total momentum in the $x$ and $y$ direction. We clearly observe an almost threefold degenerate ground state (the energy splitting between these three states is $3.1 \times 10^{-5}$). {\it Right panel:} Low energy spectrum for the FQHE with $N=8$ fermions and $N_{\Phi}=24$ on a torus. The Hamiltonian that we have used is the hollow core interaction for which the Laughlin state is the exact zero energy ground state.}\label{fig:kagomespectrum}
\end{figure}

For the interacting case, we consider $N$ spinless fermions on a lattice made of $N_x$ unit cells in the $\boldsymbol e_1$ and $N_y$ unit cells in the $\boldsymbol e_2$ with periodic boundary conditions. The filling factor is defined as $\nu=\frac{N}{N_x N_y}$. The simplest repulsive interaction that can be used for spinless fermions is just the nearest neighbor repulsive interaction
\begin{eqnarray}
H_{\rm int}&=&U \sum_{<i,j>}: n_i n_j :\label{eq:FCIInteraction}
\end{eqnarray}
where $<i,j>$ denotes the sum over nearest neighboring sites. Projecting this interaction onto the lowest band and using the flat-band limit, the total effective Hamiltonian is just given by the projected interaction, similar to the FQHE case in Eq.~\ref{eq:ExactHamiltonian}. Exact diagonalizations can be performed to probe this system. A typical energy spectrum for the interacting Kagome lattice at filling factor $\nu=1/3$ is shown in Fig.~\ref{fig:kagomespectrum}a. Similar to the FQHE on a torus (see Fig.~\ref{fig:kagomespectrum}b), we observe an (almost) threefold degenerate ground state clearly separated from the higher energy excitations. Note that the ground state is not exactly degenerate, as expected for the FQHE phase on a torus such as the Laughlin state. This is a consequence of the absence of an exact magnetic translation symmetry\cite{Parameswaran-2012PhRvB..85x1308P,goerbig-2012epjb,Bernevig-2012PhysRevB.85.075128} as opposed to the FQHE.

Since Chern insulators are equivalent to quantum Hall systems without an external magnetic field, one could have imagined that FCIs should have given rise to topological phases analogous to those exhibited by the FQHE. As stated previously and contrary to expectations, not all CI models\cite{Wu-2012PhysRevB.85.075116} were found to exhibit such `fractional' phases. For the time being, the emergence of FQH-like phases for a given model can only be probed through numerical simulations of that given model. Moreover, many of the signatures obtained through the energy spectrum could also be obtained for charge density waves (CDW) in finite size calculations. As we will now discuss, the concept of entanglement spectrum has proved to be a power tool to probe these systems.

\subsection{ES for FCI}

\begin{figure}
\centering
\includegraphics[width=0.55\linewidth]{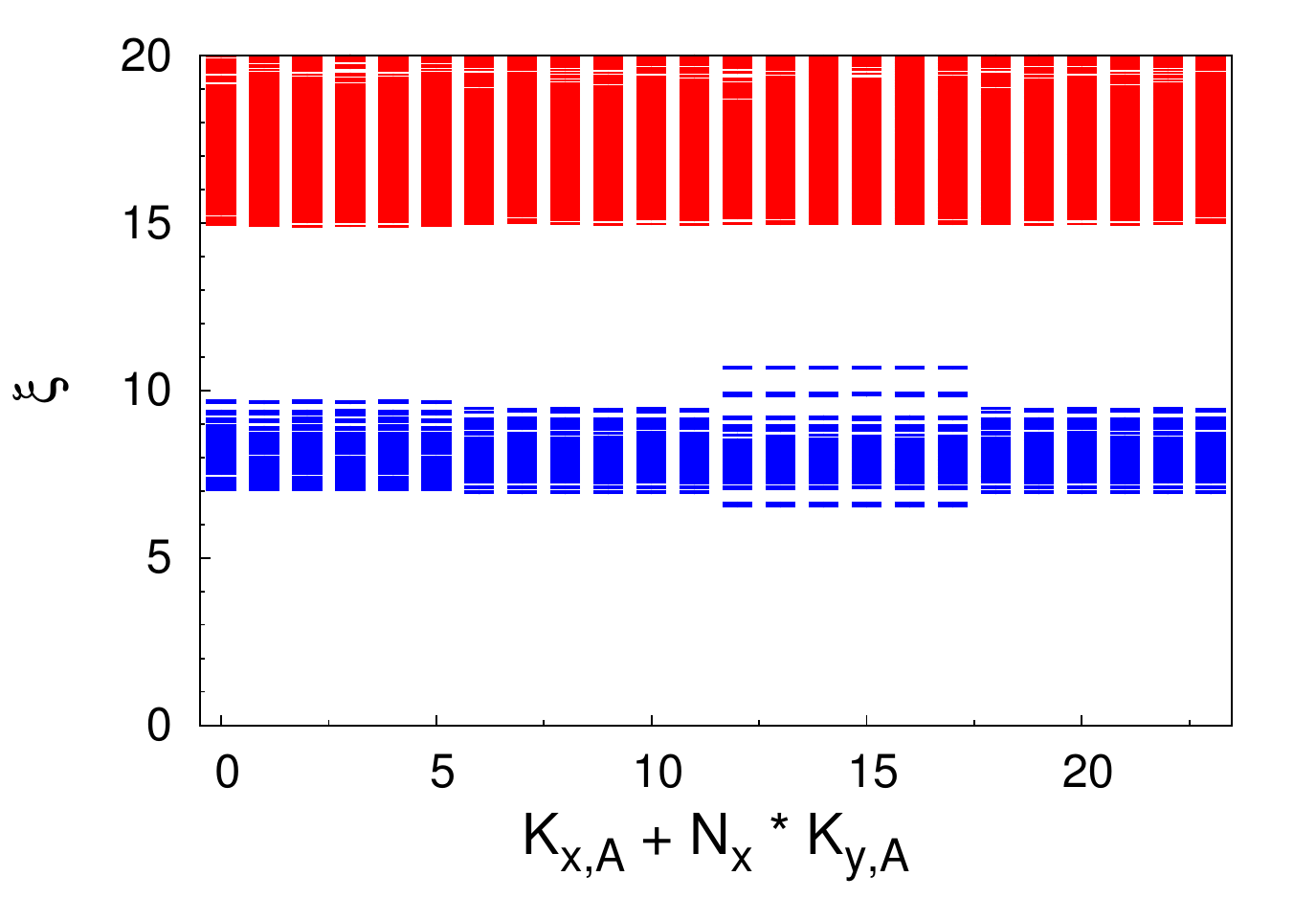}
\caption{PES for using the three lowest energy states of the Kagome FCI model for $N_x \times N_y= 6 \times 4$, keeping $N_A=4$ particles. $K_{x,A}$ and $K_{y,A}$ denote the total momentum in the $x$ and $y$ direction, and are good quantum numbers when performing the PES. There is a clear entanglement gap below which the number of levels (in blue) exactly matches the counting of Laughlin quasihole excitations on a system with 4 fermions on a system with $6 \times 4=24$ flux quanta. The counting per momentum sector below the entanglement gap matches the one predicted by the folding formula of Ref.~\cite{Bernevig-2012PhysRevB.85.075128}.}\label{fig:kagomepes}
\end{figure}

FCI are lattice models and thus one could expect that performing a real space partition is rather trivial. The projection onto the flatten lowest band, which is done in momentum space, makes such a calculation rather non-trivial. Fortunately, the particle entanglement spectrum does not suffer from this issue and can be performed using this specific representation. Indeed, one can apply the same procedure than the FQHE on the torus that we have described in Sec.~\ref{sec:PES}. We will use the same definition for the total density matrix than in Eq.~\ref{eq:densitymatrixtorus}, even if for FCI the degeneracy of the ground state is not exact. In Fig.~\ref{fig:kagomepes}, we present the PES for the almost threefold degenerate ground state of the Fig.~\ref{fig:kagomespectrum}a Kagome system. This PES, that can be plotted as a function of the momenta in both $x$ and $y$ directions, exhibits a clear and large entanglement gap. This PES is reminiscent of the one in Fig.~\ref{fig:pestorus13}b for the Coulomb ground state for the FQHE on a torus. The counting below this gap is exactly the one predicted for a Laughlin-like phase. One could wonder if an overlap calculation would have identified a Laughlin-like state. Writing the Laughlin state on FCI in a suitable way for numerical simulations was a difficult task\cite{qi-PhysRevLett.107.126803,Wu-2012PhRvB..86h5129W}. The results that have been obtained, confirmed what was already concluded from the PES. 

\begin{figure}
\centering
\includegraphics[width=0.53\linewidth]{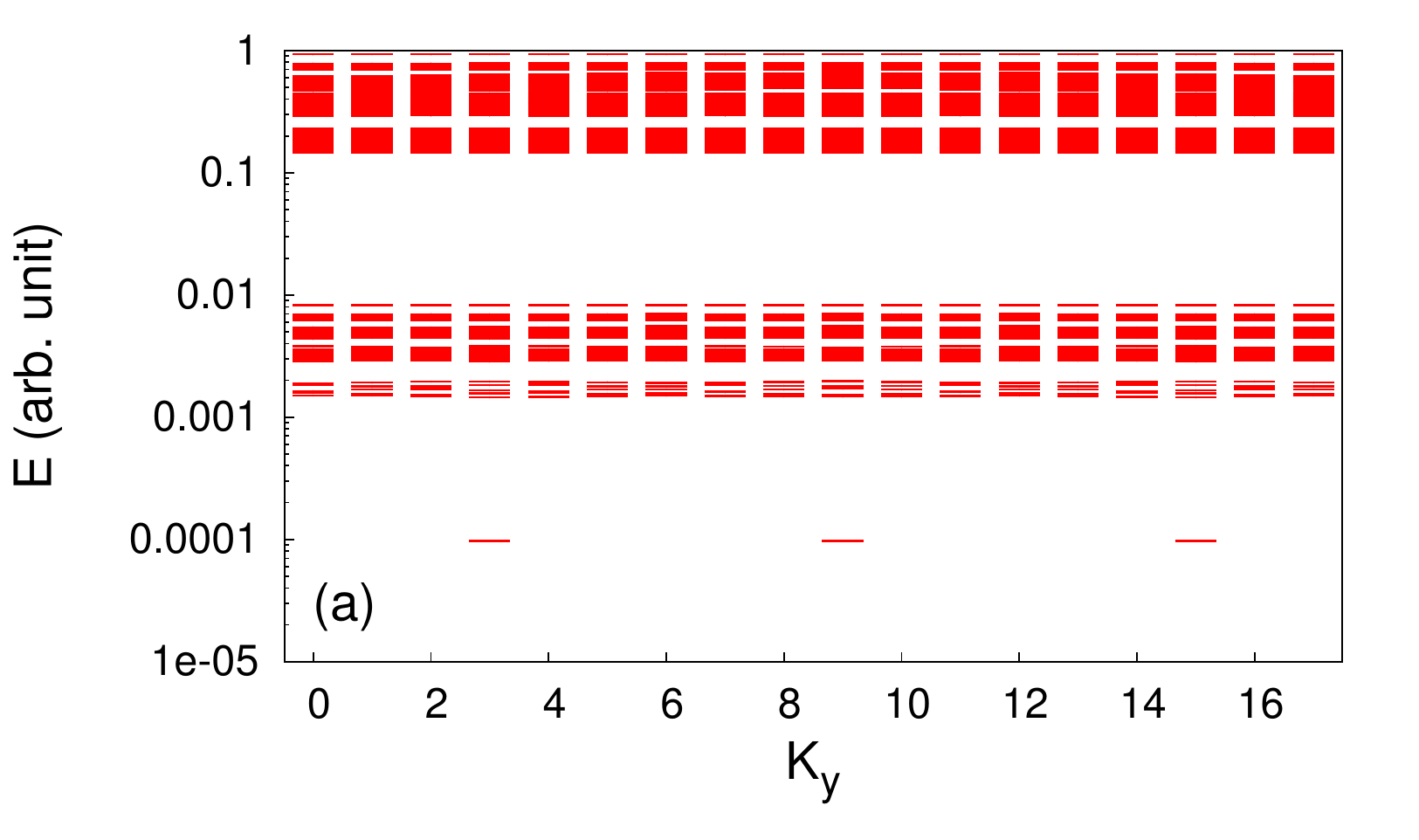}
\includegraphics[width=0.46\linewidth]{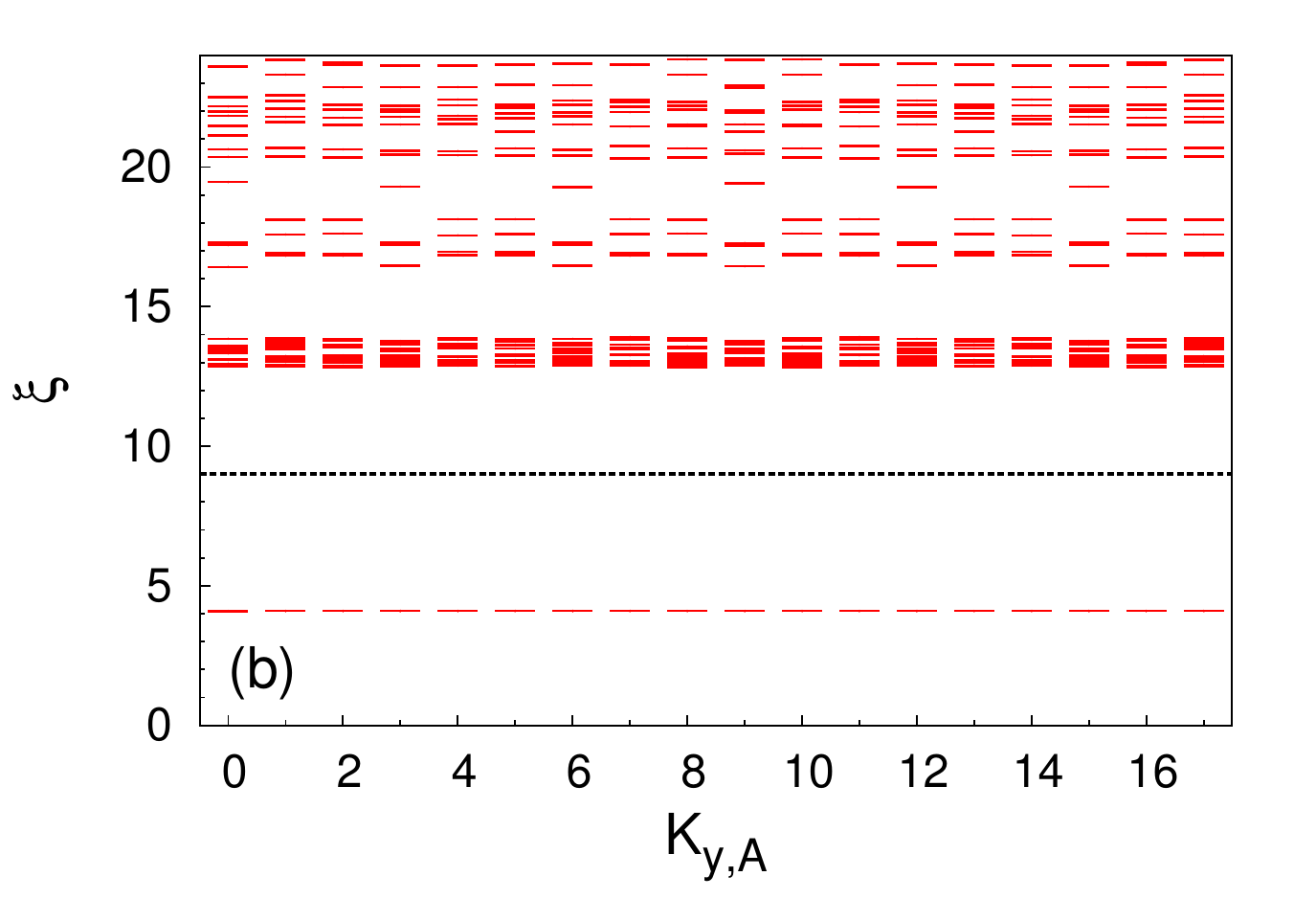}
\caption{{\it Upper panel:} The Kagome model in the one dimensional limit (i.e. $N_x=1$) for $N=6$ and $N_y=18$. While we use a logarithmic scale for the energy, we till observe an almost threefold degenerate ground state. {\it Lower panel:} PES for this almost threefold degenerate ground state, keeping $N_A=3$ particles. There are 59 states per momentum sector below the entanglement gap (depicted by the black dashed line) as expected for a CDW (a Laughlin state would give 2530 states per momentum sector)}\label{fig:kagomethintorus}
\end{figure}

Since these systems could also host CDW-like phases, one can wonder if such a phase would be detected by the PES. A simple way to force the system into a CDW phase consists to consider the one dimensional limit of a FCI\cite{Bernevig-2012arXiv1204.5682B}, keeping only one unit cell in one direction (let say $x$ here). For such a case, the signature from the energy spectra is actually quite similar to the one of a regular, two dimensional, FCI. For example, we still observe a threefold degenerate ground state at filling factor $\nu=1/3$ (see Fig.~\ref{fig:kagomethintorus}a). Performing the PES gives a completely different perspective. As observed in Fig.~\ref{fig:kagomethintorus}b, there is still a large entanglement gap but the counting does not match the one expected for a Laughlin-like state. Indeed, it has been shown that this counting is the one of a CDW\cite{Bernevig-2012arXiv1204.5682B}. 

As a last remark about ES for FCI, we point out that this technique was again quite successful to probe unusual phases. While FCIs share many common features with the FQHE, some striking differences make these systems host some new physics. The most remarkable example is that a single band can have a Chern number $C$ higher than $1$. Indeed, an usual single Landau level carries a Chern number equal to $1$ and thus a completely filled Landau level has a Hall conductance equal to $h/e^2$. This restriction does not apply to Chern insulators. The physics of non-interacting $C>1$ is actually similar to $C$ copies of a Landau level. These systems have been recently investigated numerically\cite{Wang-PhysRevB.86.201101,Yang-PhysRevB.86.241112,Liu-PhysRevLett.109.186805,Sterdyniak-PhysRevB.87.205137}. But the studies through ES \cite{Sterdyniak-PhysRevB.87.205137,Wu-PhysRevLett.110.106802} have revealed that the picture of a simple multi-component FQH-like system breaks down when strong interactions are enabled.

\section{Conclusion}

In these lecture notes, we have discussed some basic concepts of the entanglement spectroscopy and we have shown some of its features. The most remarkable result of the ES is its ability to reveal how much information is encoded within many of the quantum ground states, even for finite size systems. Of course, relations between the ground state (the bulk of the system) and the low energy excitations (the edge modes) were already pointed out before the introduction of the ES. For FQH model wave functions built from a CFT, the equivalence between the CFT of the bulk and the one associated with the edge was already conjectured in Ref.~\cite{Moore:1991p165}. In a similar manner, using the reduced density matrix in strongly correlated systems is at the heart of the density matrix renormalization group\cite{White-PhysRevLett.69.2863}. The fundamental step made by Li and Haldane was to look at the data stored in the reduced density matrix in the right way, guided by the idea that the ES should mimic the energy spectrum of the edge modes.

The fractional quantum Hall effect was a nice sandbox where the concept of ES have been developed and tested. We have seen that several types of bipartition could allow to extract different types of information about the system excitations. While part of these results are still empirical, several steps have been made to give them a more robust analytical support. Maybe the most intriguing concept is still the one of the entanglement gap. For the FQH phases, there is a good understanding of the universal (or low entanglement energy) part. On the other hand, the ``non-universal part'' has also its own structure, related to neutral excitations. But there is still missing a quantitative understanding of the entanglement gap. How large should it be for a phase to be driven by the universal part? Future studies should address this issue.

In the early days of the ES, most of the results were derived from situations where many properties were already known (such as the case of model states). The recent works on fractional Chern insulators have proved that ES can be used as a tool to probe new systems. It helped to discriminate between different phases, especially when no expression for model states was available. Since computing the ES is generally a relatively straightforward numerical calculation, it should now be part of the toolbox used to analyze quantum systems. By picking the right quantum numbers, ES can be a powerful way to unveil the physics hidden in gigabytes of data.

\section*{Acknowledgements}

I would like to acknowledge the organizers of the Les Houches 2014 summer school ``Topological aspects of condensed matter physics'' and the GGI Florence school ``SFT 2015 - Lectures on Statistical Field Theories''. Part of the text is based on my ``habilitation \`a diriger des recherches'' manuscript. Thus I am grateful to N. Cooper, D. Poilblanc, B. Dou\c{c}ot, J. Dalibard, V. Pasquier and P. Lecheminant for being part of my habilitation committee and their comments. I thank A. Sterdyniak for his useful comments about an earlier version of these lecture notes. I was supported by the Princeton Global Scholarship..

\bibliography{es_lecturenotes.bib}

\end{document}